\documentclass[preprint,12pt]{elsarticle}

\usepackage{amssymb}
\usepackage{amsmath}
\usepackage{graphicx}%
\usepackage{multirow}%
\usepackage[title]{appendix}%
\usepackage{xcolor}%
\usepackage{textcomp}%
\usepackage{manyfoot}%
\usepackage{booktabs}%
\usepackage{algorithm}%
\usepackage{algorithmicx}%
\usepackage{algpseudocode}%
\usepackage{listings}%
\usepackage{caption}
\usepackage{subcaption}
\usepackage[bookmarksnumbered=true]{hyperref}
\usepackage{longtable} 
\usepackage{tikz}
\usetikzlibrary{positioning}
\usepackage{geometry}
\journal{arXiv}

\begin{document}

\begin{frontmatter}




\title{Principal Component Stochastic Subspace Identification for Output-Only Modal Analysis}


\author[label1]{Biqi Chen} 

\author[label1]{Jun Zhang} 
\affiliation[label1]{organization={School of Intelligent Civil and Marine Engineering,  Harbin Institute of Technology, Shenzhen},
           addressline={Taoyuan Street}, 
          city={Shenzhen},
            postcode={518055}, 
            state={Guangdong},
            country={China}}

\author[label1,label2,label3]{Ying Wang} 

\affiliation[label2]{organization={Guangdong Provincial Key Laboratory of Intelligent and Resilient Structures for Civil Engineering,  Harbin Institute of Technology, Shenzhen},
            addressline={Taoyuan Street}, 
           city={Shenzhen},
            postcode={518055}, 
           state={Guangdong},
           country={China}}

\affiliation[label3]{organization={School of Sustainability, Civil and Environmental Engineering,  University of Surrey},
           city={Guildford},
            postcode={ GU2 7XH}, 
           state={Surrey},
           country={United Kingdom}}

\ead{yingwang@hit.edu.cn}

\begin{abstract}
Stochastic Subspace Identification (SSI) is widely used in modal analysis of engineering structures, known for its numerical stability and high accuracy in modal parameter identification. SSI methods are generally classified into two types: Data-Driven (SSI-Data) and Covariance-Driven (SSI-Cov), which have been considered to originate from different theoretical foundations and computational principles. In contrast, this study demonstrates that SSI-Cov and SSI-Data converge to the same solution under the condition of infinite observations, by establishing a unified framework incorporating instrumental variable analysis.  
Further, a novel modal identification approach, Principal Component Stochastic Subspace Identification (PCSSI), is proposed based on this framework. This method employs Principal Component Analysis (PCA) to extract key components of the signal subspace and project the observed data onto this space, enhancing modal identification stability while significantly reducing computational complexity. 
Through 5000 Monte Carlo numerical simulations, the statistical analysis shows that PCSSI consistently outperforms traditional SSI methods in terms of numerical stability and noise reduction, demonstrating clear advantages over both SSI-Cov and SSI-Data. Its effectiveness is further validated using experimental data from a scaled bridge model. Compared to conventional SSI approaches, PCSSI demonstrates superior robustness under complex engineering conditions, especially when dealing with limited data or high noise levels, underscoring its strong potential for practical applications.

\end{abstract}

\begin{graphicalabstract}

A schematic representation of the Principal Component Stochastic Subspace Identification method is shown below. The subscript \( s \) denotes the signal component, while \( n \) represents the noise component. The principal signal subspace is extracted by truncating small singular values via singular value decomposition (SVD). Further exploiting the orthogonality between the signal and noise subspaces, the projection of \( Y_f \) onto the denoised subspace \( Y_p^{\text{denoised}} \) using the projection matrix \( V_s V_s^\top \) enables asymptotic noise elimination as the number of observations tends to infinity, thereby improving the robustness and accuracy of the subsequent modal analysis.

\vspace{1em} 
\includegraphics[width=0.95\linewidth]{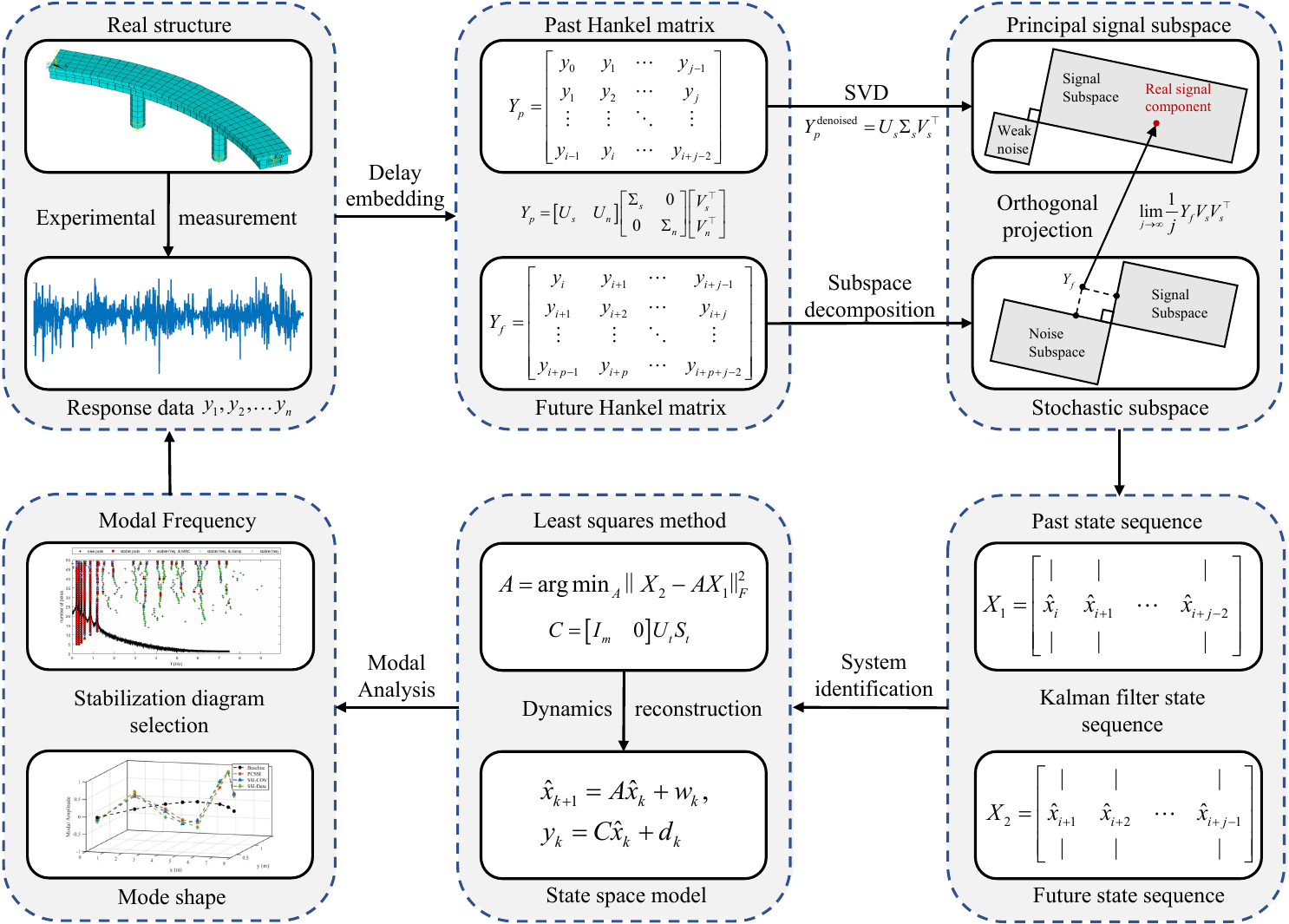}

\end{graphicalabstract}


\begin{highlights}
  \item \textbf{A Unified Stochastic Subspace Identification Framework:} Establishes a theoretical link between SSI-Cov and SSI-Data based on instrumental variables and 
  subspace orthogonality, proving their convergence under infinite observations.

  \item \textbf{A New Stochastic Subspace Identification Method:} Introduces the Principal Component Stochastic Subspace Identification (PCSSI) algorithm, which integrates 
  Principal Component Analysis (PCA) into the stochastic subspace identification process, significantly enhancing noise reduction and achieving lower variance in frequency
   and mode shape estimations.

  \item \textbf{Performance Evaluation under Limited Data:} Evaluates the performance of PCSSI, SSI-Cov, and SSI-Data under limited data conditions based on three perspectives: \textbf{estimation variance} (Monte Carlo simulations), \textbf{numerical stability} (matrix condition numbers), and \textbf{computational efficiency} (computational complexity). 
  The effectiveness of the proposed method is validated through 5000 Monte Carlo simulations and experimental tests on a scaled bridge model.
\end{highlights}


\end{frontmatter}




\section{Introduction}

Structural Health Monitoring (SHM) is essential for assessing the condition of civil engineering structures by analyzing response data to identify both local and 
global structural states \cite{gharehbaghi2022critical}. 
The core challenge is to establish a reliable relationship between these observed responses and the inherent physical parameters—such as stiffness, mass, and damping—that govern structural dynamic behavior. Accurately estimating these parameters is a complex inverse problem, further complicated by environmental variability, sparse sensor placement, and uncertainties in external excitations \cite{chen5097818adaptive}.

Modal parameters, including natural frequencies, damping ratios, and mode shapes, provide an effective and practical means of characterizing the dynamic behavior of 
structures \cite{brownjohn2011vibration}. 
Over the past decades, they have been extensively employed in system identification frameworks for assessing structural conditions and detecting early damage—particularly in bridge engineering. 
Among various modal analysis algorithms, Operational Modal Analysis (OMA) has become a widely studied research direction, comprising a range of methods aimed at identifying modal parameters under ambient or operational excitations. OMA is particularly attractive in situations where applying controlled excitations is either impractical or prohibitively expensive \cite{brownjohn2018bayesian, zahid2020review}.

In general, OMA techniques can be classified into two main categories: frequency-domain methods and time-domain methods \cite{zahid2020review}. Frequency-domain methods, such as Peak-Picking \cite{felber1994development} and Frequency Domain Decomposition (FDD) \cite{brincker2000modal}, typically utilize frequency response functions to derive modal parameters from structural responses. Although these methods demonstrate excellent performance under stationary excitation conditions, their accuracy may degrade significantly when structures are subjected to complex, non-stationary, or environmentally influenced excitations. In contrast, time-domain methods can be summarized as a process of system identification and the determination of modal parameters from the identified system parameters\cite{reynders2008uncertainty}. System identification methods, including Stochastic Subspace Identification (SSI) \cite{peetersREFERENCEBASEDSTOCHASTICSUBSPACE}, the Natural Excitation Technique (NExT) \cite{james1993natural}, and the Eigensystem Realization Algorithm (ERA) \cite{juang1985eigensystem}, analyze response time histories or correlation functions. As a result, they are more suitable for operational conditions where direct measurements of structural excitations are unavailable or challenging.

ERA constructs a Hankel matrix using impulse or free-response data and then applies Singular Value Decomposition (SVD) to obtain a minimal-order realization of the system, from 
which modal parameters are extracted \cite{juang1985eigensystem}. While ERA has demonstrated excellent accuracy in controlled laboratory conditions, its applicability in operational environments is often severely limited due to its strict requirement for well-defined impact excitations.
The NExT method addresses this limitation by computing correlation functions from output-only responses under ambient excitations, thereby obtaining structural responses equivalent 
to those under impact excitation. As a result, NExT is often used in conjunction with ERA. However, NExT requires the input to be wide-sense stationary, and violations of this 
assumption can lead to identification inaccuracies.

Among time-domain OMA approaches, SSI has obtained widespread recognition due to its robustness, computational efficiency, and superior accuracy in modal parameter 
estimation \cite{reynders2016uncertainty}. SSI methods can be broadly categorized into two main variants: covariance-driven SSI (SSI-Cov) and data-driven SSI (SSI-Data). 
The SSI-Cov method utilizes the covariance matrix of output responses, making it particularly suitable for scenarios where response data exhibit wide-sense 
stationary statistical properties. In contrast, SSI-Data relies on QR decomposition to project the future outputs onto the space of past outputs, followed by SVD-based system identification. By avoiding the stationarity assumption required by SSI-Cov, SSI-Data becomes highly effective in non-stationary environments.

It is commonly considered that SSI-Cov and SSI-Data are fundamentally distinct due to their different underlying assumptions and implementation procedures. 
However, based on the instrumental variable-based stochastic subspace method framework presented in Section \ref{IV_SSI}, it can be shown that the fundamental differences between 
SSI-Cov and SSI-Data primarily arise from their respective choices of instrumental variables. Under the theoretical assumption of infinite-time observations, both methods 
converge to identical parameter estimates, as detailed in Section \ref{sec:SSI}.

\begin{figure}[h]
  \centering
  \includegraphics[width=0.9\textwidth]{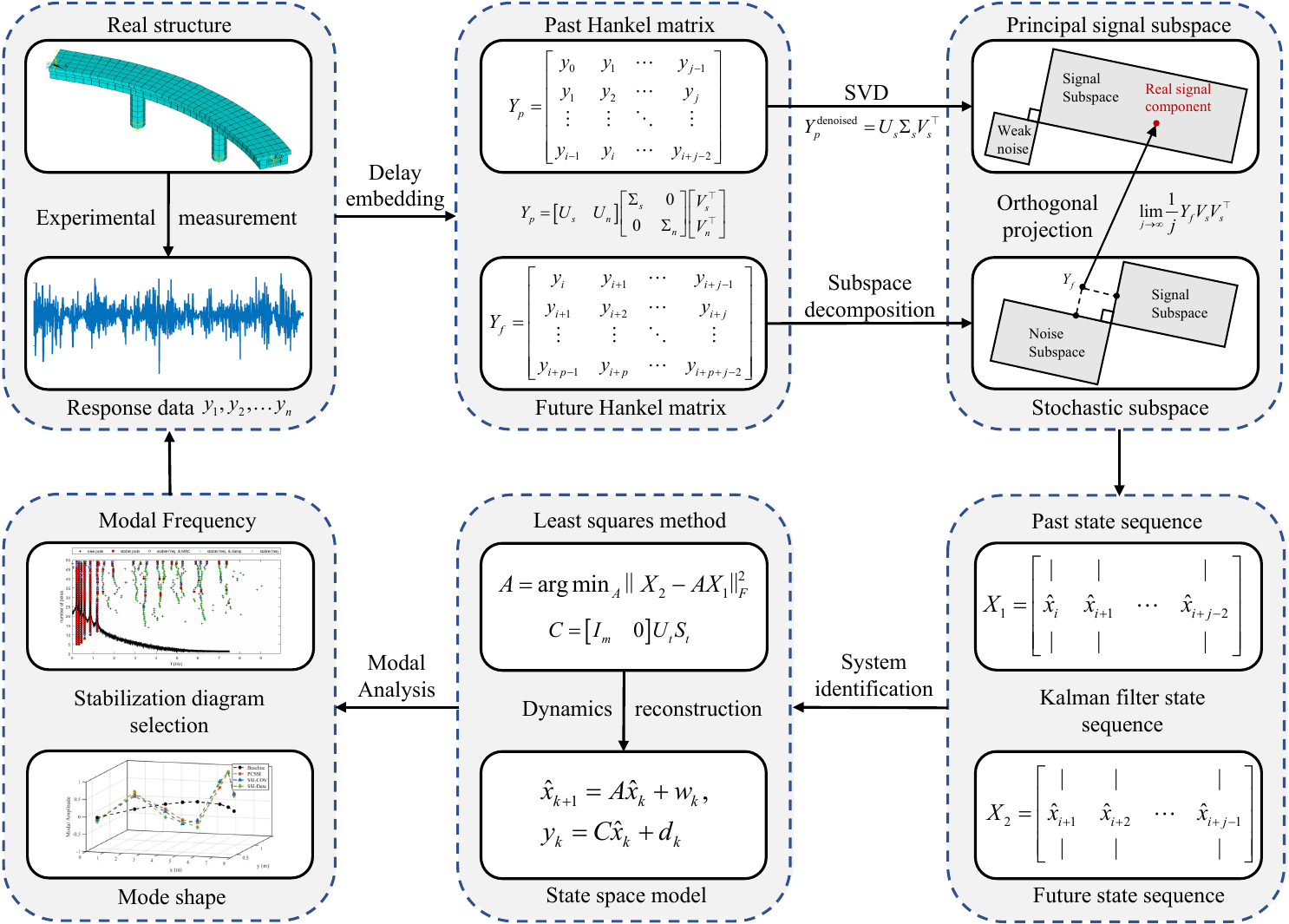}
  \caption{A schematic representation of the Principal Component Stochastic Subspace Identification method. The subscript \( s \) denotes the signal component, while \( n \) represents the noise component. The principal signal subspace is extracted by truncating small singular values via SVD. Further exploiting the orthogonality between the signal and noise subspaces, the projection of \( Y_f \) onto the denoised subspace \( Y_p^{\text{denoised}} \) using the projection matrix \( V_s V_s^\top \) enables asymptotic noise elimination as the number of observations tends to infinity, thereby improving the robustness and accuracy of the subsequent modal analysis.}  
  \label{fig:pcssi}
\end{figure}

Building upon this unified framework, Section \ref{sec:PCSSI} introduces a novel modal identification algorithm called Principal Component Stochastic Subspace Identification (PCSSI). 
The proposed PCSSI method carefully selects instrumental variables by projecting future observations onto the principal signal subspace of past data, significantly enhancing noise 
rejection capabilities (see Figure \ref{fig:pcssi}).
As a result, PCSSI not only retains the robust performance and computational efficiency advantages of classical SSI methods but also exhibits improved stability in frequency 
estimation and mode shape reconstruction, particularly when handling limited monitoring data.

Section \ref{sec:Case} systematically investigates and compares the performance of PCSSI, SSI-Cov, and SSI-Data in modal parameter identification through 5000 Monte Carlo simulations and scaled model experimental data. The results indicate that PCSSI consistently outperforms traditional SSI methods by significantly reducing variance and alleviating issues such as mode underestimation and spurious frequency identification. These findings highlight the distinct theoretical and practical advantages of PCSSI, making it particularly well-suited for real-world applications characterized by complex environmental excitations and high noise levels. Furthermore, Section \ref{numerical_stability} presents a comparative analysis of PCSSI, SSI-Cov, and SSI-Data under limited data conditions, focusing on variance, computational efficiency, and numerical stability. An open-source MATLAB implementation of PCSSI is available on \href{https://github.com/Chen861368/Principal-Component-Stochastic-Subspace-Identification-for-Output-Only-Modal-Analysis}{GitHub}.

\section{Instrumental Variable-Based Stochastic Subspace Identification Framework}\label{IV_SSI}
\subsection{Innovations Form of the State-Space Model}
In this section, a modified version of the classical state-space equation, referred to as the innovations form of the state-space model, is introduced. This form is the system 
model typically identified in the general SSI-Data setting. The innovations form treats the Kalman filter (KF) estimates as the state vectors of the equation.
To begin, we consider the classical state-space model, which is given by:
\begin{equation}\label{eqc_{1}}
    x_{k+1} = A x_k + w_k, \quad y_k = C x_k + d_k,
\end{equation}
where \( x_k \in \mathbb{R}^{n} \) represents the system state vector, and \( y_k \in \mathbb{R}^{m} \) denotes the observation vector, typically obtained from sensors. 
The system matrix \( A \in \mathbb{R}^{n \times n} \) describes the system dynamics, while \( C \in \mathbb{R}^{m \times n} \) represents the observation matrix. 
The terms \(w_k \in \mathbb{R}^{n}\) and \(d_k \in \mathbb{R}^{m}\) 
represent the process noise and measurement noise, respectively.

Given the challenges associated with acquiring complete external load data in civil engineering applications, and considering that external excitations are typically modeled as ambient excitation spectra, both the process noise and measurement noise in this study are modeled as zero-mean Gaussian white noise:
\begin{equation}
    w_k \sim \mathcal{N}(0, W_k), 
    \quad 
    d_k \sim \mathcal{N}(0, D_k),
\end{equation}
where \(W_k \in \mathbb{R}^{n \times n}\) and \(D_k \in \mathbb{R}^{m \times m}\) are the respective covariance matrices of the process noise and measurement noise. For the subsequent derivations of SSI, the state-space equations are reformulated into the innovations form.

The core concept of the KF can be understood as a projection-based state estimation approach\cite{kalman1960new}. Specifically, the KF state estimate can be viewed as the orthogonal projection of the system state \(x_k\) onto the subspace spanned by the observed data \(y_k\). Mathematically, this can be expressed as:
\begin{equation}
    \hat{x}_{k \mid k} = \mathbb{E}\bigl[x_k \mid y_1, \dots, y_k\bigr] = \operatorname{Proj}_{\{y_1, \dots, y_k\}}\bigl(x_k\bigr),
\end{equation}
where \(\hat{x}_{k \mid k}\) represents the KF estimate of the state \(x_k\) based on the measurements \(y_1, \dots, y_k\), and \(\operatorname{Proj}_{\{y_1, \dots, y_k\}}(\cdot)\) denotes the projection of a random variable onto the subspace spanned by \(y_1, \dots, y_k\).
Similarly, for predicting the next state, the estimate is given by:
\begin{equation}
    \hat{x}_{k+1 \mid k} = \mathbb{E}\bigl[x_{k+1} \mid y_1, \dots, y_k\bigr] = \operatorname{Proj}_{\{y_1, \dots, y_k\}}\bigl(x_{k+1}\bigr),
\end{equation}
where \(\hat{x}_{k+1 \mid k}\) denotes the KF estimate of the state \(x_{k+1}\) based on the measurements \(y_1, \dots, y_k\).
As a recursive state estimation framework, the KF optimally combines predictions from the system's dynamic model with real-time measurement updates.
 This fusion of predictions and measurements enables the filter to continuously refine its state estimates, ensuring they remain as close as possible to the true values.

The filtering process is divided into two primary steps: the prediction step and the correction step. In the prediction step, the system dynamics are used to predict the next state and output:
\begin{equation}
    \hat{x}_{k+1|k} = A \hat{x}_{k|k}, \quad \hat{y}_{k|k-1} = C \hat{x}_{k|k-1}.
\end{equation}
Here, \(\hat{x}_{k+1|k}\) represents the KF prior state estimate, \(\hat{x}_{k|k}\) is the KF posterior state estimate, and \(\hat{y}_{k|k-1}\) is the KF predicted output. 
In the correction step, the measurement \(y_k\) is used to update the predicted state estimate, thereby reducing the error between the prediction and the actual observation. The correction formula is given by:
\begin{equation}
    \hat{x}_{k|k} = \hat{x}_{k|k-1} + S_k e_k, \quad e_k \doteq y_k - \hat{y}_{k|k-1}
\end{equation}
where $e_k$ is the innovation term, representing the deviation between the predicted model and the actual observation. \(S_k\) is the Kalman gain matrix, which determines the weight of the correction term in the state estimation.
Combining the prediction and correction steps, the state update equation can be expressed as:
\begin{equation}
    \hat{x}_{k+1|k} = A\hat{x}_{k|k} = A (\hat{x}_{k|k-1} + S_k e_k) = A\hat{x}_{k|k-1} + AS_k e_k.
\end{equation}
This equation combines the dynamic characteristics of the system model with measurement corrections, enabling a recursive estimation 
of the system state. For simplicity in notation, we define \(K_k \doteq AS_k\), which represents the Kalman gain for the innovations form. 
Thus, the equation can be expressed as\cite{reynders2012system}:
\begin{equation}
    \hat{x}_{k+1|k} = A\hat{x}_{k|k-1} + K_k e_k, \quad y_k = C \hat{x}_{k|k-1} + e_k.
\end{equation}
This formulation represents a state-space model where the state is the KF prior estimate, the measurement noise is white noise, 
and the process noise is related to the observation noise through the Kalman gain matrix \(K_k\).
Based on the above state-space model, the recursive representation of the output variable \( y_k \) can be derived. 
The recursive expansion of the output variable can be expressed as:
\begin{align}
y_k &= C \hat{x}_{k|k-1} + e_k, \\
y_{k+1} &= C(A\hat{x}_{k|k-1} + K_k e_k) + e_{k+1}, \\
y_{k+2} &= C(A^2 \hat{x}_{k|k-1} + A K_k e_k + K_{k+1} e_{k+1}) + e_{k+2}, \\
&\vdots \\
y_{k+i-1} &= C(A^{i-1} \hat{x}_{k|k-1} + A^{i-2} K_k e_k + A^{i-3} K_{k+1} e_{k+1} + \cdots + K_{k+i-2} e_{k+i-2}) + e_{k+i-1}.
\end{align}
More generally, for any \( i \in \mathbb{Z} \), the output variable \( y_{k+i} \) can be expressed as:
\begin{equation}\label{eqc_{100}}
    y_{k+i} = CA^i \hat{x}_{k|k-1} + \sum_{j=0}^{i-1} CA^{i-1-j} K_{k+j} e_{k+j} + e_{k+i}.
\end{equation}
Here, \( CA^i \hat{x}_{k|k-1} \) describes the evolution of the estimated system state \( \hat{x}_{k|k-1} \) under the influence of the system dynamics matrix \( A \).  
The term \( \sum_{j=0}^{i-1} CA^{i-1-j} K_{k+j} e_{k+j} \) represents the propagation of noise \( e_k \) through \( K_k \) and \( A \), capturing its cumulative effect on the system response.

\subsection{Noise Modeling and Analysis}

Stochastic subspace methods estimate the system's state-space model by representing the output data in the form of a Hankel matrix \cite{verhaegen1994identification}. 
To facilitate subsequent analysis, the recursive expressions derived earlier in equation \eqref{eqc_{100}} are reformulated into a more compact matrix representation.
We begin by defining the data Hankel matrix \( Y_{k} \), which is constructed from the observed output data \( y_{k}, \ldots, y_{k+i+j-2} \). The structure of this matrix is as follows:
\begin{equation}
     Y_{k} = \begin{bmatrix}
        y_{k} & y_{k+1} & \cdots & y_{k+j-1} \\
        y_{k+1} & y_{k+2} & \cdots & y_{k+j} \\
        \vdots & \vdots & \ddots & \vdots \\
        y_{k+i-1} & y_{k+i} & \cdots & y_{k+i+j-2}
        \end{bmatrix}
\end{equation}
where \( Y_{k} \in \mathbb{R}^{im \times j} \), with each column representing \( i \) onsecutive observations that can be further expressed in a recursive form based on the parameters of the state-space model.
For simplicity, \(\hat{x}_{k|k-1}\) is abbreviated as \(\hat{x}_k\), allowing the 
observation data to be expressed in matrix form as\cite{reynders2021uncertainty}:
\begin{equation}
    \underbrace{
\begin{bmatrix}
y_k \\
y_{k+1} \\
\vdots \\
y_{k+i-1}
\end{bmatrix}}_{y_i[k]}
=
\underbrace{
\begin{bmatrix}
C \\
CA \\
\vdots \\
CA^{i-1}
\end{bmatrix}
}_{O_i}
\hat{x}_k
+
\underbrace{
\begin{bmatrix}
I & 0 & 0 & \cdots & 0 \\
CK_k & I & 0 & \cdots & 0 \\
\vdots & \vdots & \vdots & \ddots & \vdots \\
CA^{i-2} K_k & CA^{i-3} K_{k+1} & \cdots & CK_{k+i-2} & I
\end{bmatrix}
}_{F_i}
\underbrace{
\begin{bmatrix}
e_k\\
e_{k+1} \\
\vdots \\
e_{k+i-1}
\end{bmatrix}}_{e_i[k]}.
\end{equation}
For simplicity, the above expression can be compactly written as:
\begin{equation}
    y_i[k] = O_i \hat{x}_k + F_i e_i[k], \quad Y_k = \begin{bmatrix}
        y_i[k], 
        y_i[k+1],
        \cdots,
        y_i[k+j-1]
    \end{bmatrix},
\end{equation}
Here, \(y_i[k] \in \mathbb{R}^{im}\) represents the extended observation vector, spanning from \(y_k\) to \(y_{k+i-1}\). \(O_i \in \mathbb{R}^{im \times n}\) denotes the extended observation matrix, while \(F_i \in \mathbb{R}^{im \times im}\) represents the noise influence matrix. The vector \(e_i[k] \in \mathbb{R}^{im}\) captures the extended noise components.
To further express this, the KF estimates and the noise data can be organized into the following matrix form:
\begin{equation}
    \hat{X}_k = 
    \begin{bmatrix}
    \hat{x}_k & \hat{x}_{k+1} & \hat{x}_{k+2} & \cdots & \hat{x}_{k+j-1}
    \end{bmatrix}, \quad 
    E_k = 
    \begin{bmatrix}
    e_i[k] & e_i[k+1] & \cdots & e_i[k+j-1]
    \end{bmatrix}. 
\end{equation}
Here, \(\hat{X}_k \in \mathbb{R}^{n \times j}\) represents the KF state sequence matrix, while \(E_k \in \mathbb{R}^{im \times j}\) represents the noise matrix. 
 Under the assumption of the innovations form of the state-space model, the output 
 Hankel matrix of the system can be expressed as:
\begin{align}
    Y_{k} &= O_i \hat{X}_{k} + F_i E_{k}\\
    &= (O_i  T )(T^{-1}  \hat{X}_{k}) + F_i E_{k}\\
    &=  \begin{bmatrix}
        C \\
        CA \\
        \vdots \\
        CA^{i-1}
            \end{bmatrix}   T  \cdot T^{-1}   \begin{bmatrix}
            \hat{x}_k & \hat{x}_{k+1} & \hat{x}_{k+2} & \cdots & \hat{x}_{k+j-1}
            \end{bmatrix}+ F_i E_{k} 
\end{align}
where \( T \in \mathbb{R}^{n \times n} \) is an invertible linear transformation. From the last row, it is clear 
that \( O_i \hat{X}_{k} \) can be decomposed in multiple ways through \( T \) and can also be expressed as the product of two  matrices.
And it is not difficult to see that\cite{van2012subspace}:
\begin{equation}
    \text{column space}(O_i \hat{X}_{k}) = \text{column space}(O_i),\quad \text{row space}(O_i \hat{X}_{k}) =\text{row space}( \hat{X}_{k})   .
\end{equation}
Therefore, at this point, \( \text{rank}(O_i \hat{X}_{k}) \) should remain \( n \), exhibiting a low-rank characteristic:
\begin{equation}
    \text{rank}(O_i) = \text{rank}(\hat{X}_{k}) = \text{rank}(O_i \hat{X}_{k}) = n,\quad  \text{rank}(Y_k) = \text{rank}(F_i E_k ) = \min(im, j).
\end{equation}
Here, \( \text{rank}(Y_k) \) is primarily controlled by the noise, resulting in a high-rank characteristic.
For simplicity in subsequent analysis, we assume that \(e_t\)  are zero-mean white 
noises with an ergodic property, characterized by\cite{verhaegen1994identification}:
\begin{equation}
    \lim_{N \to \infty} \frac{1}{N} \sum_{i=0}^{N} e_{j+i} e_{k+i}^\top = \mathbb{E}[e_j e_k^\top] = E_{jk} \delta_{jk},
\end{equation}
where \( E_{jk} \) is the covariance matrix of the noise, and \( \delta_{jk} \) is 
the Kronecker delta, implying that the noise is uncorrelated over 
time (i.e., \( E_{jk} = 0 \) for \( j \neq k \)) and its covariance is constant 
for \( j = k \).

\subsection{Instrumental Variable Stochastic Subspace Identification}
To ensure consistency with the notation used in the classical SSI method described 
in \cite{peetersREFERENCEBASEDSTOCHASTICSUBSPACE}, we define:
\begin{equation}\label{eqc_{2}}
    Y_p  =Y_0= O_i \hat{X}_{0} + F_i E_{0}  , \quad Y_f = Y_i= O_i \hat{X}_{i} + F_i E_{i}
\end{equation}
where \(Y_p\) represents the past outputs and \(Y_f\) represents the future outputs.
For simplicity of notation, this paper does not explicitly define the 
reference output \(y_k^{\text{ref}} \in \mathbb{R}^{r \times 1}\) as used 
in the classical SSI framework\cite{peetersREFERENCEBASEDSTOCHASTICSUBSPACE}. Instead, we directly define \(Y_p\) to represent the aggregated past outputs. However, it will be evident in the subsequent analysis that this simplification does not affect the conclusions of the analysis.

\begin{figure}[h]
    \centering
    \includegraphics[width=0.7\textwidth]{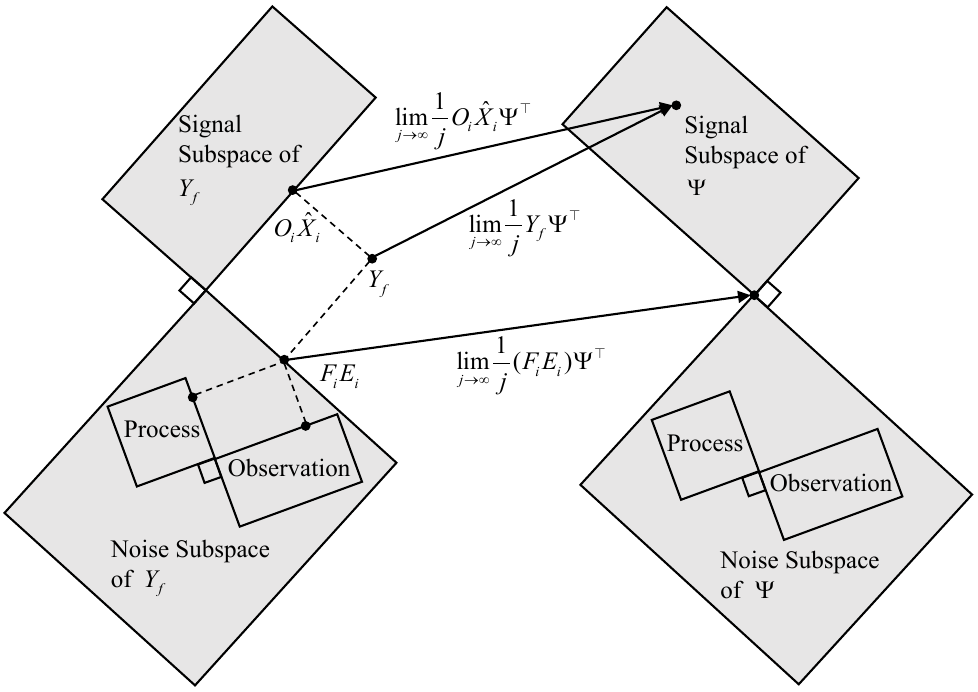}
    \caption{A schematic representation of the effect of instrumental variables on 
    stochastic subspaces.}\label{fig:ssk1}
\end{figure}

In the formulation described in Equation \eqref{eqc_{2}}, although all variables on the right-hand side of \( Y_f \) are unknown, it is 
clear that  the signal 
subspace \(\text{span}( O_i \hat{X}_i)\) and the noise subspace \(\text{span}(F_i E_i)\) are statistically independent. Leveraging this 
relationship between the signal and noise subspaces, the instrumental variable (IV) 
method can be applied to eliminate the noise term 
\( F_i E_i \) on the right-hand side, thereby isolating the primary signal components 
of the system\cite{ottersten1994subspace,li2012tracking,gustafsson2001subspace,viberg1997analysis}.
Specifically, an instrumental variable matrix \( \Psi^\top \in \mathbb{R}^{j \times p} \) is introduced, satisfying the following constraints:
\begin{equation}\label{eqc_{6}}
    \mathbb{E}\left [ F_i E_i \Psi^\top \right ]  = 0, \quad \text{rank}\left(\mathbb{E}\left [ O_i \hat{X}_i \Psi^\top \right ] \right) = n.
\end{equation}
These constraints ensure that the subspace \(\text{span}(\Psi^\top)\) is statistically orthogonal to the noise subspace \(\text{span}(F_iE_i)\), and that the instrumental variable matrix captures the full information of the signal subspace \(\text{span}(O_i \hat{X}_i)\). Geometrically, the operation \(Y_f \cdot \Psi^\top\) effectively projects the noise components onto zero, leaving only the signal subspace \(\text{span}(O_i \hat{X}_i)\) in the final results.
Considering the ergodicity assumption of the white noise process, the instrumental variable assumptions can be reformulated as:
\begin{equation}\label{eqc_{17}}
    \lim_{j \to \infty}\frac{1}{j} (F_iE_i) \Psi^\top = 0, \quad \text{rank}\left(\lim_{j \to \infty} \frac{1}{j} O_i \hat{X}_i  \Psi^\top\right) = n.
\end{equation}
The independence of the noise components \( e_i \) at different sampling points allows for the selection of observation vectors \( y_i \) outside the interval \( Y_f \). If the subspace \(\Psi\) is chosen 
to satisfy equation \eqref{eqc_{17}}, then as \( j \to \infty \), 
the influence of noise on the normalized projection \(\frac{1}{j}(Y_f \Psi^\top)\) 
becomes negligible, leaving only the signal subspace.
Figure \ref{fig:ssk1} illustrates the relationships among various subspaces. In the figure, the orthogonality between subspaces represents statistical independence. 

\begin{figure}[htbp]
    \centering
    \resizebox{\linewidth}{!}{
    \begin{tikzpicture}[>=stealth]

        \draw[->,thick] (-0.8,0) -- (13,0) node[anchor=north] {Time};

        \foreach \x/\label in {0.5/0, 2.5/$\dots$,   4.5/$i-1$, 6/$i$,8/$\dots$,10/$2i-1$,11.5/$\dots$}
            \draw[thick] (\x,0.1) -- (\x,-0.1) node[anchor=north] {\label};

        \draw[blue,thick,->,bend left=50] (0.5,0) to (4.5,0);
        \node[blue,anchor=south] at (2.5,0.9) {past};

        \draw[blue,thick,->,bend left=50] (6,0) to (10,0);
        \node[blue,anchor=south] at (8,0.9) {future};

        \node at (2.5,-0.8) {$Y_p$};
        \node at (8,-0.8) {$Y_f$};
    \end{tikzpicture}
    }
    \caption{Segmentation of observation time intervals. }
    \label{fig26}
\end{figure}
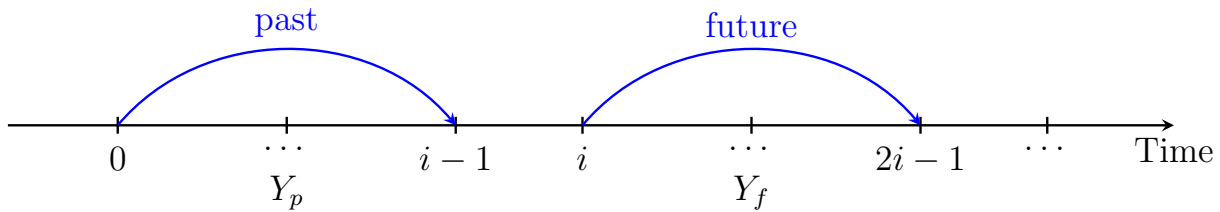

This provides a conceptual understanding of how instrumental variables interact with stochastic subspaces.
In practice, the subspace \( Y_f \) can be decomposed into two mutually independent components:
\begin{align}
    \text{span}(Y_f)&=  \text{signal subspace }(Y_f) \oplus   \text{noise subspace }(Y_f)  \\
    &=\text{span}(O_i \hat{X}_i ) \oplus \text{span}(F_iE_i) 
\end{align}
where \( \oplus \) denotes the direct sum of subspaces.
Similarly, the subspace \(\text{span}(\Psi)\) can be decomposed as:
\begin{equation}
    \text{span}(\Psi) = \text{signal subspace}(\Psi) \oplus   \text{noise subspace}(\Psi)
\end{equation}
Moreover, based on equation \eqref{eqc_{6}}, the signal subspace within \( Y_f \) aligns with that of \(\Psi\), 
while the noise subspace of \( Y_f \) is orthogonal to \(\Psi\). Mathematically, this relationship can be expressed as:
\begin{equation}
  \text{span}(O_iX_i) =   \text{signal subspace}(\Psi),\quad   \text{span}(F_iE_i) \perp   \text{span}(\Psi)
\end{equation}
As shown in Figure \ref{fig26}, the elements of \( \Psi \) can be arbitrarily selected from observation vectors outside the future 
time interval, provided that the resulting subspace \(\text{span}(\Psi)\) fully captures the information contained in \( O_i \hat{X}_i  \).

\section{Theoretical Relationship between SSI-Cov and SSI-Data}\label{sec:SSI}
In this section, we utilize the previously introduced instrumental variable-based SSI framework to explain the relationship between SSI-Cov and SSI-Data, demonstrating why both methods theoretically converge to a unique solution when the number of observations approaches infinity. 
\subsection{Covariance-Driven Stochastic Subspace Identification}
\begin{figure}[h]
    \centering
    \includegraphics[width=0.7\textwidth]{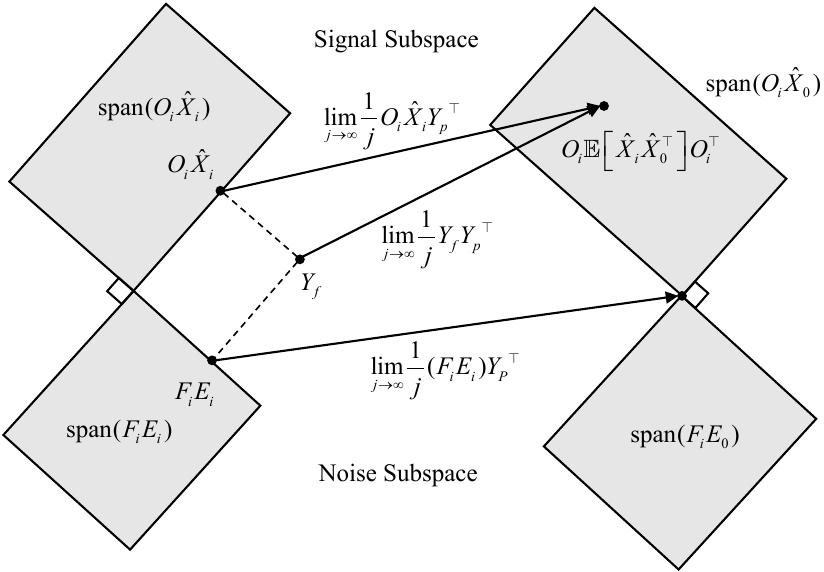}
    \caption{A schematic representation of the instrumental variable-based SSI-Cov method.}\label{fig:ssk}
\end{figure}

To establish the connection between the classical SSI-Cov method and the instrumental variable framework, we select the sequentially 
arranged matrix \( Y_p \) as the instrumental variable matrix \( \Psi \):  
\begin{equation}
    \Psi = Y_p=\begin{bmatrix}
        y_{0} & y_{1} & \cdots & y_{j-1} \\
        y_{1} & y_{2} & \cdots & y_{j} \\
        \vdots & \vdots & \ddots & \vdots \\
        y_{i-1} & y_{i} & \cdots & y_{i+j-2}
        \end{bmatrix}=O_i \hat{X}_{0} + F_i E_{0} 
\end{equation}
The data construction begins with \( y_0 \), ensuring that all elements in \( Y_p \) are 
independent of \( F_i E_i \) \cite{verhaegen1994identification}. 
Notably, \( Y_p \) contains a significant amount of redundant information. To reduce 
computational complexity, the classical SSI-Cov method selects a subset of reference 
vectors from \( Y_p \) to form \( Y_p^{\text{ref}} \), rather than utilizing the 
entire data matrix.
Despite this reduction, \( Y_p^{\text{ref}} \) is still required to satisfy the condition specified in equation \eqref{eqc_{6}}. Therefore, the 
following proof 
applies equally to \( Y_p^{\text{ref}} \).
To reinterpret the meaning of \( Y_f Y_p^\top \) in the classical SSI-Cov method within the framework of instrumental variables, we proceed as follows:  
\begin{align}
    \lim_{j \to \infty} \frac{1}{j}Y_f\Psi^\top &= \lim_{j\to \infty} \frac{1}{j} (O_i \hat{X}_i  + F_iE_i)Y_p^\top \\
    &= \lim_{j \to \infty} \frac{1}{j} (O_i \hat{X}_i  + F_iE_i)(O_i \hat{X}_0 + F_iE_{0})^\top \\
    &= \lim_{j \to \infty} \frac{1}{j} \Big(
    \underbrace{O_i \hat{X}_i \hat{X}_0^\top O_i^\top}_{T_1} + 
    \underbrace{O_i \hat{X}_i E_0^\top F_i^\top + F_i E_i \hat{X}_0^\top O_i^\top }_{T_2} + 
    \underbrace{F_i E_i E_0^\top F_i^\top}_{T_3} \Big).
\end{align}
The first term represents the signal component. Assuming that the state 
vector \(\hat{x}_k\) satisfies 
ergodicity properties, the time average can replace the spatial average, leading to:
\begin{equation}
    T_1 = \lim_{j \to \infty}\frac{1}{j}O_i \hat{X}_i \hat{X}_0^\top O_i^\top 
     = O_i \left( \lim_{j \to \infty} \frac{1}{j} \sum_{t=0}^{j-1} \hat{x}_t \hat{x}_{i+t}^\top \right)O_i^\top 
     = O_i \mathbb{E} \left[ \hat{X}_i \hat{X}_0^\top \right]O_i^\top
\end{equation}
This result follows from Birkhoff's Ergodic Theorem\cite{petersen1989ergodic}, which 
ensures that the infinite-time average of \(\hat{x}_k\) is equivalent to their spatial 
average over the state space.
Since the noise is assumed to have zero mean and is uncorrelated with the signal, 
the noise-signal cross terms vanish:
\begin{align}
    T_2  &= \lim_{j \to \infty} \frac{1}{j} \Big( O_i \hat{X}_i E_0^\top F_i^\top + F_i E_i \hat{X}_0^\top O_i^\top\Big) \\
    &= O_i \left( \lim_{j \to \infty} \frac{1}{j} \hat{X}_i E_0^\top \right) F_i^\top 
    + F_i \left( \lim_{j \to \infty} \frac{1}{j} E_i \hat{X}_0^\top \right) O_i^\top  \\
    &= O_i \, \mathbb{E} \big[ \hat{X}_i E_0^\top \big] F_i^\top + F_i \, \mathbb{E} \big[ E_i \hat{X}_0^\top \big] O_i^\top \\
    &= 0.
\end{align}
For the noise-noise term, the independence and zero-mean assumptions imply:
\begin{align}
    T_3 &=\lim_{j \to \infty}\frac{1}{j} \left( F_i E_i E_0^\top F_i^\top\right) =   F_i \left( \lim_{j \to \infty}\frac{1}{j}   E_i E_0^\top \right)     F_i^\top               \\
    & =  F_i \cdot  \lim_{j \to \infty}   \frac{1}{j} \sum_{k=0}^{j-1}       \begin{bmatrix}
        e_{i+k} \\ e_{i+k+1} \\ \vdots \\ e_{i+k+i-1}
        \end{bmatrix}
        \cdot
        \begin{bmatrix}
        e_k^\top & e_{k+1}^\top & \cdots & e_{k+i-1}^\top
        \end{bmatrix} \cdot  F_i^\top         \\
&=  F_i   
\begin{bmatrix}
    \mathbb{E}[e_i e_0^\top] & \mathbb{E}[e_i e_1^\top] & \cdots & \mathbb{E}[e_i e_{i-1}^\top] \\
    \mathbb{E}[e_{i+1} e_0^\top] & \mathbb{E}[e_{i+1} e_1^\top] & \cdots & \mathbb{E}[e_{i+1} e_{i-1}^\top] \\
    \vdots & \vdots & \ddots & \vdots \\
    \mathbb{E}[e_{2i-1} e_0^\top] & \mathbb{E}[e_{2i-1} e_1^\top] & \cdots & \mathbb{E}[e_{2i-1} e_{i-1}^\top]
    \end{bmatrix}   F_i^\top                       \\
    &= 0.
\end{align}
In conclusion, in the large sample limit, the result retains only the signal component:
\begin{equation}\label{eqc_{16}}
    \lim_{j \to \infty} \frac{1}{j}Y_fY_p^\top  = O_i \mathbb{E}  \left[  \hat{X}_i  \hat{X}_0^\top   \right]O_i^\top
\end{equation}
This analysis shows that as \(j \to \infty\), the instrumental variable \(Y_p\) 
effectively eliminates all cross terms and noise components, leaving only the component 
governed by the true system dynamics.This remaining component 
can then be used to identify the system parameters.
 The above analysis is illustrated in Figure \ref{fig:ssk}.
    
Due to the conditions specified in Equations \eqref{eqc_{6}} 
and \eqref{eqc_{16}}, the application of the matrix \( Y_p\) 
ensures that the information within the signal subspace is preserved. 
Consequently, the column space of \( O_i \) remains unchanged.
\begin{equation}
    \text{column space}(O_i ) = \text{column space}(O_i \hat{X}_i \hat{X}_0^\top O_i^\top). 
\end{equation}
To further analyze the subspace, similar to the classical SSI method, we consider a compact SVD of the 
term \( \frac{1}{j} Y_f Y_p^\top \) \cite{peetersREFERENCEBASEDSTOCHASTICSUBSPACE}:  
\begin{equation}
    \frac{1}{j}Y_fY_p^\top = U \Sigma V^\top =
    \begin{bmatrix}
    U_1 & U_2
    \end{bmatrix}
    \begin{bmatrix}
    \Sigma_1 & 0 \\ 
    0 & 0
    \end{bmatrix}
    \begin{bmatrix}
    V_1^\top \\ 
    V_2^\top
    \end{bmatrix}
    = U_1 \Sigma_1 V_1^\top.  
\end{equation}
Here, \( U_1 \in \mathbb{R}^{im \times n} \) and \( V_1^\top \in \mathbb{R}^{n \times j} \) are the left and right singular matrices corresponding to the 
nonzero singular values, while \( \Sigma_1 \in \mathbb{R}^{n \times n} \) is a diagonal matrix containing the \( n \) largest singular values. Since the 
column spaces of \( O_i \) and \( O_i \hat{X}_i \hat{X}_0^\top O_i^\top \) are equivalent, \( O_i \) and \( U_1 \) differ only by an invertible linear 
transformation \( T \). For consistency with the classical SSI-Cov method, the decomposition method used in SSI is adopted here, leading to the following 
representation\cite{peetersREFERENCEBASEDSTOCHASTICSUBSPACE}:
\begin{equation}\label{eqc_{11}}
    O_i  =\begin{bmatrix}
        C, 
        CA, 
        \cdots, 
        CA^{i-1}
    \end{bmatrix}^\top  =U_1 \Sigma_1^{1/2}
\end{equation}
After obtaining the extended observability matrix \( O_i \), the system parameters 
can be extracted using an algorithm consistent with the SSI-Cov method. 
The specific details of the algorithm are not elaborated further in this paper.  
Additionally, the classical SSI method assumes that the state variables \( x_k \) are zero-mean stationary stochastic processes and derives the 
system matrices \( A \) and \( C \) based on this assumption from the perspective of the covariance matrix.

However, although this study assumes only that the noise is Gaussian white noise, it achieves stronger conclusions under this more relaxed assumption. 
The results indicate that the state-space model derived through \(\lim_{j \to \infty} \frac{1}{j} Y_f Y_p^\top\) ultimately attains an estimation accuracy 
equivalent to that obtained from the state sequence generated by Kalman filtering.

\subsection{Data-Driven Stochastic Subspace Identification}
This section provides a deeper explanation of why the Kalman filter estimate appears in the classical SSI-Data 
method \cite{peetersREFERENCEBASEDSTOCHASTICSUBSPACE} and explores its connection with the instrumental variable framework.  
To begin, we present an alternative perspective for analyzing the instrumental variable approach:  
\begin{align}
    \frac{1}{j} Y_f\Psi^\top &= \frac{1}{j} (O_i \hat{X}_i + F_iE_i)\Psi^\top \\  
    &= \frac{1}{j} (O_i \hat{X}_i \Psi^\top) + O_j(\varepsilon),  
\end{align}  
where \( O_j(\varepsilon) \) is a bounded matrix of appropriate dimensions with norm \( \varepsilon \), which vanishes as \( j \to \infty \).
By utilizing the pseudoinverse to eliminate \(\Psi^\top\) from the right-hand 
side of the equation, it can be rewritten as:
\begin{equation}
    \frac{1}{j} Y_f\Psi^\top (\Psi^\top)^{\dagger} = \frac{1}{j} O_i \hat{X}_i\Psi^\top (\Psi^\top)^{\dagger}  + O_j(\varepsilon)(\Psi^\top)^{\dagger}.
\end{equation}  
Here, the matrix \( \Psi^\top (\Psi^\top)^{\dagger} \in \mathbb{R}^{j \times j} \) acts 
as a projection matrix that effectively projects the rows of \( Y_f \) onto the row 
space of \( \Psi \). Additional details can be found in  \ref{projection}.  
Taking the limit on both sides yields:  
\begin{equation}\label{eqc_{12}}
    \lim_{j \to \infty} \frac{1}{j} Y_f\Psi^\top (\Psi^\top)^{\dagger} = O_i \hat{X}_i,
\end{equation}  
This result can be regarded as a generalization of the classical SSI-Data method. To establish its connection with \cite{peetersREFERENCEBASEDSTOCHASTICSUBSPACE}, 
we consider \( Y_p \) as \( \Psi \) and apply the SVD decomposition (using the analysis in \ref{projection}), leading to:  
\begin{equation}\label{eqc_{14}}
      Y_p^\top (Y_p^\top)^{\dagger}  = V_p V_p^\top= \Pi_{Y_p} , \quad Y_p = U_p S_p V_p^\top.
\end{equation}  
Here, \( U_p \in \mathbb{R}^{im \times n} \) and \( V_p^\top \in \mathbb{R}^{n \times j} \) are the compact left and right singular matrices of \( Y_p \), 
while \( \Sigma_p \in \mathbb{R}^{n \times n} \) is a diagonal matrix containing the largest \( n \) nonzero singular values of \( Y_p \).
The matrix \( \Pi_{Y_p} \) represents the projection matrix, which projects the row space of \( Y_f \) onto \( Y_p \), as detailed in \ref{projection}. 
It can be expressed as:  
\begin{equation}
    \Pi_{Y_p} = Y_p^\top (Y_p Y_p^\top)^\dagger Y_p=V_p V_p^\top
\end{equation}
Substituting this into equation \eqref{eqc_{12}}, we get:  
\begin{equation}\label{eqc_{13}}
    \lim_{j \to \infty} \frac{1}{j}Y_f\Pi_{Y_p} = \lim_{j \to \infty} \frac{1}{j} Y_f V_p V_p^\top  = O_i \hat{X}_i.
\end{equation}  
Apart from notation differences, this result is entirely consistent with the classical SSI formulation. Therefore, we can also consider a compact SVD decomposition as follows:
\begin{equation}
    \frac{1}{j} Y_f\Pi_{Y_p} = \frac{1}{j} Y_f V_p V_p^\top = U_1 \Sigma_1 V_1^\top.
\end{equation}  
Here, \( U_1 \in \mathbb{R}^{im \times n} \) and 
\( V_1^\top \in \mathbb{R}^{n \times j} \) represent the left and right singular 
matrices, while \( \Sigma_1 \) is a diagonal matrix containing the \( n \) 
largest singular values. 
Thus, we can express \( \hat{X}_i \) as:  
\begin{equation}
    \hat{X}_i = \begin{bmatrix}
        \hat{x}_i & \hat{x}_{i+1} & \hat{x}_{i+2} & \cdots & \hat{x}_{i+j-1}
    \end{bmatrix} = \Sigma_1^{1/2} V_1^\top.
\end{equation}  
The subsequent steps for analysis and derivation can be found in the classical SSI paper. Based on the above analysis, we can see that as \( j \to \infty \), noise 
interference is completely eliminated, making the performance of the SSI-Cov and SSI-Data methods fundamentally equivalent.  

Thus, the core principle of the SSI method lies in leveraging the orthogonality between different subspaces. Theoretically, there is considerable freedom in the 
choice of instrumental variables. However, in practice, only a limited amount of monitoring data \( j \) is available, which may lead to numerical instability. 
Consequently, the selection of \( \Psi \) and the numerical algorithms used can slightly influence the method’s performance, as discussed in \ref{numerical_stability}.

\section{Principal Component Stochastic Subspace Identification}\label{sec:PCSSI}
To improve the performance of classical SSI methods, PCSSI is proposed in this study, which aims to enhance its noise reduction capability and computational efficiency under limited data conditions.
Clearly, \( Y_p \) contains both signal and noise subspaces, where larger singular 
 values correspond to the signal and smaller ones correspond to noise. Therefore, 
 the SVD decomposition in Equation \eqref{eqc_{12}} can be further truncated to 
 retain only the top \( k \) singular values, capturing the most significant signal 
 features and producing a denoised version:
\begin{equation}
    Y_p^{\text{denoised}} = 
\begin{bmatrix}
U_s & U_n
\end{bmatrix}
\begin{bmatrix}
\Sigma_s & 0 \\ 
0 & 0
\end{bmatrix}
\begin{bmatrix}
V_s^\top \\ 
V_n^\top
\end{bmatrix}
= U_s \Sigma_s V_s^\top,
\end{equation}
where \( \Sigma_s \in \mathbb{R}^{k \times k} \), and \( k \) represents the 
possible maximum model order chosen. Due to the effect of noise, it is generally 
true that \( n \leq k \ll \text{rank}(Y_p) \). \( U_s \in \mathbb{R}^{im \times k} \) 
and \( V_s^\top \in \mathbb{R}^{k \times j} \) are the left and right singular
matrices corresponding to the larger singular values.

Although identifying an appropriate truncation threshold for the singular values can 
be challenging when noise levels are high, within the framework of instrumental 
variables, we do not require \( Y_p^{\text{denoised}} \) to contain only noise-free 
signal subspace information. What we need is for \( \Psi \) to retain the complete 
signal subspace information while minimizing the noise components as much as possible. 
Therefore, we can select \( V_s^\top \) as \( \Psi \), which also satisfies the 
condition in equation \eqref{eqc_{6}}. Based on equation \eqref{eqc_{12}}, it can be 
computed as:
\begin{equation}
    \lim_{j \to \infty} \frac{1}{j} Y_f V_s V_s^\top = O_i \hat{X}_i.
\end{equation}
Here, \( V_s V_s^\top \in \mathbb{R}^{j \times j} \) represents a projection 
matrix similar to \( \Psi^\top (\Psi^\top)^{\dagger} \).
Based on this expression, an SVD can be performed:
\begin{equation}
    \frac{1}{j} Y_f V_s V_s^\top = U_k S_k V_k^\top =
\begin{bmatrix} u_1 & u_2 & \cdots & u_k \end{bmatrix}
\begin{bmatrix}
\sigma_1 & 0 & \cdots & 0 \\ 
0 & \sigma_2 & \cdots & 0 \\ 
\vdots & \vdots & \ddots & \vdots \\ 
0 & 0 & \cdots & \sigma_k \\ 
\end{bmatrix}
\begin{bmatrix} 
v_1^\top \\ 
v_2^\top \\ 
\vdots \\ 
v_k^\top 
\end{bmatrix}.
\end{equation}
Here, \( U_k \in \mathbb{R}^{im \times k} \) and \( V_k^\top 
\in \mathbb{R}^{k \times j} \) are the left and right singular matrices, 
and \( \Sigma_k \in \mathbb{R}^{k \times k} \) contains the largest singular 
values.The classical SSI method employs a matrix decomposition similar to equation \eqref{eqc_{11}}, 
where \( O_i = U_k S_k^{1/2} \) and \( \hat{X}_i = \Sigma_k^{1/2} V_k^\top \).  
To reduce computational effort when constructing the state-space equation, we adopt a matrix decomposition approach different from the classical SSI method. 
Specifically, we use the following decomposition:  
\begin{equation}
    O_i = U_k S_k, \quad \hat{X}_i = \begin{bmatrix}
        \hat{x}_i & \hat{x}_{i+1} & \hat{x}_{i+2} & \cdots & \hat{x}_{i+j-1}
    \end{bmatrix} = V_k^\top.
\end{equation}
By selecting the first \( j-1 \) columns and the last \( j-1 \) columns of \( V_k^\top \), we can obtain:
\begin{equation}
    X_1 = \begin{bmatrix}
        \hat{x}_i & \hat{x}_{i+1} & \hat{x}_{i+2} & \cdots & \hat{x}_{i+j-2}
    \end{bmatrix}, \quad 
    X_2 = \begin{bmatrix}
        \hat{x}_{i+1} & \hat{x}_{i+2} & \hat{x}_{i+3} & \cdots & \hat{x}_{i+j-1}
    \end{bmatrix},
\end{equation}
where \( X_1, X_2 \in \mathbb{R}^{k \times (j-1)} \). Using the least squares algorithm, the system matrix \( A \) can be estimated as:
\begin{equation}
    A = \arg\min \| X_2 - AX_1 \|_F^2 = X_2 X_1^\dagger = X_2 X_1^\top.
\end{equation}
The observation matrix \( C \) can be estimated as:
\begin{equation}
    C = \begin{bmatrix} I_m & 0 \end{bmatrix} O_i,
\end{equation}
where \( A \in \mathbb{R}^{k \times k} \) and \( C \in \mathbb{R}^{m \times k} \)
 represent the state-transition matrix and the first \( m \) rows of \( O_k \),
  respectively.
  Since \( \hat{X}_i = V_k^\top \), there is no need to compute the pseudoinverse of \( X_1 \); a simple transpose operation is sufficient. Overall, 
  compared to the classical SSI-Data method, this approach significantly reduces computational complexity.
To construct a stabilization diagram, for instance, when calculating system models from \(m \sim k\), the following components can be directly selected:  
\begin{equation}
    U_t =  \begin{bmatrix} u_1 & u_2 & \cdots & u_t\end{bmatrix}, \quad 
   S_t =
    \begin{bmatrix}
    \sigma_1 & 0 & \cdots & 0 \\ 
    0 & \sigma_2 & \cdots & 0 \\ 
    \vdots & \vdots & \ddots & \vdots \\ 
    0 & 0 & \cdots & \sigma_t \\ 
    \end{bmatrix}, \quad 
    V_t^\top = \begin{bmatrix} 
    v_1^\top \\ 
    v_2^\top \\ 
    \vdots \\ 
    v_t^\top 
    \end{bmatrix},
\end{equation}
Where \( m \leq t \leq k \), \( U_t \) represents the first \( t \) columns of \( U_k \), 
\( S_t \) contains the first \( t \) singular values, and \( V_t^\top \) corresponds 
to the first \( t \) rows of \( V_k^\top \).
Using the previously described algorithm, the system matrix 
\( A \in \mathbb{R}^{t \times t} \) and the observation matrix 
\( C \in \mathbb{R}^{m \times t} \) can be obtained by selecting the first \( m \) 
rows of \( U_t S_t \).
To enhance the flexibility of the PCSSI algorithm, we made a slight modification to the definition of $Y_p$ based on previous work. Now, the number of rows is $p$, which 
does not necessarily have to match the number of rows $i$ of $Y_p$. This modification is summarized in Algorithm \ref{algo:improved_ssi}.

\begin{algorithm} [H]
  \caption{Principal Component Stochastic Subspace Identification Method}\label{algo:improved_ssi}
  \begin{algorithmic}[1]
  \Require Observational data \( y_i \) and parameter \( k \) (maximum model order)
  \Ensure Modal parameters of the system obtained via stabilization diagram  
  
  \State \textbf{Step 1}: Construct Hankel data matrices \( Y_f \) and \( Y_p \):
  \[
      Y_p = \begin{bmatrix}
          y_{0} & y_{1} & \cdots & y_{j-1} \\ 
          y_{1} & y_{2} & \cdots & y_{j} \\ 
          \vdots & \vdots & \ddots & \vdots \\ 
          y_{i-1} & y_{i} & \cdots & y_{i+j-2}
          \end{bmatrix}, \quad Y_f = \begin{bmatrix}
      y_i & y_{i+1} & \cdots & y_{i+j-1} \\ 
      y_{i+1} & y_{i+2} & \cdots & y_{i+j} \\ 
      \vdots & \vdots & \ddots & \vdots \\ 
      y_{i+p-1} & y_{i+p} & \cdots & y_{i+p+j-2}
      \end{bmatrix}.
  \]
  
  \State \textbf{Step 2}: Perform economy SVD on \( Y_p \in \mathbb{R}^{p \times j}\), retaining the top \( k \) singular values:
  \[
  Y_p = U_p \Sigma_p V_p^\top, \quad Y_p^{\text{denoised}} = U_s \Sigma_s V_s^\top,
  \]
  
  \State \textbf{Step 3}: Choose the instrumental variable \( V_s V_s^\top\) and compute:
  \[
  \frac{1}{j} Y_f V_s V_s^\top = U_k S_k V_k^\top.
  \]
  
  \For{$t = m$ to $k$}
      \State Construct the extended observability matrix and KF state sequence:
      \[
      O_t = U_t S_t, \quad \hat{X}_i = \begin{bmatrix}
          \hat{x}_i & \hat{x}_{i+1} & \hat{x}_{i+2} & \cdots & \hat{x}_{i+j-1}
      \end{bmatrix} = V_t^\top.
      \]
      \State Construct data matrices \( X_1 \) and \( X_2 \) by selecting columns of \( \hat{X}_i \):
      \[
      X_1 = 
      \begin{bmatrix}
      \hat{x}_i & \hat{x}_{i+1} & \cdots & \hat{x}_{i+j-2}
      \end{bmatrix}, \quad 
      X_2 = 
      \begin{bmatrix}
      \hat{x}_{i+1} & \hat{x}_{i+2} & \cdots & \hat{x}_{i+j-1}
      \end{bmatrix}.
      \]
      \State Estimate the system dynamics matrices:
      \[
      A = \arg\min_{A} \| X_2 - AX_1 \|_F^2 = X_2 X_1^\top, \quad 
      C = \begin{bmatrix} I_m & 0 \end{bmatrix} U_t S_t.
      \]
      \State Perform modal decomposition based on \( A \) and \( C \).
  \EndFor
  
  \State \textbf{Step 4}: Use the modal decomposition results from \( t \in [m, k]  \) to construct a stabilization diagram.
  
  \end{algorithmic}
  \end{algorithm}

Notably, based on the above analysis, the reference-based SSI method can essentially be regarded as a manual selection process for extracting the primary 
signal components, thereby reducing noise interference and achieving better results than traditional SSI\cite{peetersREFERENCEBASEDSTOCHASTICSUBSPACE}. In contrast, PCSSI employs the PCA method to 
automatically select the main signal components, making it more convenient and potentially yielding superior results compared to the reference-based SSI method.

\section{Validation of the Proposed Algorithm: Numerical Simulation and Experimental Data}\label{sec:Case}
Due to the inherent difficulty in obtaining true modal parameters for real structures, validation typically relies on cross-comparisons between alternative methods or 
reference solutions derived from physical models.   
Therefore, this section first presents numerical simulations with known true values to compare the performance of PCSSI, SSI-Cov, and SSI-Data. This is followed by 
experimental validation using a laboratory-scaled model to evaluate the three methods under real-world conditions. 
Additionally, a numerical simulation examining 
their performance with a large monitoring data volume \( j \) is provided in \ref{numerical_stability}.

\subsection{Numerical Validation}

To evaluate the performance of PCSSI, we use a well-established dataset generated by Cheynet et al. \cite{cheynet2016buffeting,cheynet2017damping}, based on a 
simplified model of the Lysefjord Suspension Bridge The dataset includes displacement responses recorded by five sensors distributed along the length of the bridge. 
A major challenge in modal identification for structural health monitoring is the presence of measurement noise. To simulate this, we introduce 70\% non-stationary Gaussian white 
noise into the clean data. Additionally, we conduct 5000 Monte Carlo simulations to statistically analyze the modal analysis results. This setup enables a systematic comparison 
of PCSSI, SSI-Cov, and SSI-Data.

\begin{figure}[!ht]
    \centering
    \begin{subfigure}{0.29\textwidth}
        \centering
        \includegraphics[width=\textwidth]{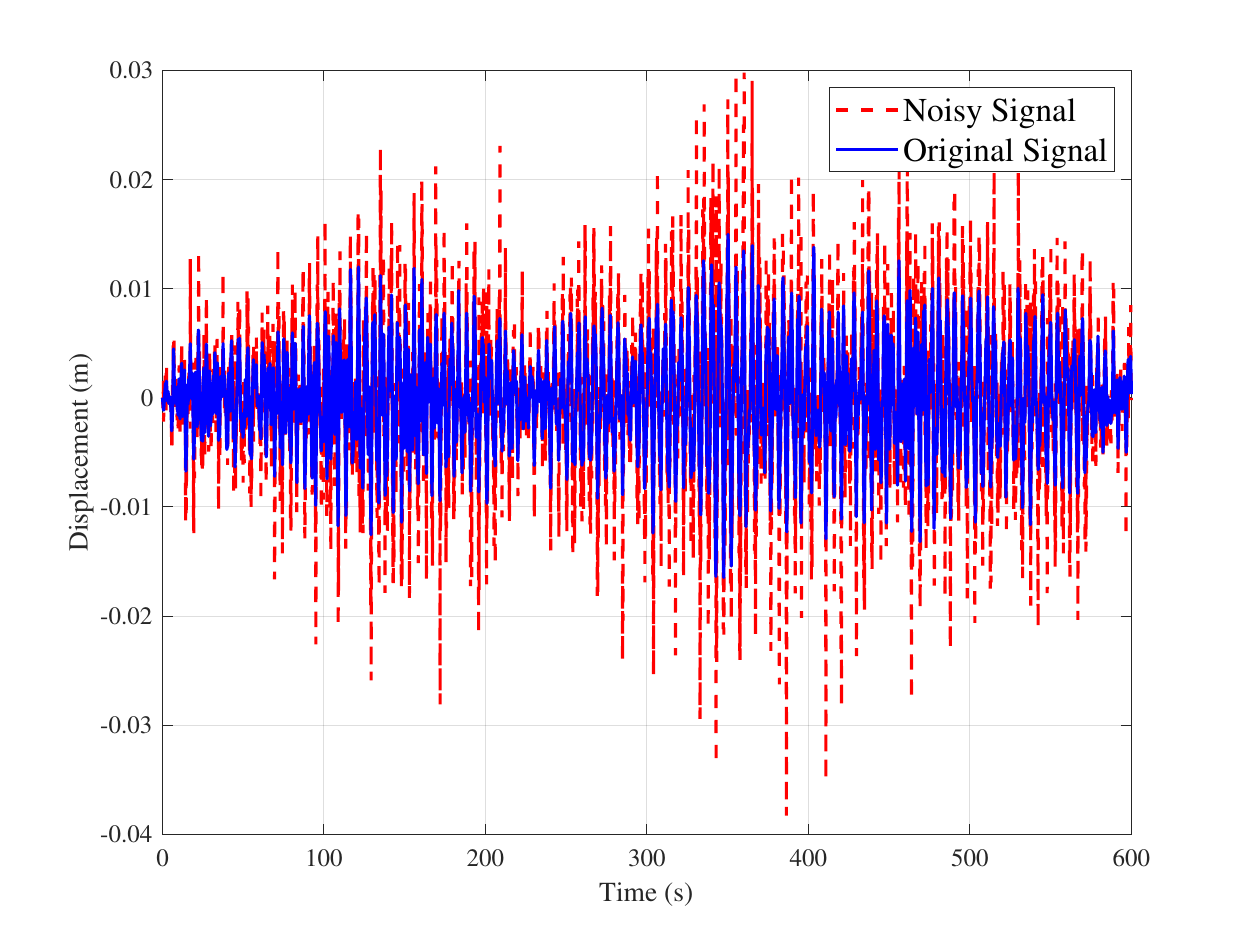}
        \caption{Displacement at Sensor 1}
    \end{subfigure}
    \hspace{2em}
    \begin{subfigure}{0.29\textwidth}
        \centering
        \includegraphics[width=\textwidth]{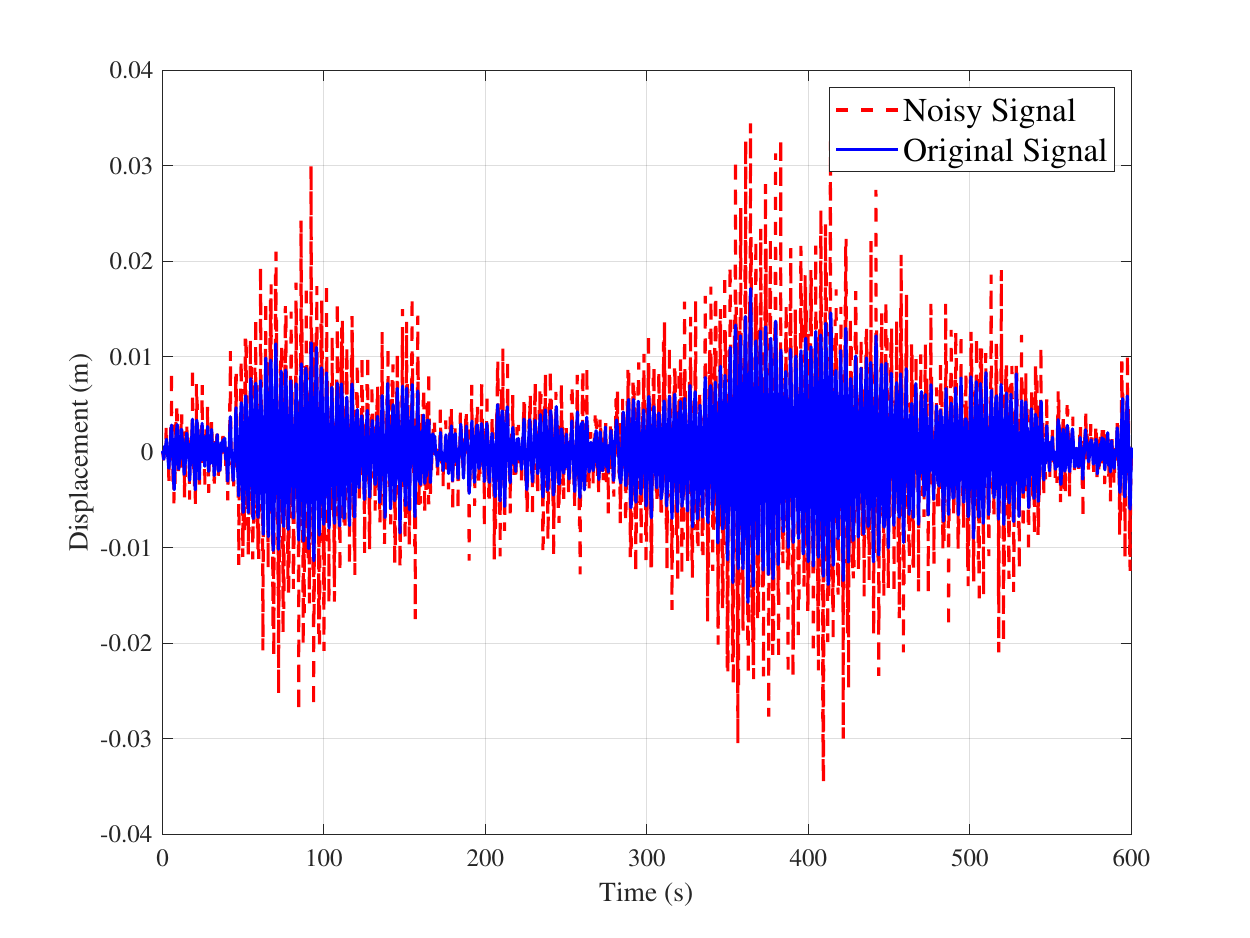}
        \caption{Displacement at Sensor 3}
      \end{subfigure}
      \hspace{2em}
      \begin{subfigure}{0.29\textwidth}
          \centering
          \includegraphics[width=\textwidth]{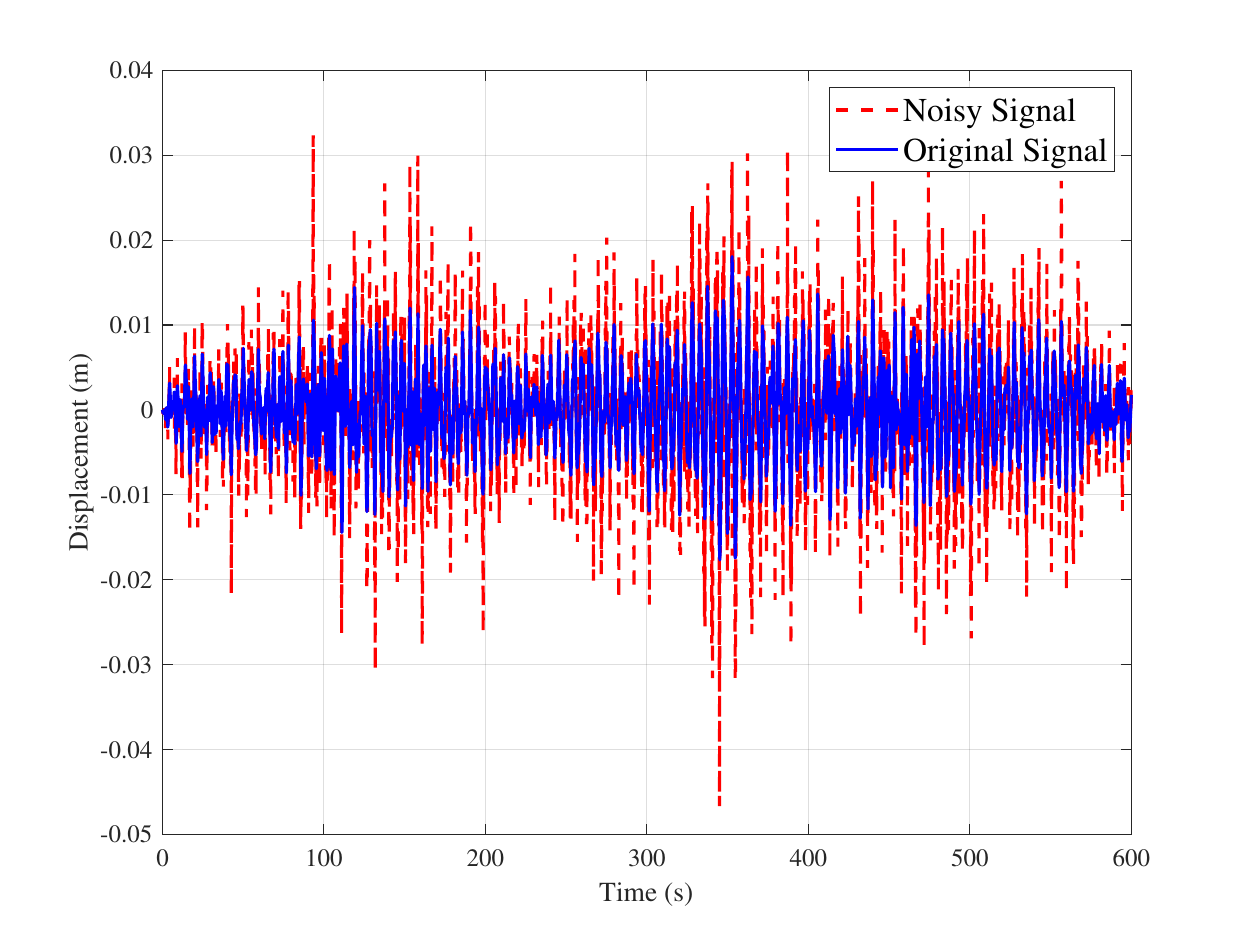}
          \caption{Displacement at Sensor 5}
    \end{subfigure}
  
    \vspace{0.5em}
    \begin{subfigure}{0.29\textwidth}
        \centering
        \includegraphics[width=\textwidth]{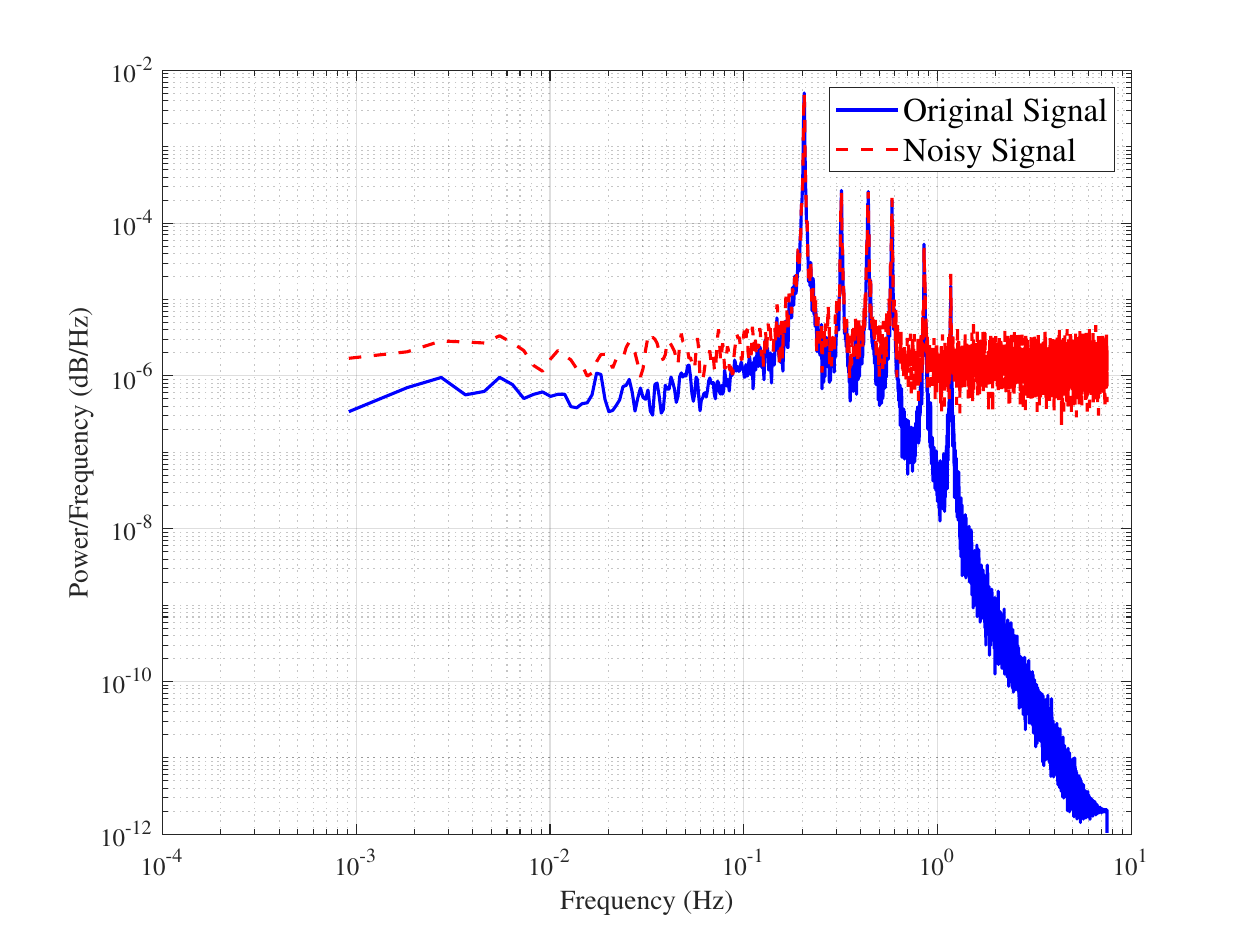}
        \caption{Power Spectrum of Sensor 1}
    \end{subfigure}
    \hspace{2em}
    \begin{subfigure}{0.29\textwidth}
        \centering
        \includegraphics[width=\textwidth]{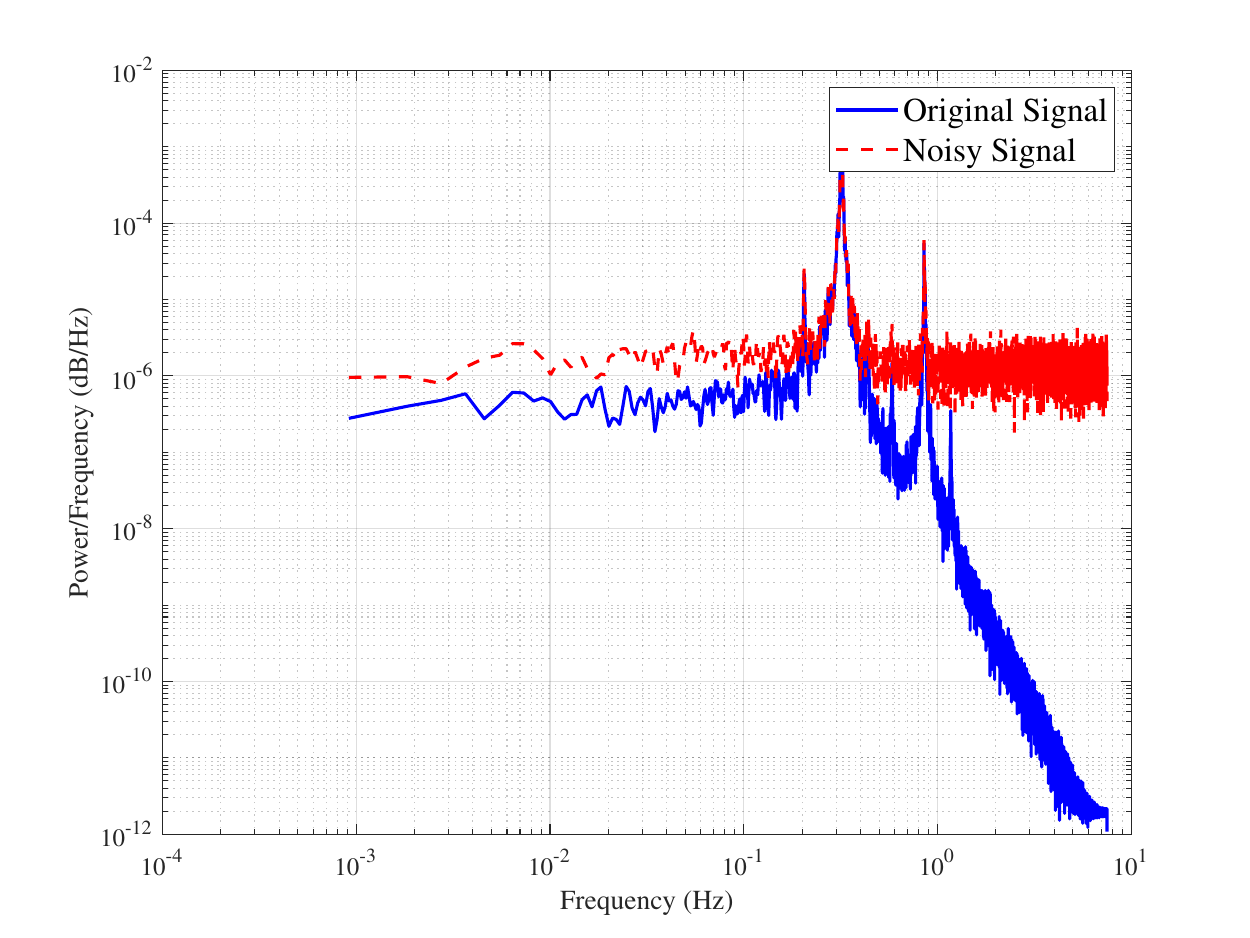}
        \caption{Power Spectrum of Sensor 3}
    \end{subfigure}
    \hspace{2em}
    \begin{subfigure}{0.29\textwidth}
      \centering
      \includegraphics[width=\textwidth]{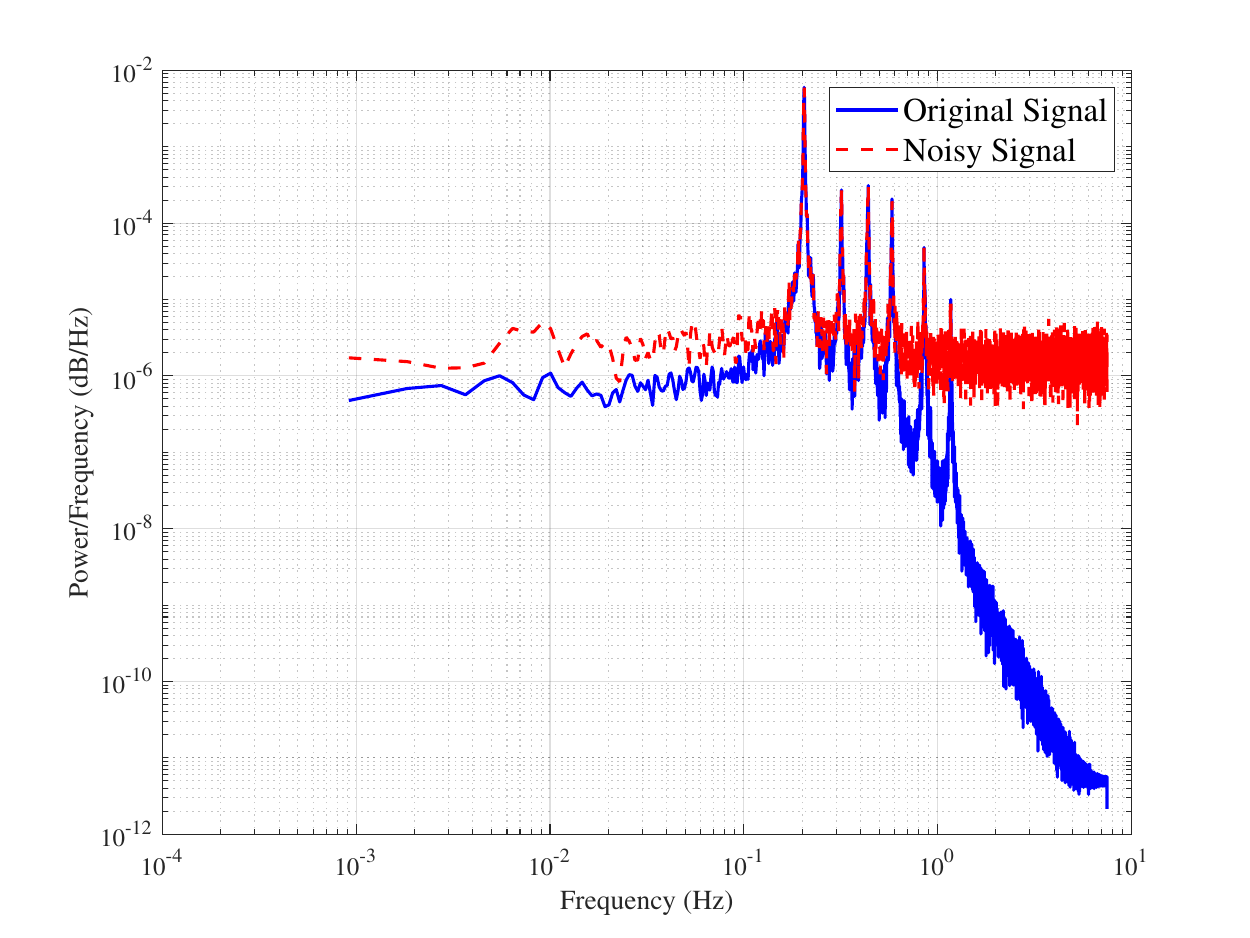}
      \caption{Power Spectrum of Sensor 5}
  \end{subfigure}
  \caption{Comparison of the original displacement signal and the noisy displacement signal: Time response and power spectrum for the Lysefjord bridge MDOF}
    \label{fig:13}
  \end{figure}

Figure \ref{fig:13} provides a comparison of time-domain and power spectral characteristics between clean and noisy signals. The original displacement records 
(subfigures a-c) exhibit consistent low-frequency oscillations, reflecting the bridge's inherent dynamic behavior. However, with added noise (subfigures d-f), the signal-to-noise 
ratio (SNR) degrades significantly, as evidenced by the dominance of broadband noise components in the power spectra. This degradation underscores the challenges faced by modal 
identification algorithms in real-world scenarios.

\subsubsection{Modal Frequency Analysis Based on Monte Carlo Simulation}

\begin{figure}[!ht]
    \centering
    \begin{subfigure}{0.29\textwidth}
        \centering
        \includegraphics[width=\textwidth]{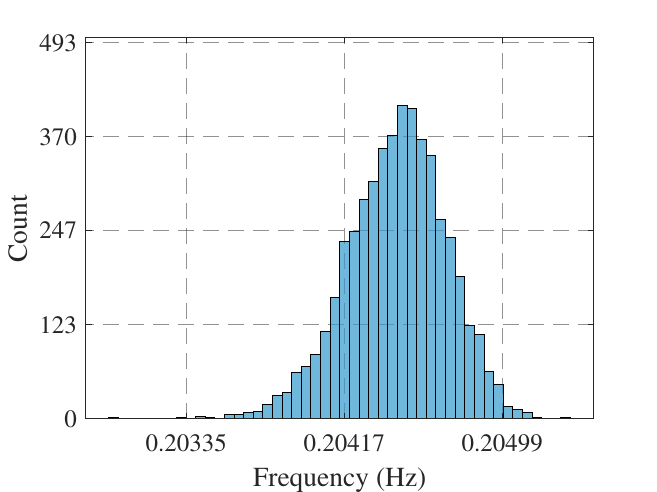}
        \caption{First-order  of PCSSI}
    \end{subfigure}
    \hspace{2em} 
    \begin{subfigure}{0.29\textwidth}
        \centering
        \includegraphics[width=\textwidth]{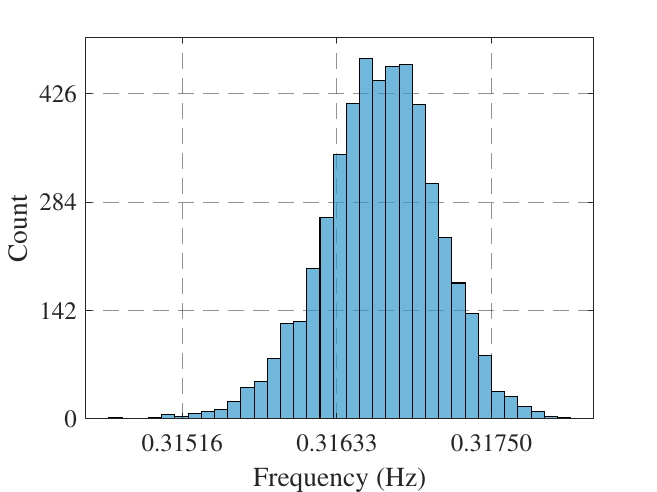}
        \caption{Second-order of PCSSI}
    \end{subfigure}
    \hspace{2em} 
    \begin{subfigure}{0.29\textwidth}
        \centering
        \includegraphics[width=\textwidth]{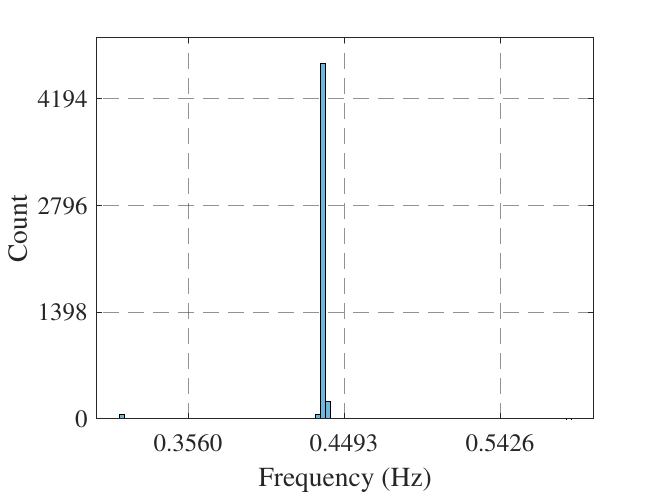}
        \caption{Third-order  of PCSSI}
    \end{subfigure}

    \vspace{0.5em} 

    \begin{subfigure}{0.29\textwidth}
        \centering
        \includegraphics[width=\textwidth]{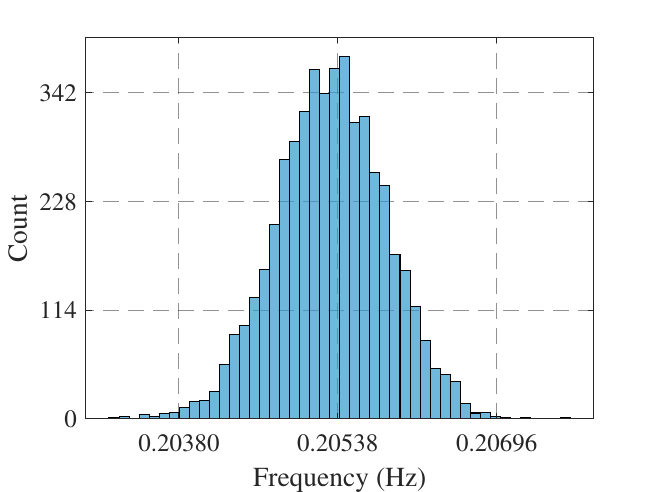}
        \caption{First-order of SSI-Cov}
    \end{subfigure}
    \hspace{2em} 
    \begin{subfigure}{0.29\textwidth}
        \centering
        \includegraphics[width=\textwidth]{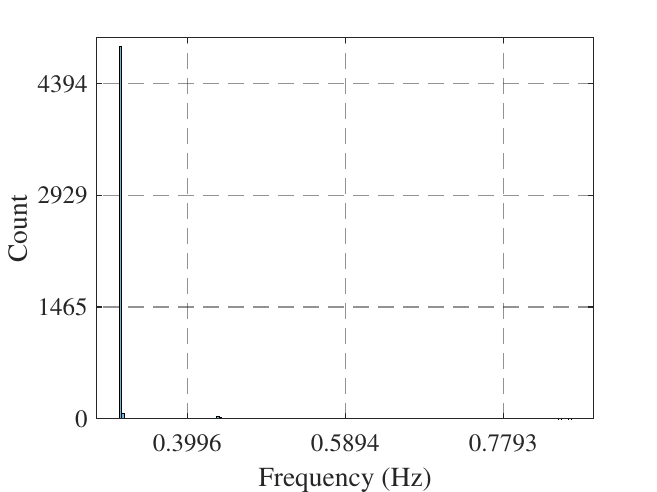}
        \caption{Second-order  of SSI-Cov}
    \end{subfigure}
    \hspace{2em} 
    \begin{subfigure}{0.29\textwidth}
        \centering
        \includegraphics[width=\textwidth]{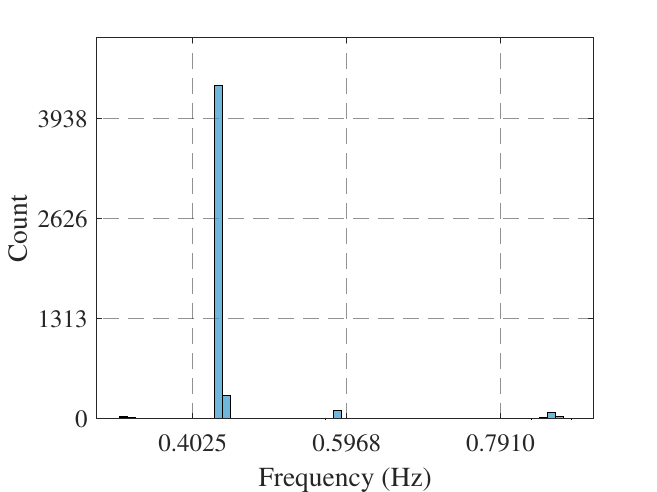}
        \caption{Third-order of SSI-Cov}
    \end{subfigure}

    \vspace{0.5em} 

    \begin{subfigure}{0.29\textwidth}
        \centering
        \includegraphics[width=\textwidth]{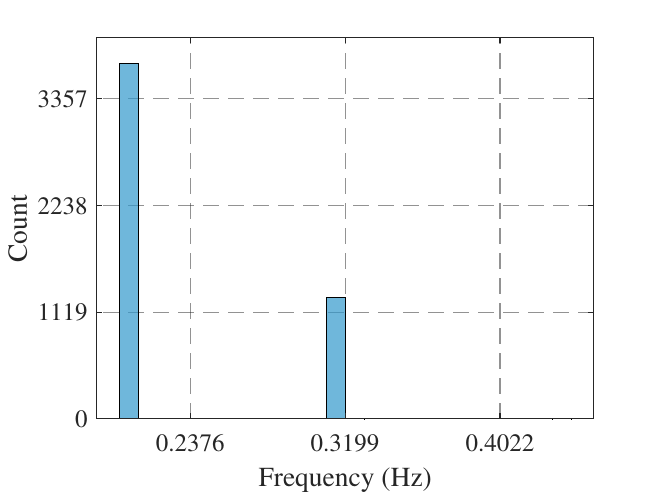}
        \caption{First-order of SSI-Data}
    \end{subfigure}
    \hspace{2em} 
    \begin{subfigure}{0.29\textwidth}
        \centering
        \includegraphics[width=\textwidth]{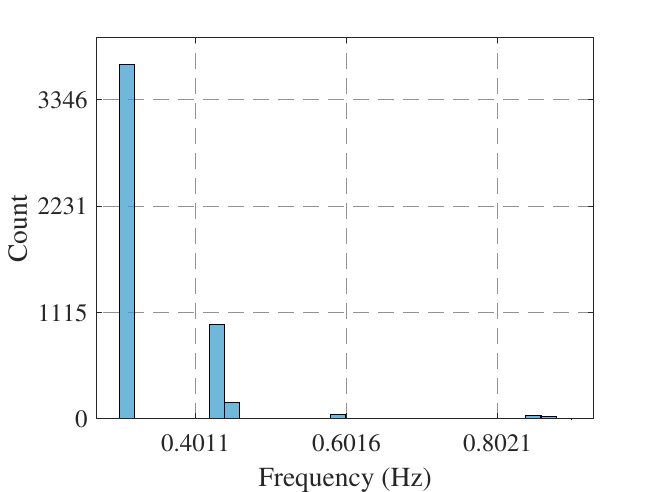}
        \caption{Second-order  of SSI-Data}
    \end{subfigure}
    \hspace{2em} 
    \begin{subfigure}{0.29\textwidth}
        \centering
        \includegraphics[width=\textwidth]{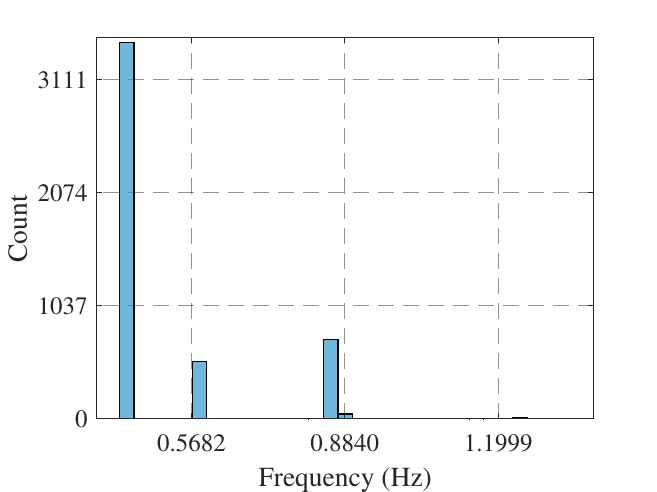}
        \caption{Third-order of SSI-Data}
    \end{subfigure}
    
\caption{Comparison of the first three frequency distributions for PCSSI, SSI-Cov, and SSI-Data. Since the estimated modal frequencies for each mode may vary significantly across different methods, different scales are used for visualization.}
    \label{fig:PCSSI_and_SSI-Cov1}
\end{figure}

\begin{figure}[!ht]
    \centering
    \begin{subfigure}{0.29\textwidth}
        \centering
        \includegraphics[width=\textwidth]{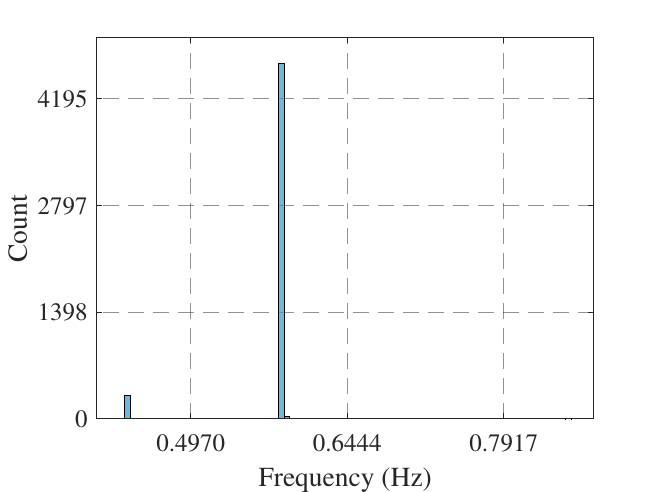}
        \caption{Fourth-order of PCSSI}
    \end{subfigure}
    \hspace{2em}
    \begin{subfigure}{0.29\textwidth}
        \centering
        \includegraphics[width=\textwidth]{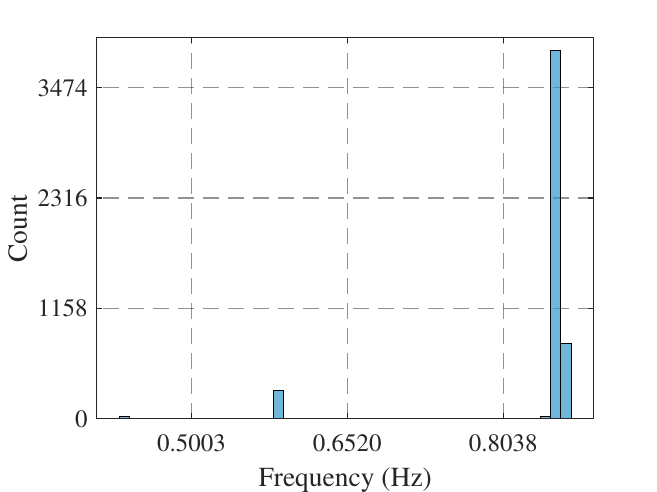}
        \caption{Fifth-order of PCSSI}
    \end{subfigure}
    \hspace{2em}
    \begin{subfigure}{0.29\textwidth}
        \centering
        \includegraphics[width=\textwidth]{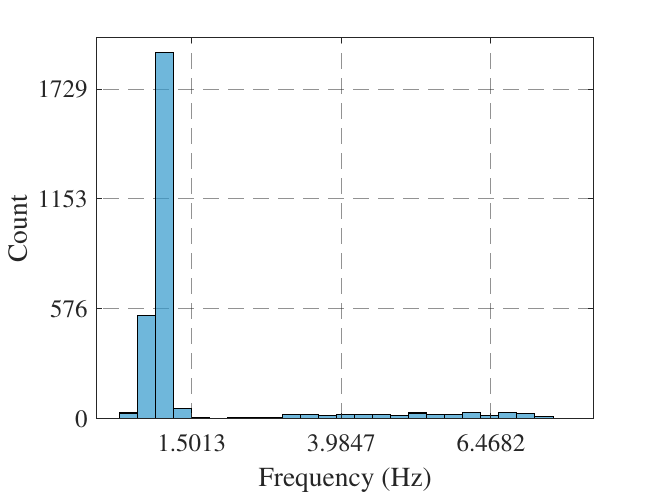}
        \caption{Sixth-order of PCSSI}
    \end{subfigure}
  
    \vspace{0.5em}

    \begin{subfigure}{0.29\textwidth}
        \centering
        \includegraphics[width=\textwidth]{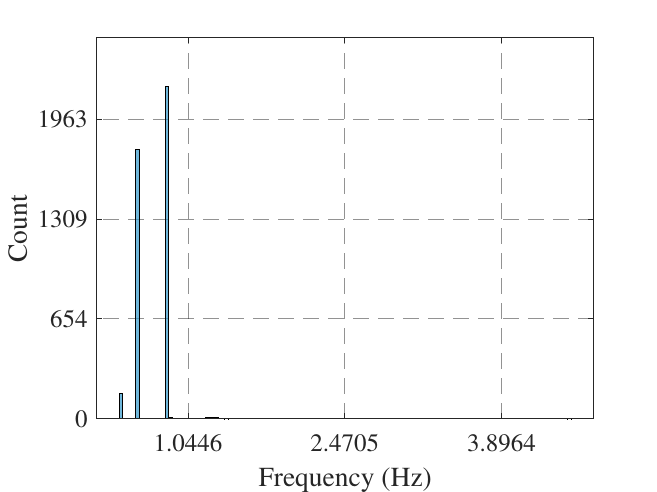}
        \caption{Fourth-order of SSI-Cov}
    \end{subfigure}
    \hspace{2em}
    \begin{subfigure}{0.29\textwidth}
        \centering
        \includegraphics[width=\textwidth]{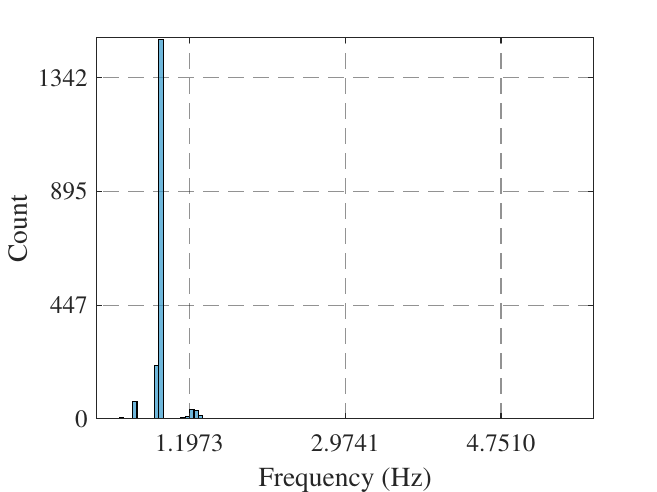}
        \caption{Fifth-order of SSI-Cov}
    \end{subfigure}
    \hspace{2em}
    \begin{subfigure}{0.29\textwidth}
        \centering
        \includegraphics[width=\textwidth]{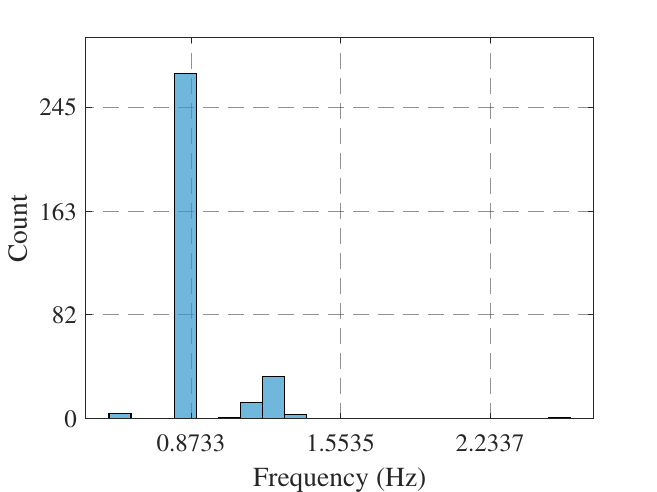}
        \caption{Sixth-order of SSI-Cov}
    \end{subfigure}

    \vspace{0.5em}

    \begin{subfigure}{0.29\textwidth}
        \centering
        \includegraphics[width=\textwidth]{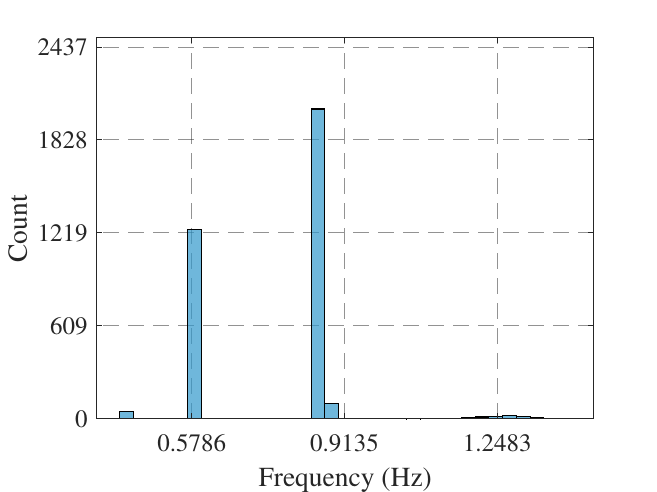}
        \caption{Fourth-order of SSI-Data}
    \end{subfigure}
    \hspace{2em}
    \begin{subfigure}{0.29\textwidth}
        \centering
        \includegraphics[width=\textwidth]{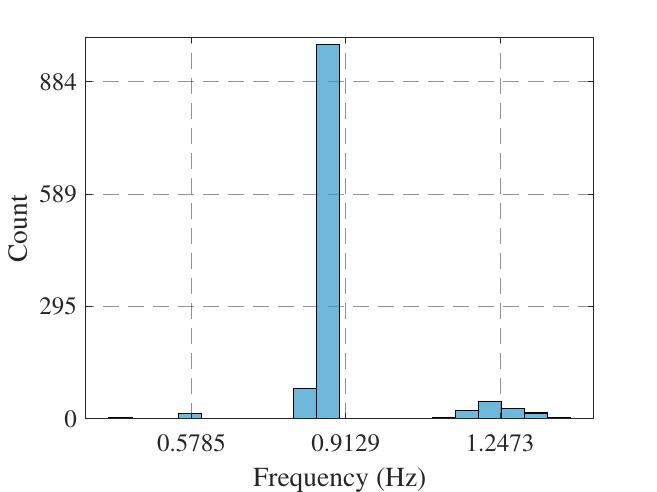}
        \caption{Fifth-order of SSI-Data}
    \end{subfigure}
    \hspace{2em}
    \begin{subfigure}{0.29\textwidth}
        \centering
        \includegraphics[width=\textwidth]{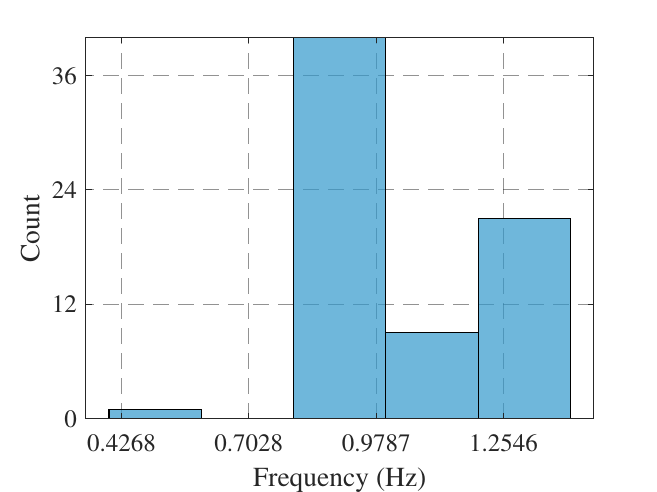}
        \caption{Sixth-order of SSI-Data}
    \end{subfigure}

    \caption{Comparison of the 4th, 5th, and 6th frequencies for PCSSI, SSI-Cov, and SSI-Data.}
    \label{fig:PCSSI_and_SSI-Cov2}
\end{figure}

While PCSSI, SSI-Cov, and SSI-Data theoretically converge under infinite observations, practical implementations with limited data exhibit significant differences in variance. 
To facilitate a fair comparison of the performance of different methods under limited monitoring data, we use the same Hankel 
matrix \( Y_f \in \mathbb{R}^{18 \times 5000} \) for all calculations. Each method then applies its respective instrumental variable projection and utilizes the 
same stabilization diagram algorithm.
To analyze the empirical variance, we conduct 5000 Monte Carlo simulations and examine the modal frequency distributions of the identified modes. The true modal frequencies, 
which serve as ground truth for validation, are listed in Table \ref{tab:1}.
Figures \ref{fig:PCSSI_and_SSI-Cov1} and \ref{fig:PCSSI_and_SSI-Cov2} illustrate the modal frequency identification results for modes 1–6 obtained from 5000 Monte Carlo simulations. 
\begin{table}[ht]
    \centering
    \caption{True Frequency Values for Modes 1 to 6}\label{tab:1}
    \begin{tabular}{|c|c|c|c|c|c|c|}
    \hline
    Mode & 1st & 2nd & 3rd & 4th & 5th & 6th \\
    \hline
    Frequency (Hz) & 0.205 & 0.319 & 0.439 & 0.585 & 0.864 & 1.194 \\
    \hline
    \end{tabular}
\end{table}

For PCSSI, the frequency estimates are tightly clustered around the true modal frequencies, with minimal spread, as indicated by the narrow peaks and low counts in the off-peak 
regions. In contrast, the identification of the 4th mode using SSI-Cov and SSI-Data exhibits noticeable multi-peaking behavior, whereas PCSSI maintains a single dominant 
peak at the true frequency. This underscores PCSSI's superior ability to suppress noise-induced spurious modes.
The primary reason for this difference is that SSI-Cov frequently produces multiple duplicate 
estimates within a single modal analysis, leading to an overestimation of the number of identified frequencies and resulting in the multi-peak phenomenon. 
On the other hand, SSI-Data tends to underestimate the number of frequencies in a single estimation, causing later modes to appear at earlier frequency positions than expected.

\begin{figure}[!ht]
    \centering
    \begin{subfigure}{0.31\textwidth}
        \centering
        \includegraphics[width=\textwidth]{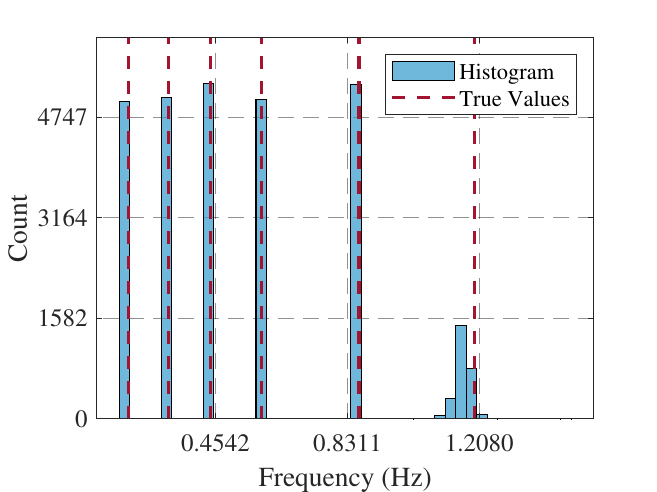}
        \caption{PCSSI}
    \end{subfigure}
    \hspace{0.5em}
    \begin{subfigure}{0.31\textwidth}
        \centering
        \includegraphics[width=\textwidth]{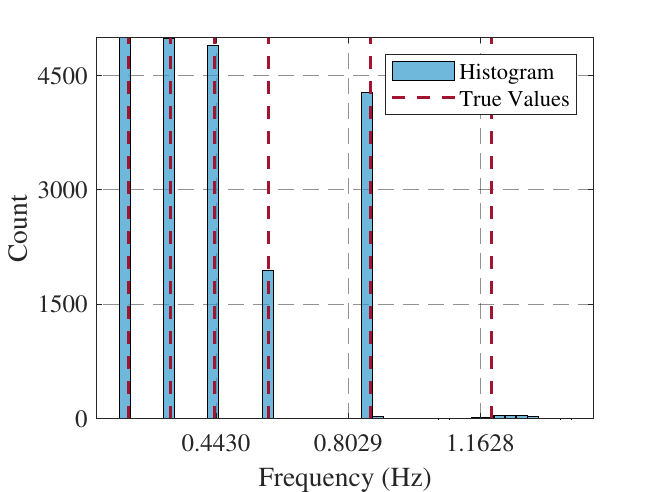}
        \caption{SSI-Cov}
    \end{subfigure}
    \hspace{0.5em}
    \begin{subfigure}{0.31\textwidth}
        \centering
        \includegraphics[width=\textwidth]{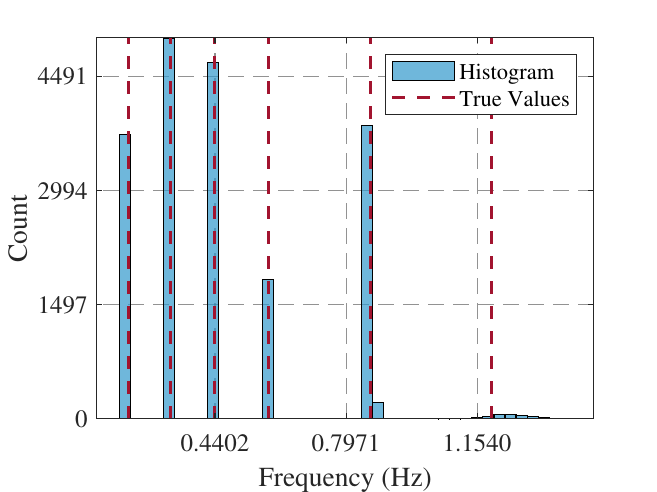}
        \caption{SSI-Data}
    \end{subfigure}
    \caption{Comparison of the identified truncated frequencies for PCSSI, SSI-Cov, and SSI-Data.}\label{fig:PCSSI_and_SSI-Cov3}
  \end{figure}

The previous analysis focused on plotting the frequencies identified for each mode individually. In Figure \ref{fig:PCSSI_and_SSI-Cov3}, the frequencies obtained from all 
5000 Monte Carlo simulations are combined into a single plot to illustrate the ability of the three methods to identify frequencies.
It is observed that for all three methods, the identified modal frequencies above 1.5 Hz are sparse, whereas frequencies below 1.5 Hz are crucial for capturing the bridge's 
low-frequency modal characteristics. Therefore, to present the results more clearly, only frequencies below 1.5 Hz are shown in Figure \ref{fig:PCSSI_and_SSI-Cov3}.

It is evident that, apart from minor errors in estimating the sixth modal frequency, PCSSI accurately identifies all modal frequencies without producing any spurious modes.  
However, all three methods exhibit certain issues in identifying the sixth modal frequency, with a general tendency toward underestimation. This issue is particularly 
pronounced in SSI-Cov and SSI-Data, where the number of successful identifications of the sixth mode is significantly limited. A possible explanation is that noise interference 
causes the stabilization diagram algorithm to misclassify the sixth mode as a spurious mode, thereby reducing its identification frequency and affecting the final recognition results.  

Additionally, although SSI-Cov statistically identifies the first three modal frequencies, it, like SSI-Data, exhibits noticeable errors in estimating the fourth modal frequency, 
with a significant decrease in the number of successful identifications. This suggests that both methods may experience instability when dealing with mid-to-high-order modal frequencies.  
Notably, in certain cases, SSI-Data fails to correctly identify the first modal frequency. This phenomenon may be attributed to the projection process being unable to effectively 
eliminate noise when the available data is limited, thereby impacting the accuracy of frequency identification. In contrast, PCSSI, by projecting onto the principal signal subspace, 
achieves more effective noise suppression, resulting in improved stability and accuracy in modal identification.

\subsubsection{Mode Shape Analysis Based on Monte Carlo Simulation}
\begin{figure}[!ht]
    \centering
    \begin{subfigure}{0.29\textwidth}
        \centering
        \includegraphics[width=\textwidth]{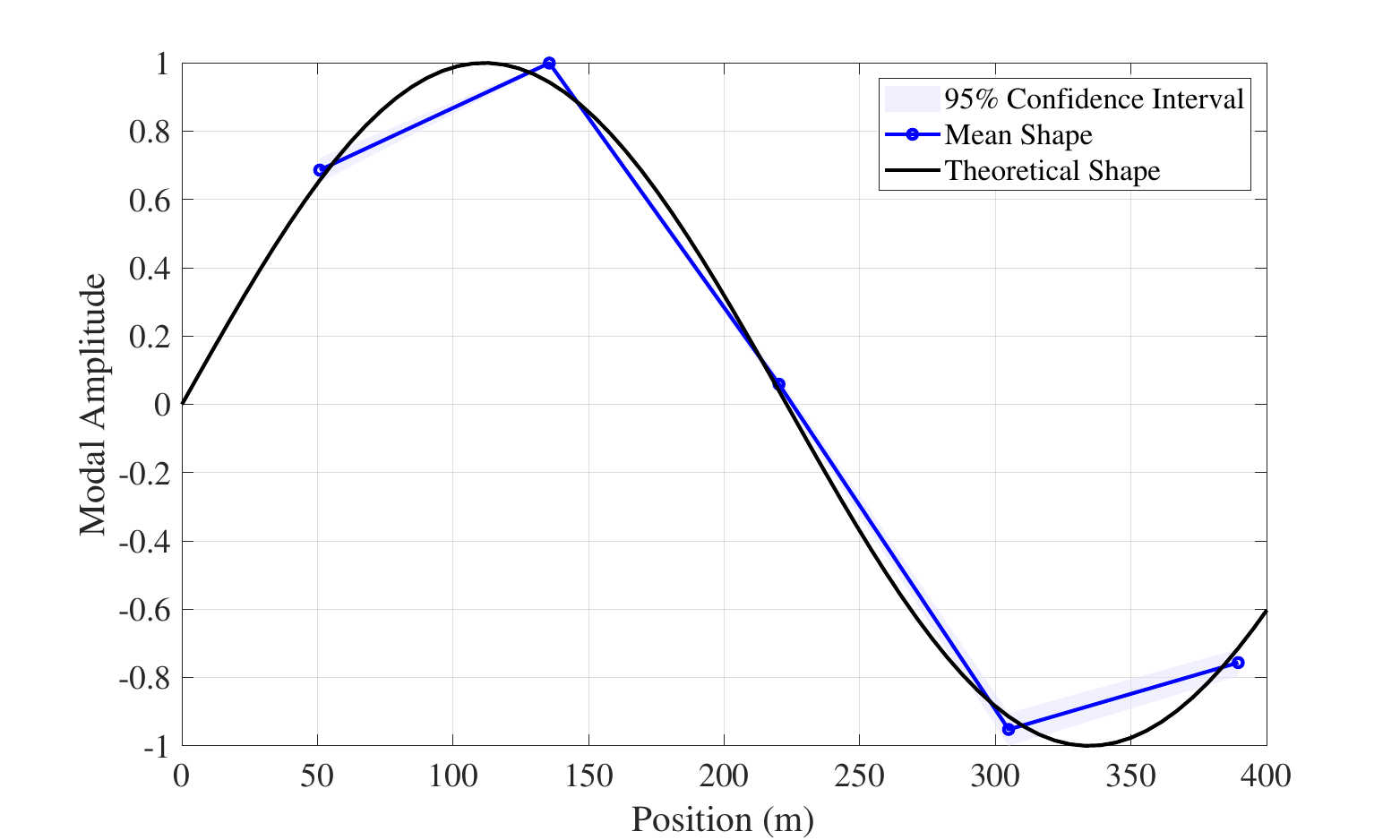}
        \caption{First-order of PCSSI}
    \end{subfigure}
    \hspace{2em}
    \begin{subfigure}{0.29\textwidth}
        \centering
        \includegraphics[width=\textwidth]{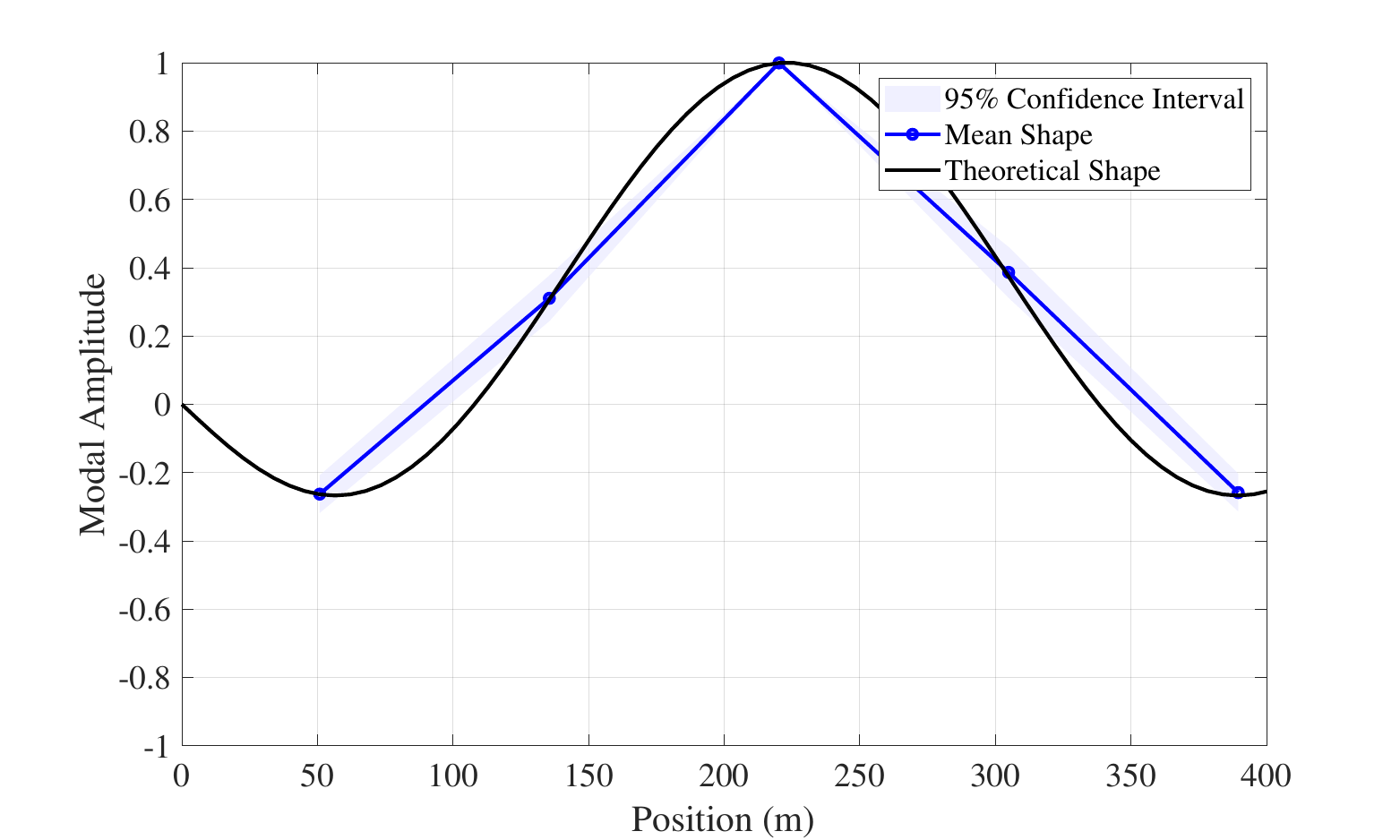}
        \caption{Second-order of PCSSI}
    \end{subfigure}
    \hspace{2em}
    \begin{subfigure}{0.29\textwidth}
        \centering
        \includegraphics[width=\textwidth]{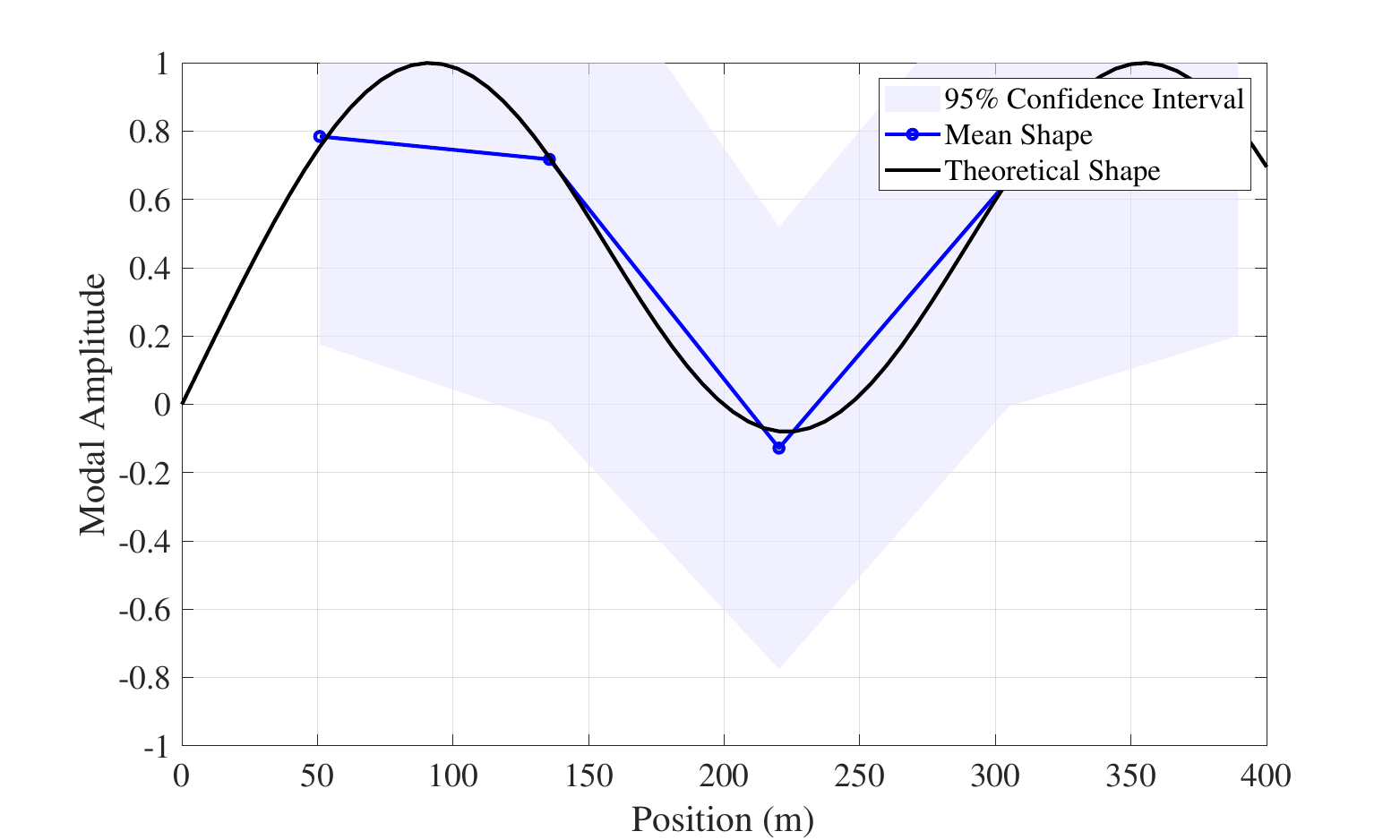}
        \caption{Third-order of PCSSI}
    \end{subfigure}
  
    \vspace{0.5em}

    \begin{subfigure}{0.29\textwidth}
        \centering
        \includegraphics[width=\textwidth]{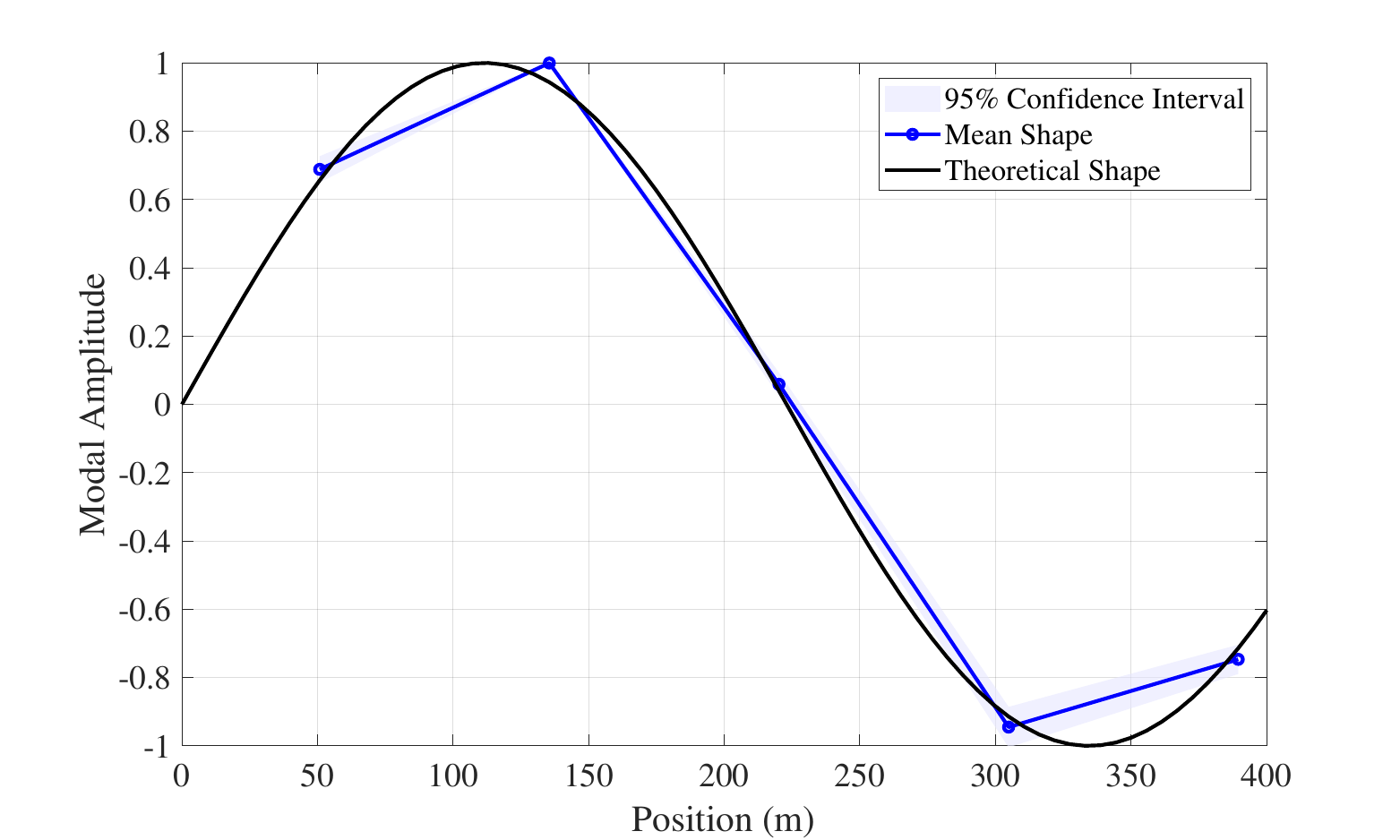}
        \caption{First-order of SSI-Cov}
    \end{subfigure}
    \hspace{2em}
    \begin{subfigure}{0.29\textwidth}
        \centering
        \includegraphics[width=\textwidth]{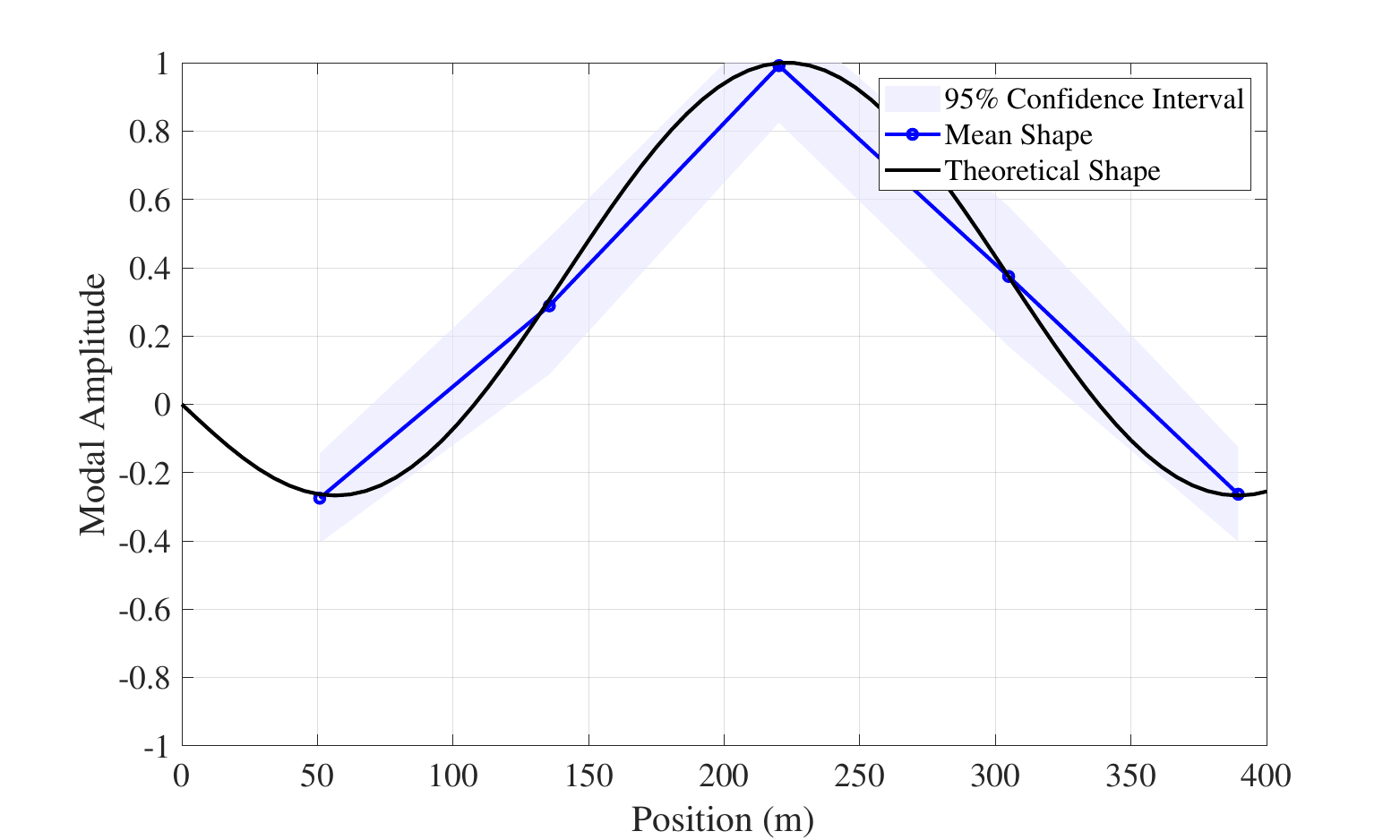}
        \caption{Second-order of SSI-Cov}
    \end{subfigure}
    \hspace{2em}
    \begin{subfigure}{0.29\textwidth}
        \centering
        \includegraphics[width=\textwidth]{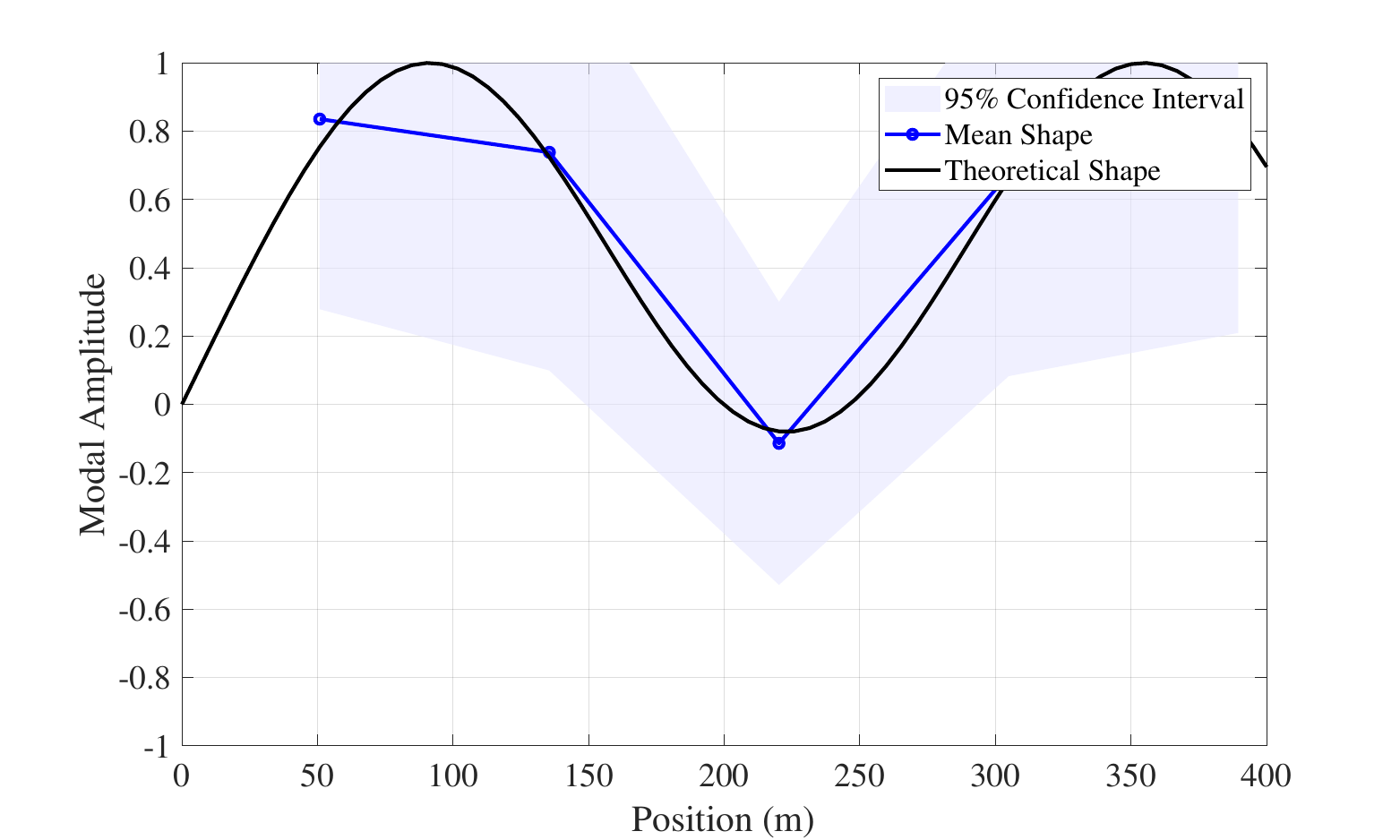}
        \caption{Third-order of SSI-Cov}
    \end{subfigure}

    \vspace{0.5em}

    \begin{subfigure}{0.29\textwidth}
        \centering
        \includegraphics[width=\textwidth]{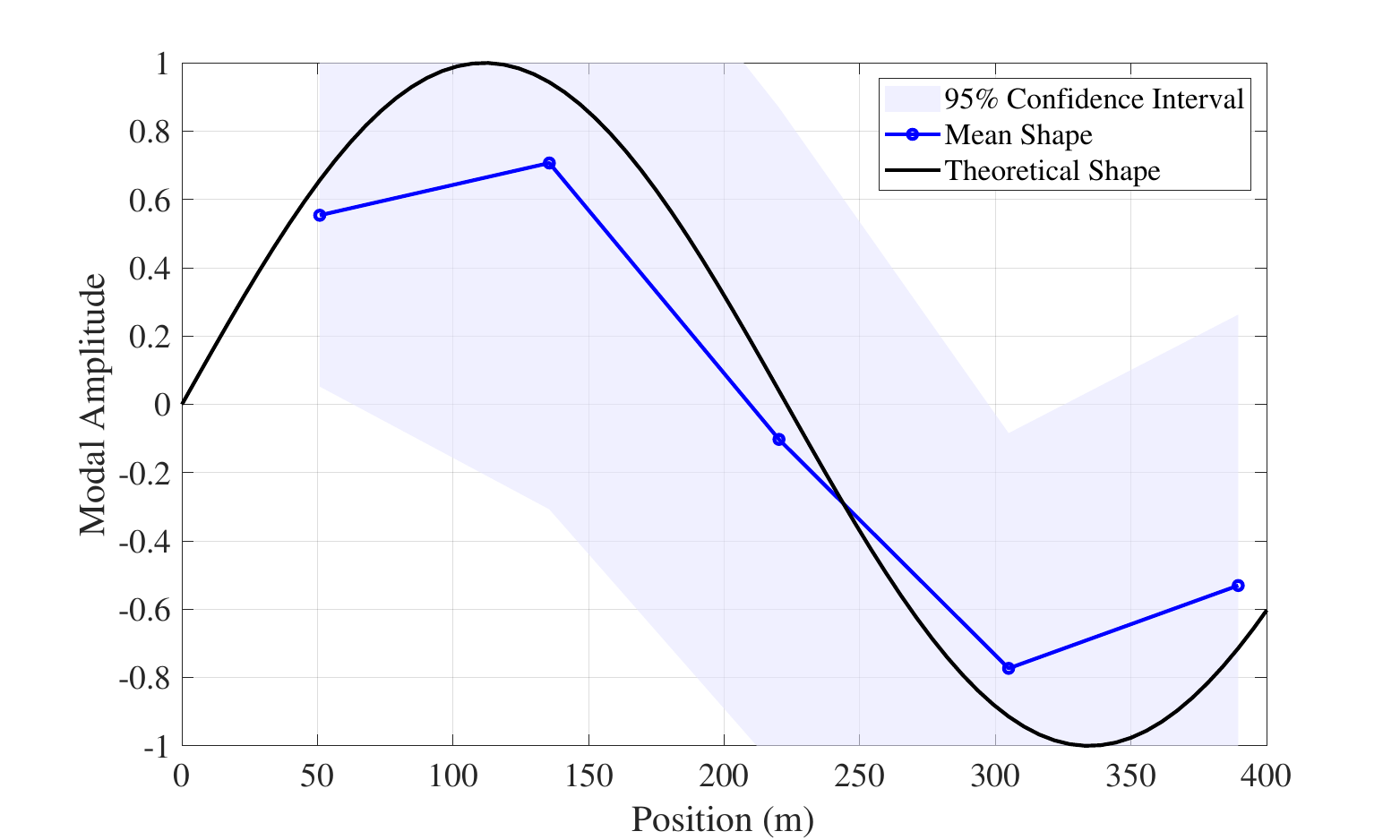}
        \caption{First-order of SSI-Data}
    \end{subfigure}
    \hspace{2em}
    \begin{subfigure}{0.29\textwidth}
        \centering
        \includegraphics[width=\textwidth]{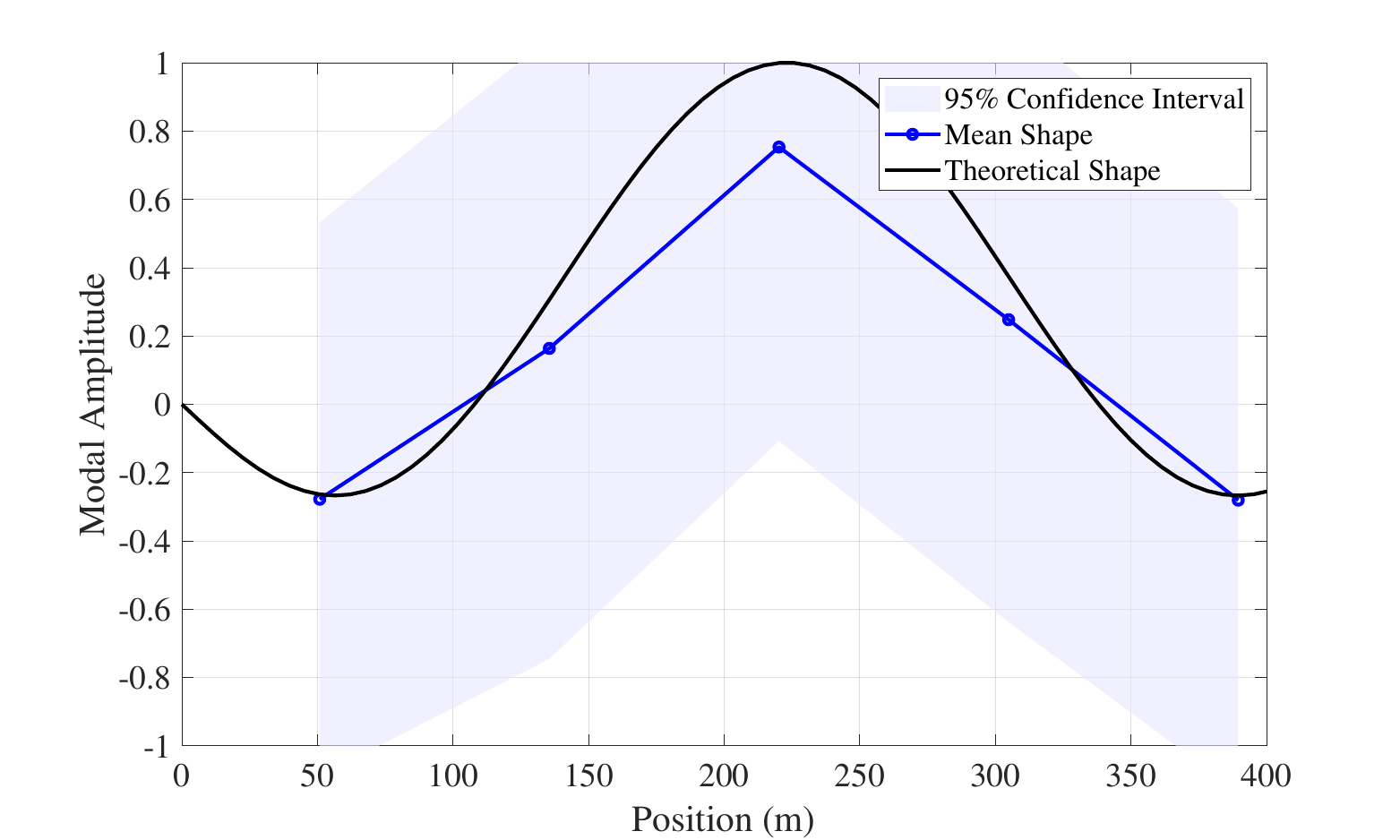}
        \caption{Second-order of SSI-Data}
    \end{subfigure}
    \hspace{2em}
    \begin{subfigure}{0.29\textwidth}
        \centering
        \includegraphics[width=\textwidth]{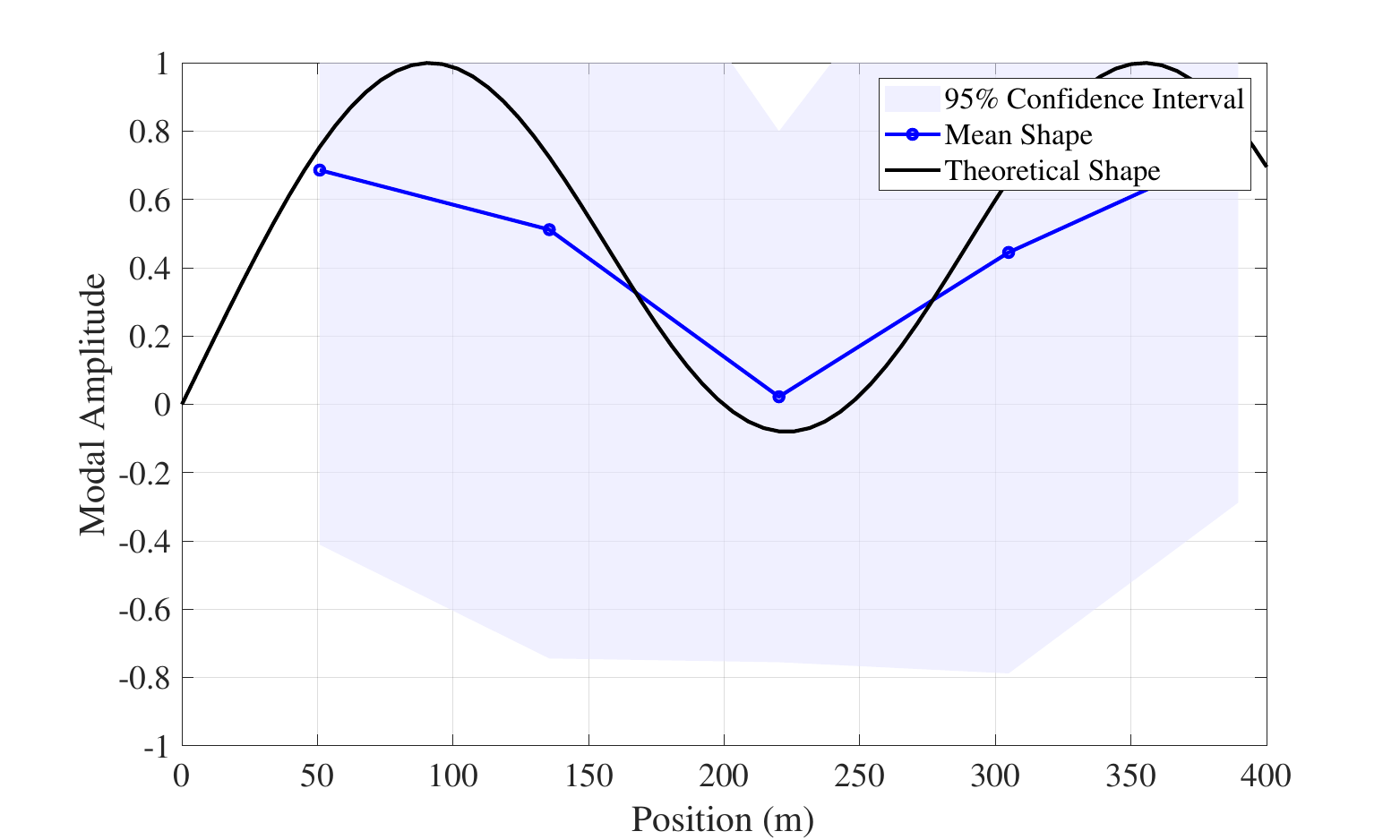}
        \caption{Third-order of SSI-Data}
    \end{subfigure}

    \caption{Comparison of the first three identified mode shapes for PCSSI, SSI-Cov, and SSI-Data.}
    \label{fig:PCSSI_and_SSI-Cov_mode_shape1}
\end{figure}

\begin{figure}[!ht]
    \centering
    \begin{subfigure}{0.29\textwidth}
        \centering
        \includegraphics[width=\textwidth]{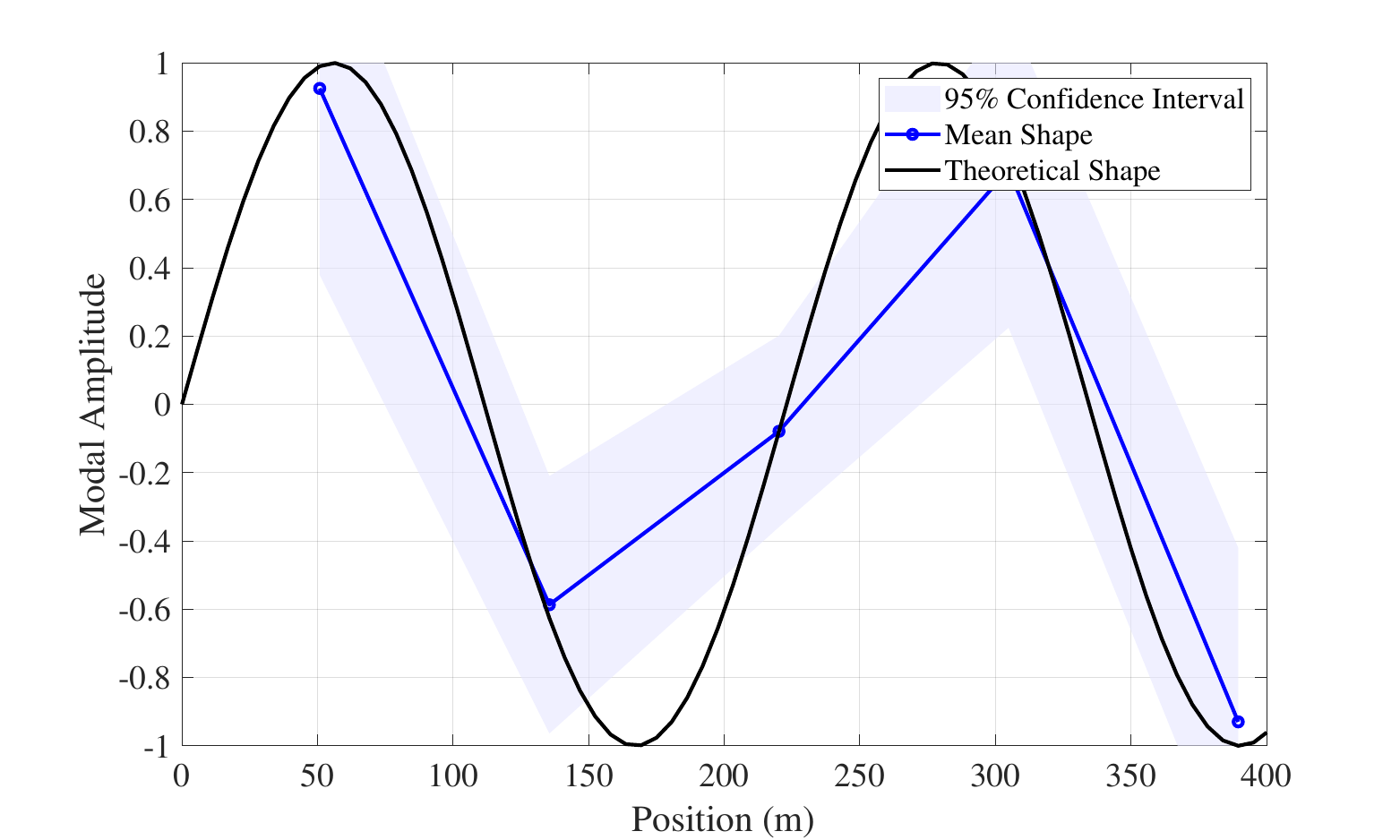}
        \caption{Fourth-order of PCSSI}
    \end{subfigure}
    \hspace{2em}
    \begin{subfigure}{0.29\textwidth}
        \centering
        \includegraphics[width=\textwidth]{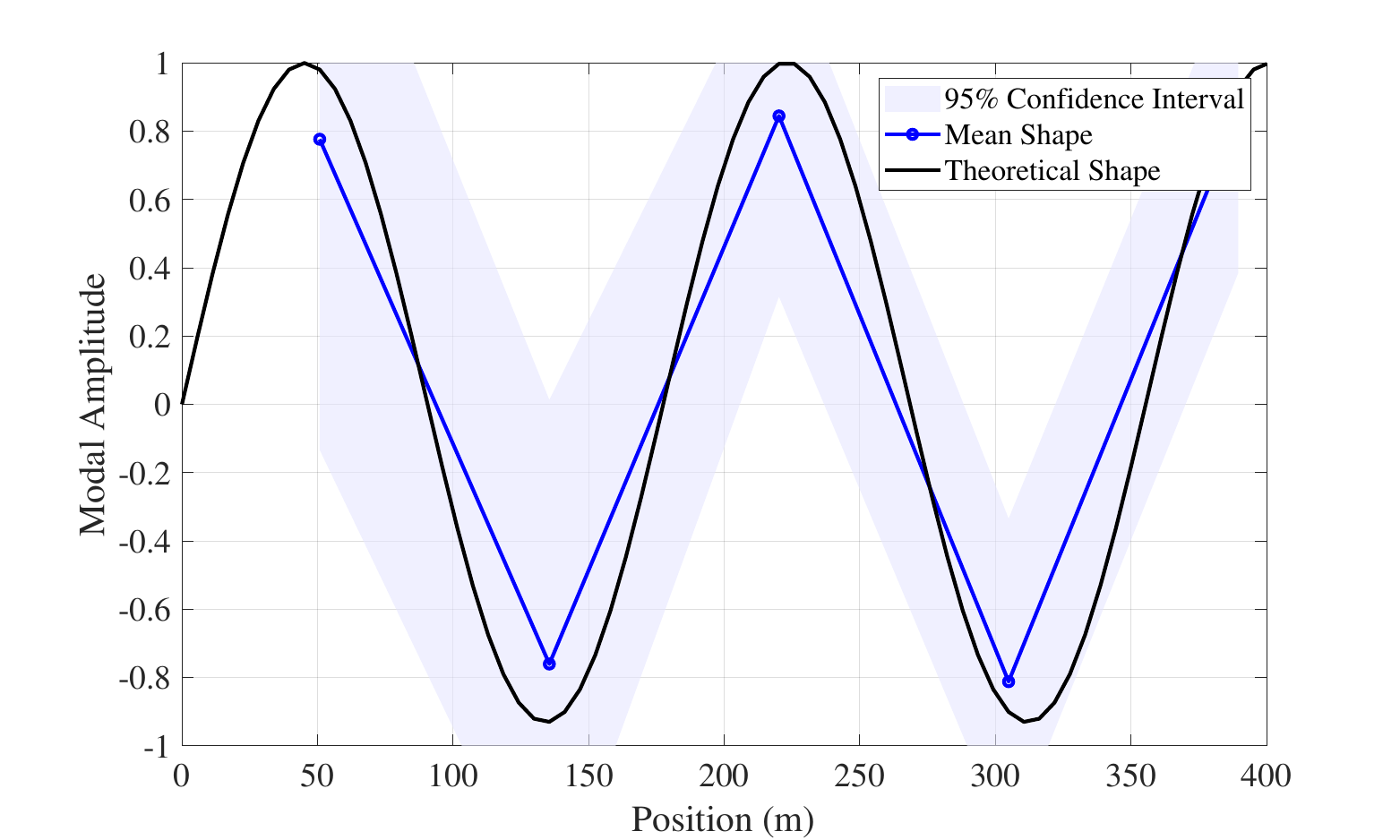}
        \caption{Fifth-order of PCSSI}
    \end{subfigure}
    \hspace{2em}
    \begin{subfigure}{0.29\textwidth}
        \centering
        \includegraphics[width=\textwidth]{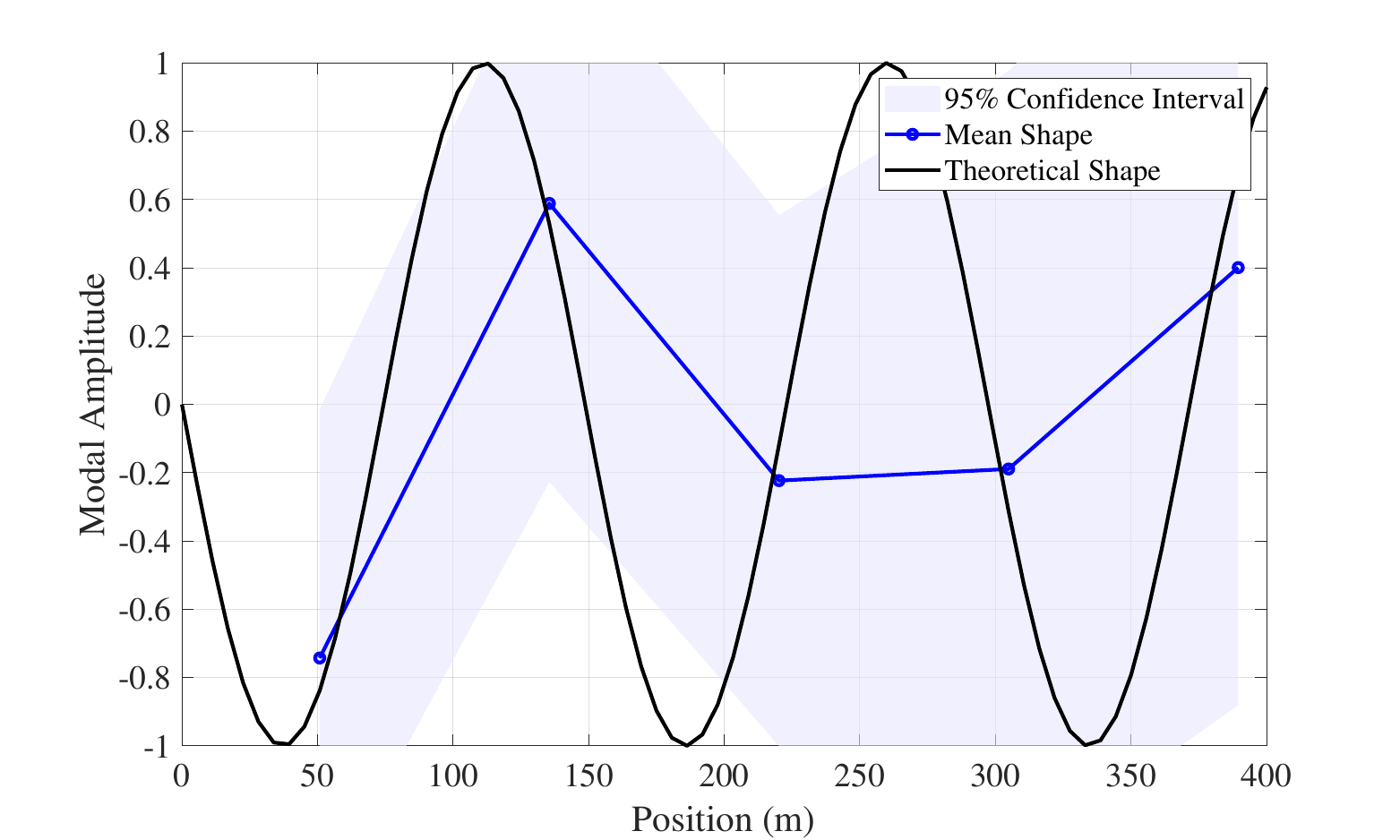}
        \caption{Sixth-order of PCSSI}
    \end{subfigure}
  
    \vspace{0.5em}

    \begin{subfigure}{0.29\textwidth}
        \centering
        \includegraphics[width=\textwidth]{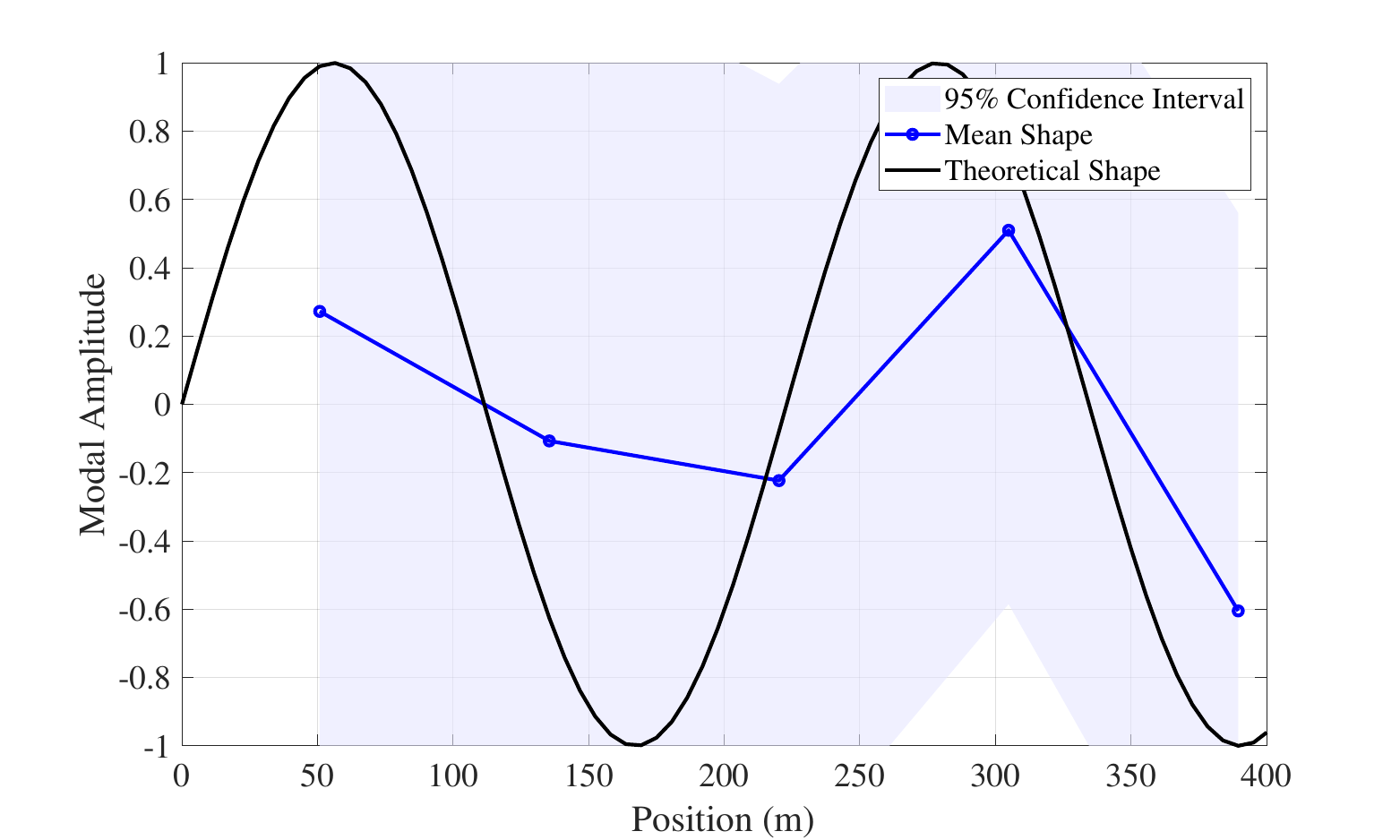}
        \caption{Fourth-order  of SSI-Cov}
    \end{subfigure}
    \hspace{2em}
    \begin{subfigure}{0.29\textwidth}
        \centering
        \includegraphics[width=\textwidth]{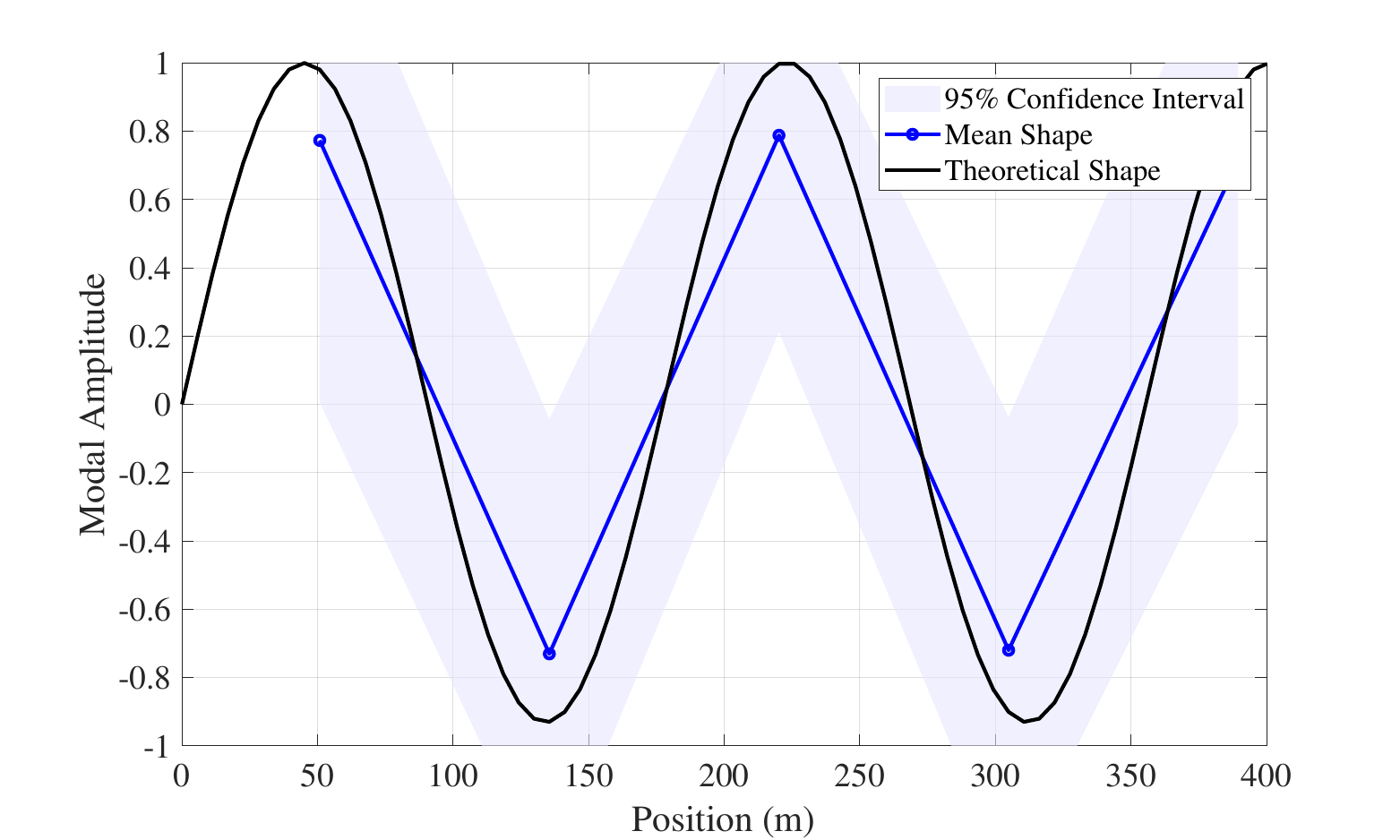}
        \caption{Fifth-order of SSI-Cov}
    \end{subfigure}
    \hspace{2em}
    \begin{subfigure}{0.29\textwidth}
        \centering
        \includegraphics[width=\textwidth]{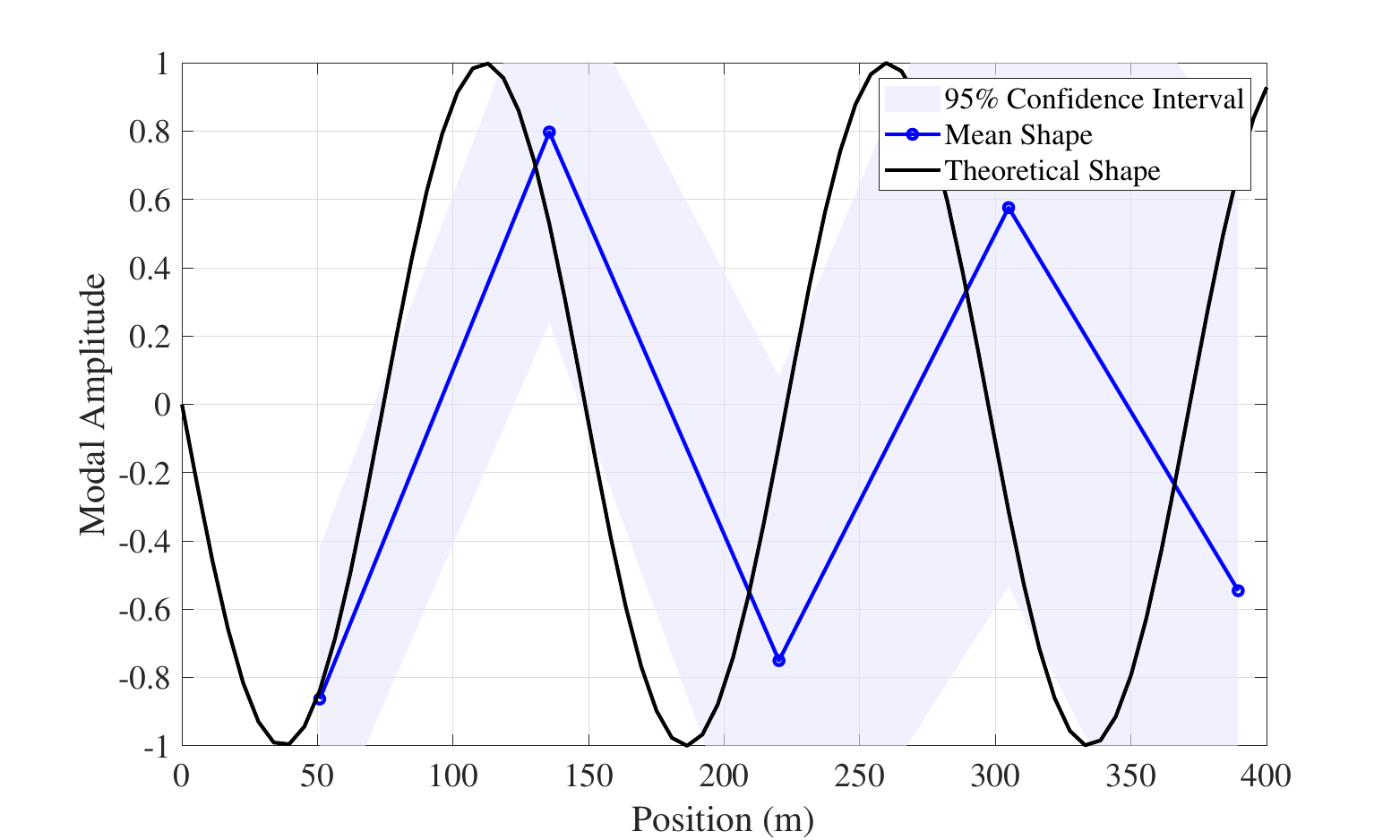}
        \caption{Sixth-order of SSI-Cov}
    \end{subfigure}

    \vspace{0.5em}

    \begin{subfigure}{0.29\textwidth}
        \centering
        \includegraphics[width=\textwidth]{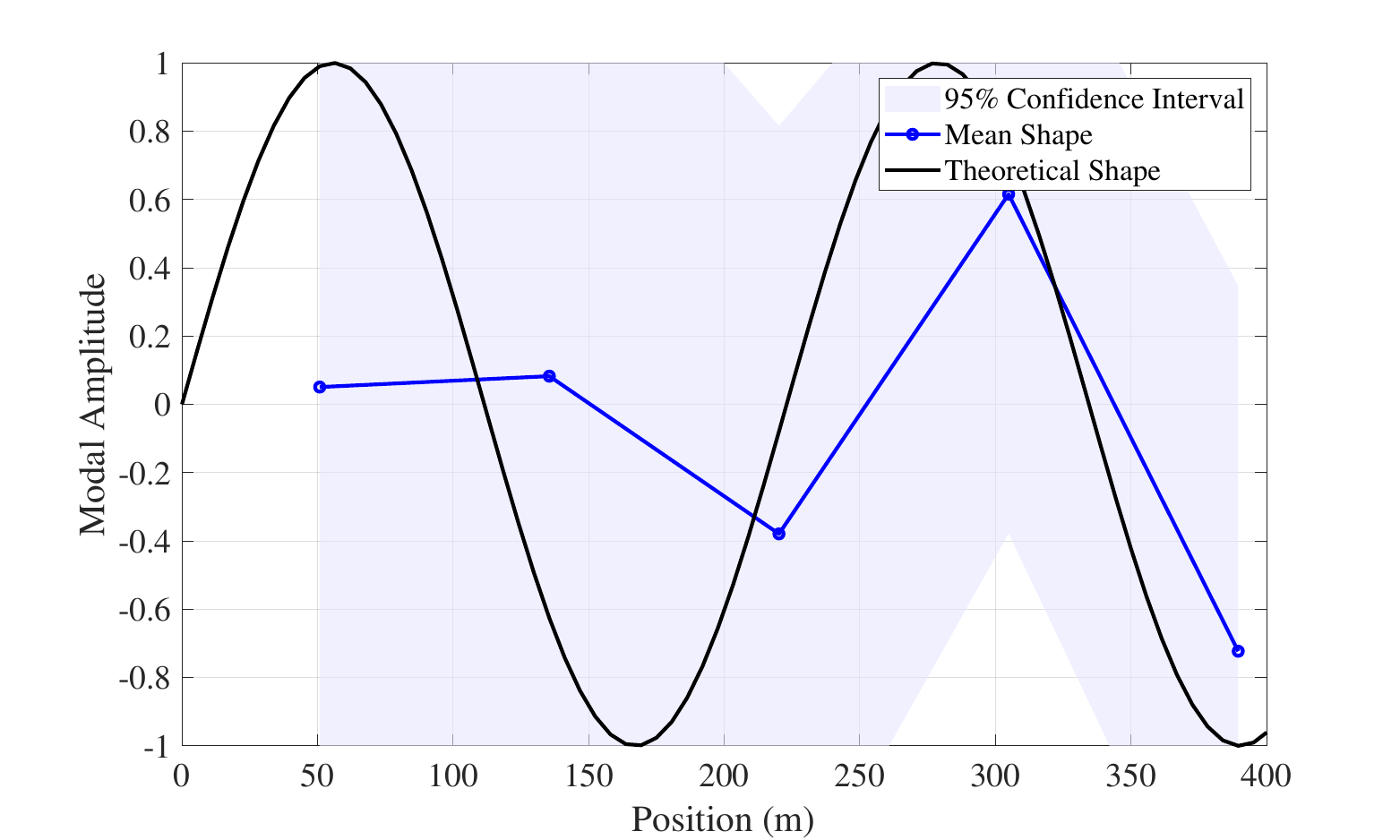}
        \caption{Fourth-order  of SSI-Data}
    \end{subfigure}
    \hspace{2em}
    \begin{subfigure}{0.29\textwidth}
        \centering
        \includegraphics[width=\textwidth]{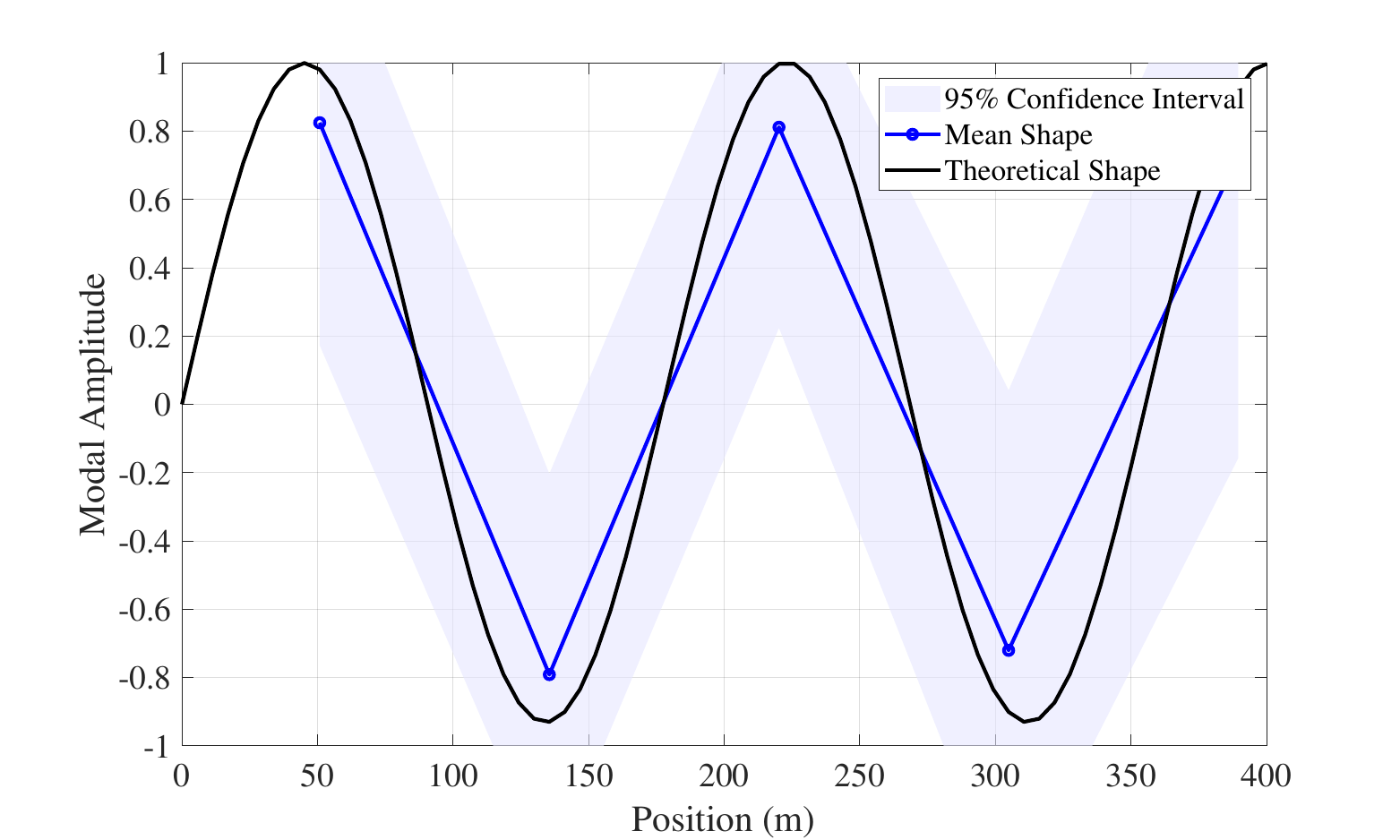}
        \caption{Fifth-order of SSI-Data}
    \end{subfigure}
    \hspace{2em}
    \begin{subfigure}{0.29\textwidth}
        \centering
        \includegraphics[width=\textwidth]{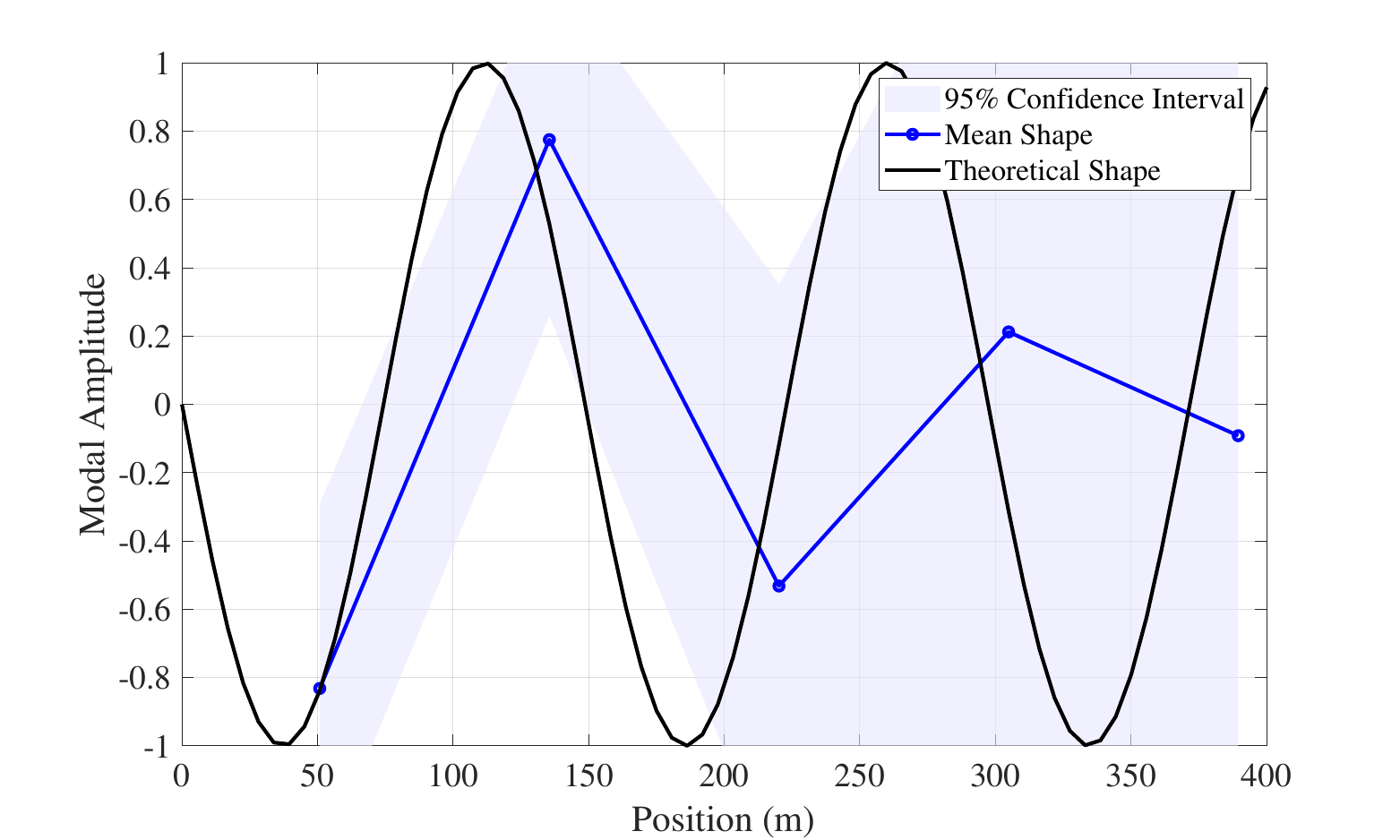}
        \caption{Sixth-order of SSI-Data}
    \end{subfigure}
    \caption{Comparison of the fourth to sixth identified mode shapes for PCSSI, SSI-Cov, and SSI-Data.}
    \label{fig:PCSSI_and_SSI-Cov_mode_shape2}
\end{figure}

Beyond frequency estimation, accurate mode shape reconstruction is crucial for structural analysis. To analyze its statistical properties, we focus on calculating 
the mean and variance of mode shapes obtained from 5000 Monte Carlo simulations, rather than using the commonly employed Modal Assurance Criterion (MAC) coefficients 
for comparison.  
Therefore, we compute the mean mode shapes and 95\% confidence intervals (CI) for the three methods and compare them with the true mode shapes, as shown in 
Figures \ref{fig:PCSSI_and_SSI-Cov_mode_shape1} and \ref{fig:PCSSI_and_SSI-Cov_mode_shape2}.

The results indicate that PCSSI exhibits significantly lower variance compared to SSI-Cov and SSI-Data. For the first two modes, both PCSSI and SSI-Cov demonstrate near-zero 
variance, with mean values closely aligning with theoretical expectations. However, even in the reconstruction of low-order modes, SSI-Data exhibits noticeable distortions in mode shapes.  
Additionally, the variance of mode shapes obtained from SSI-Cov and SSI-Data is consistently higher than that of PCSSI across all modes, indicating lower stability in 
shape reconstruction. Furthermore, both SSI-Cov and SSI-Data exhibit certain distortions in higher-order mode shapes.

\subsection{Scaled Experimental Validation}
\begin{figure}[!ht]
  \centering
  \begin{subfigure}{0.38\textwidth}
      \centering
      \includegraphics[width=\textwidth]{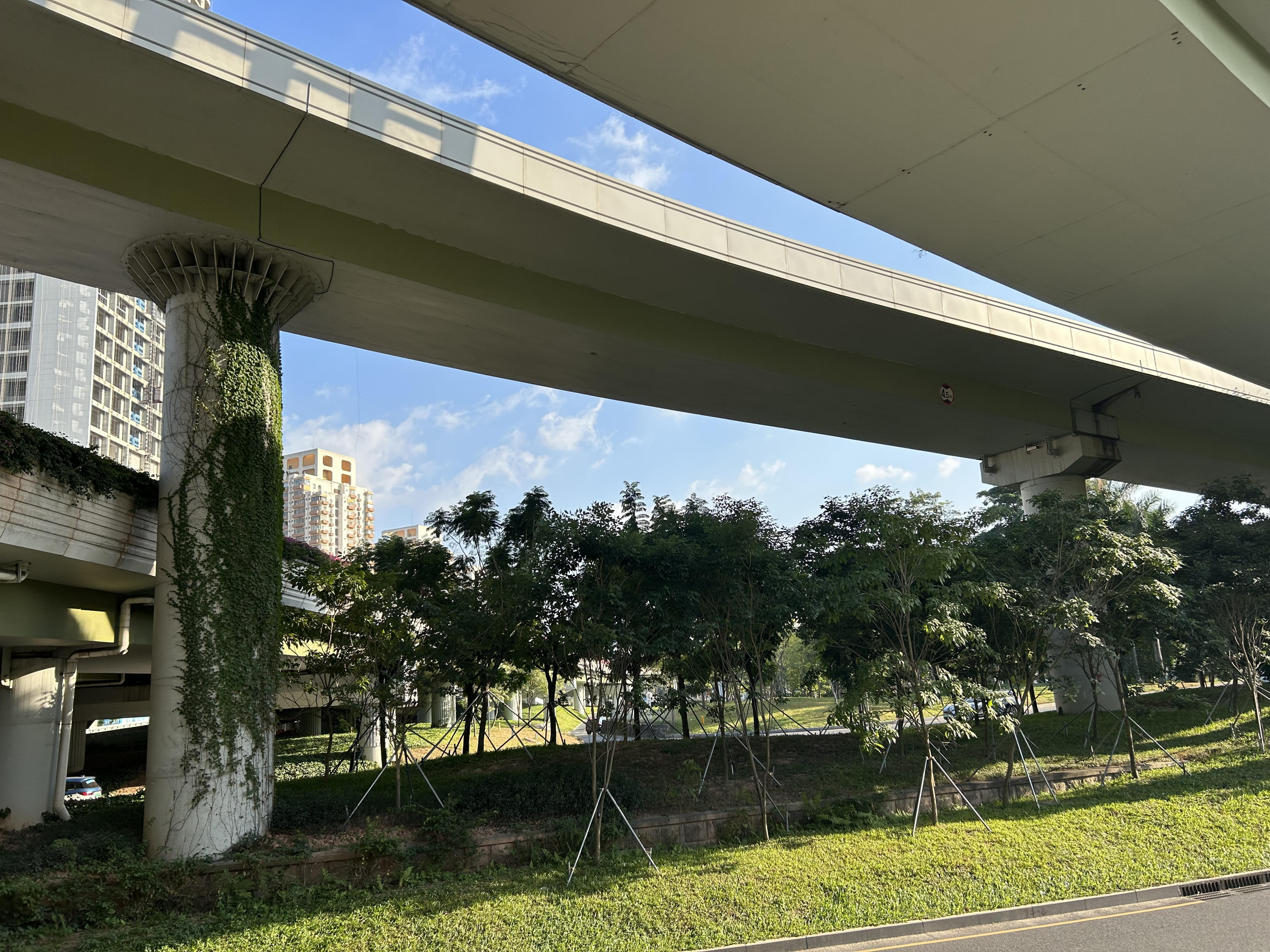}
      \caption{A Segment of an Overpass}\label{fig:3a}
  \end{subfigure}
  \hspace{2em}
  \begin{subfigure}{0.38\textwidth}
      \centering
      \includegraphics[width=\textwidth]{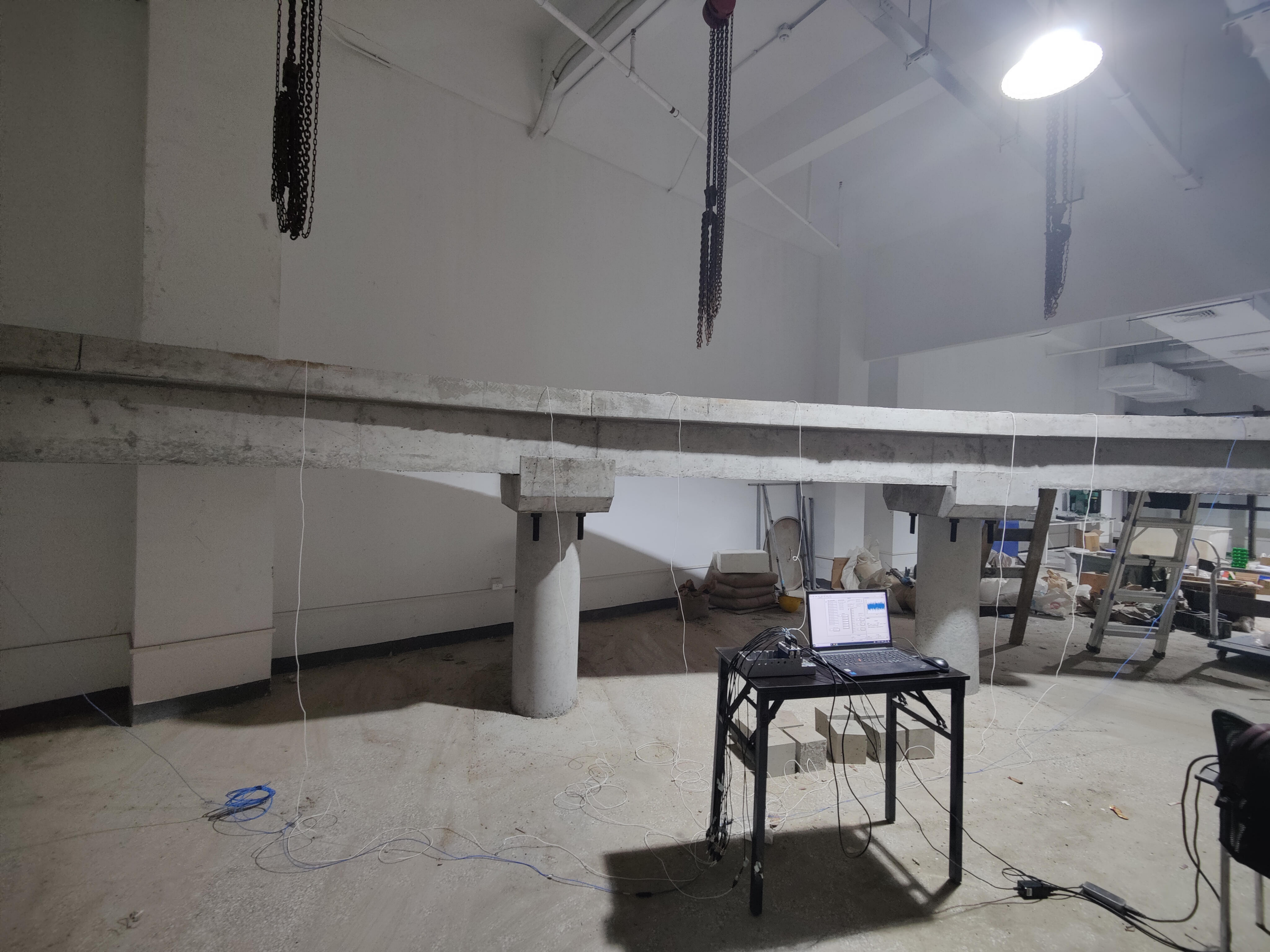}
      \caption{Scaled Experimental Model }\label{fig:3b}
  \end{subfigure}
  \caption{A Segment of an Overpass in Shenzhen and the Scaled Experimental Model}
\end{figure}

\begin{figure}[htb]
  \centering
  \includegraphics[width=0.9\linewidth]{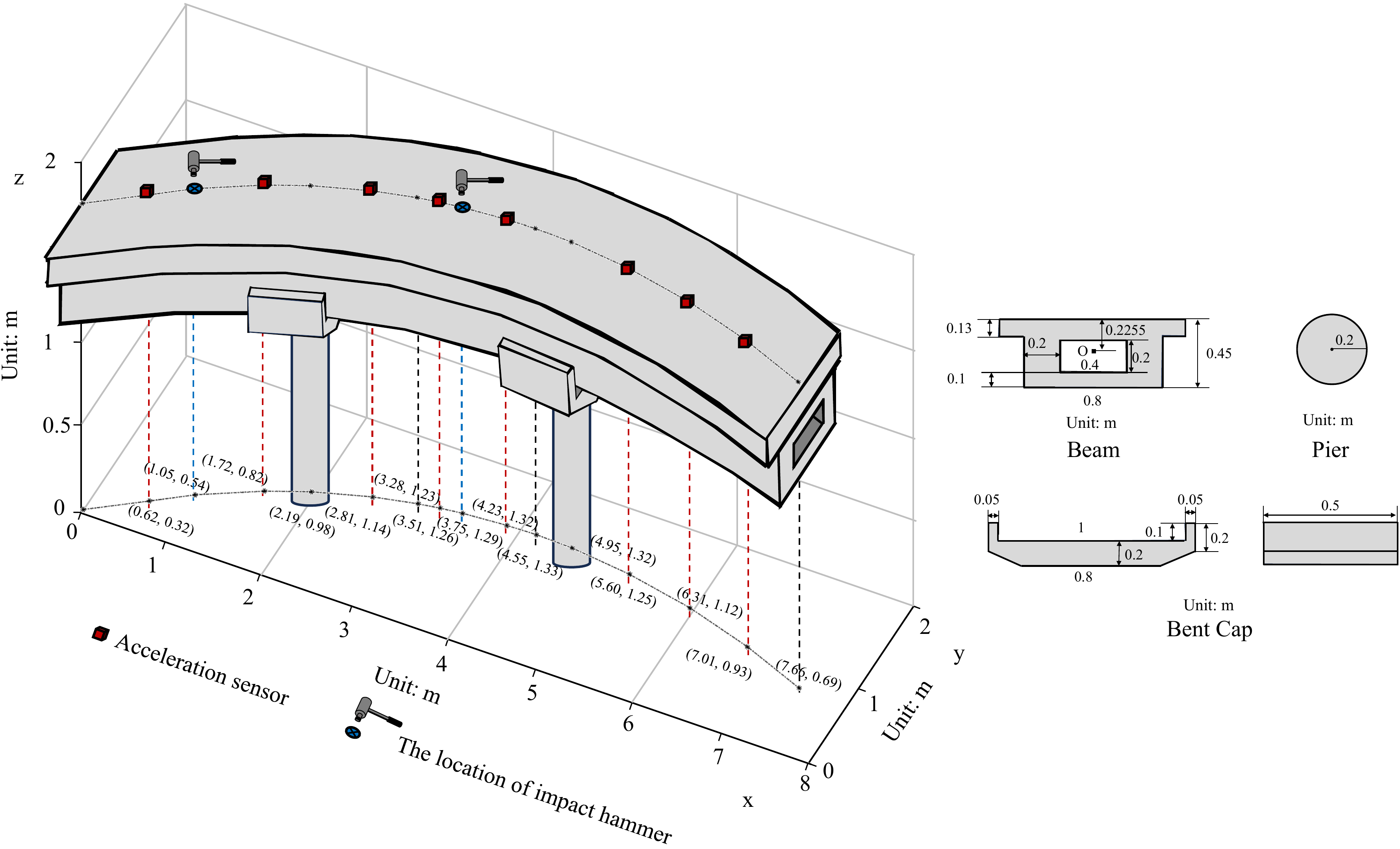}
  \caption{Scaled Experimental Model in the Laboratory}\label{fig4}
\end{figure}

To experimentally validate the PCSSI method, a scaled model of the Shenzhen Beihuan Shangbu overpass segment was constructed in the laboratory. 
The actual bridge is depicted in Figure \ref{fig:3a}. 
The bridge features a single-pier structure, a typical design for urban viaducts. The scaled model, shown in Figure \ref{fig:3b}, has a scale ratio of approximately 12:1. 
During the design phase of the experimental model, the cross-sectional dimensions of its main components were adjusted based on the principles of similarity. Figure \ref{fig4} presents 
the cross-sectional dimensions of the main components of the scaled model. All components of the experimental model were cast using C40 concrete. Foundation pits were excavated at the 
base of the two piers and backfilled with reinforced concrete blocks to serve as the structural foundation.

The entire bridge section utilized a consistent box girder cross-section. To meet the load-bearing requirements, a specific quantity of longitudinal and transverse reinforcement 
bars and stirrups were arranged in the beams, cap beams, and piers of the experimental model. The reinforcement principles were as follows: longitudinal bars were designed to meet 
flexural strength requirements, while stirrups were arranged to satisfy structural requirements (exceeding the minimum reinforcement ratio and meeting stirrup spacing requirements). 
A uniform stirrup spacing of 0.2 meters was used, with spiral stirrups in the piers and regular stirrups in the beams and cap beams.

\subsubsection{Equipment and test procedures}

\begin{figure}[!ht]
  \centering
  \begin{subfigure}{0.115\textwidth}
      \centering
      \includegraphics[width=\textwidth]{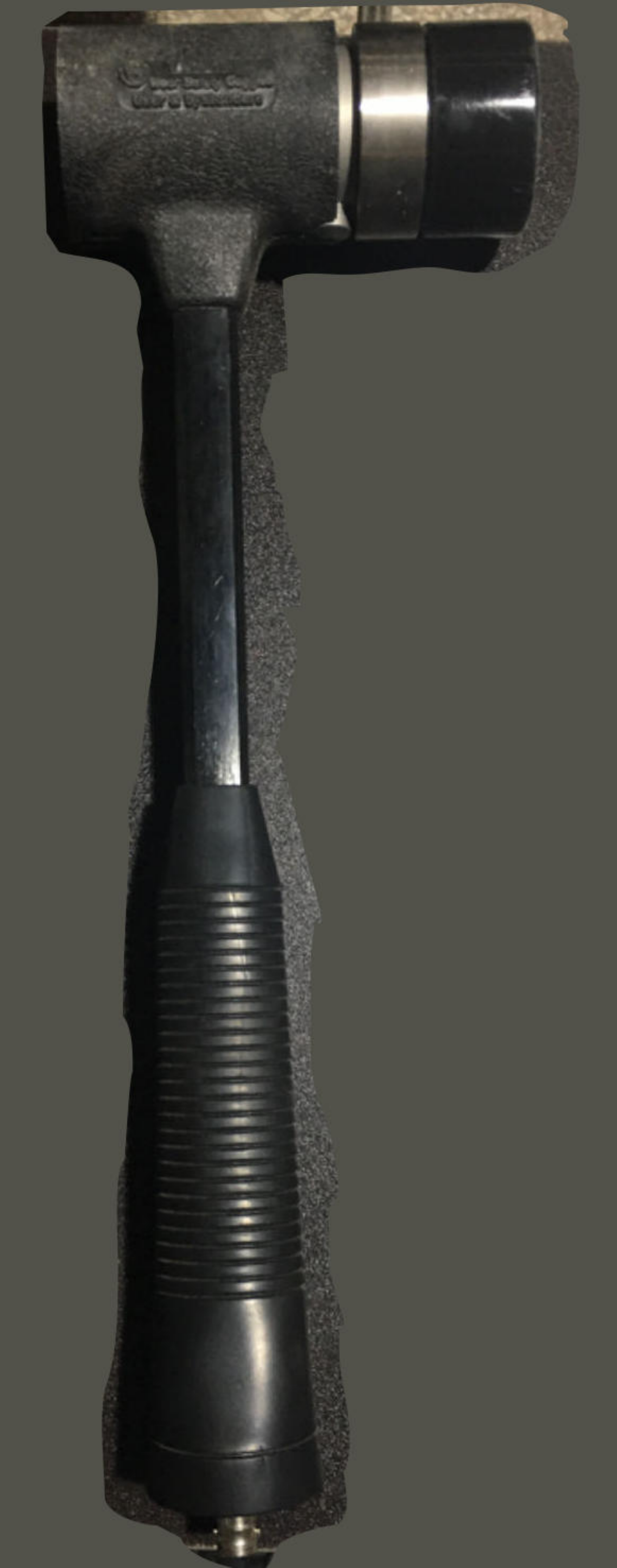}
      \caption{Hammer}
  \end{subfigure}
  \hspace{2em}
  \begin{subfigure}{0.1215\textwidth}
      \centering
      \includegraphics[width=\textwidth]{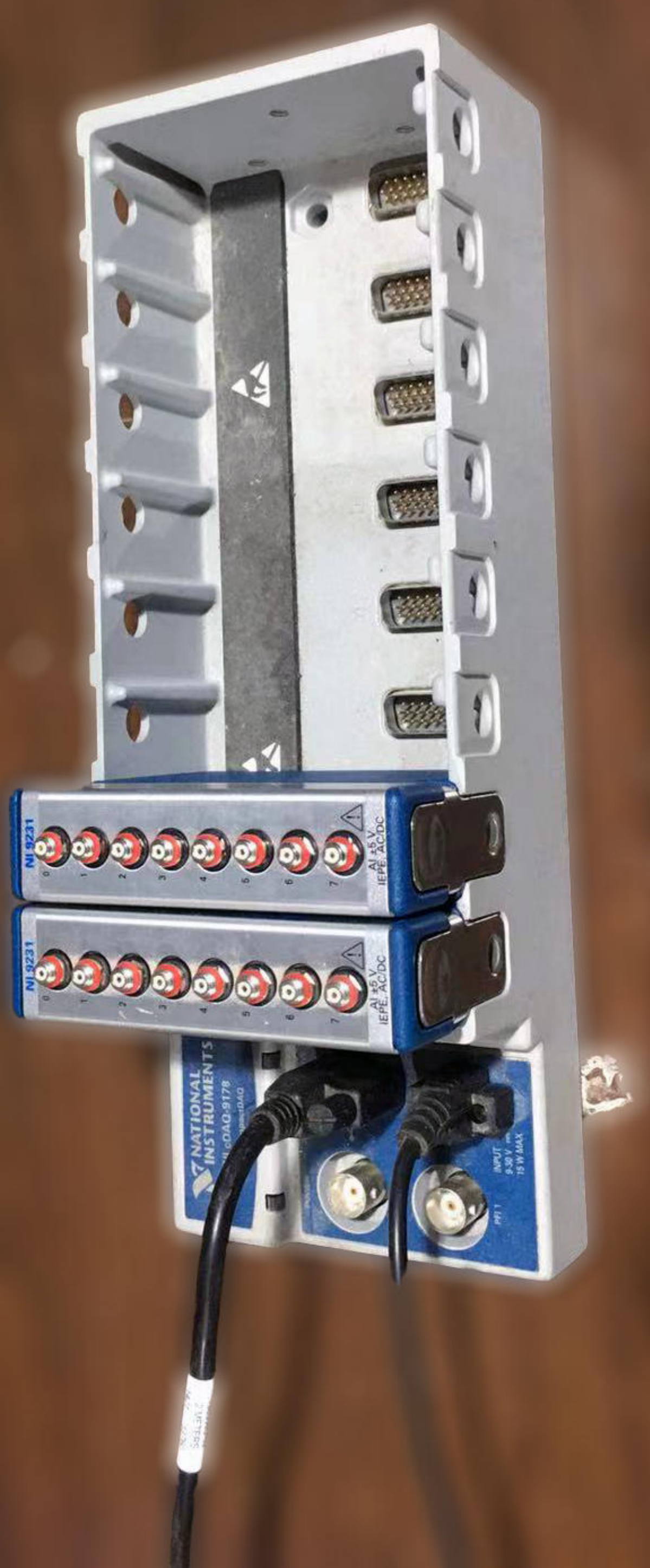}
      \caption{Card}
  \end{subfigure}
  \hspace{2em}
  \begin{subfigure}{0.357\textwidth}
      \centering
      \includegraphics[width=\textwidth]{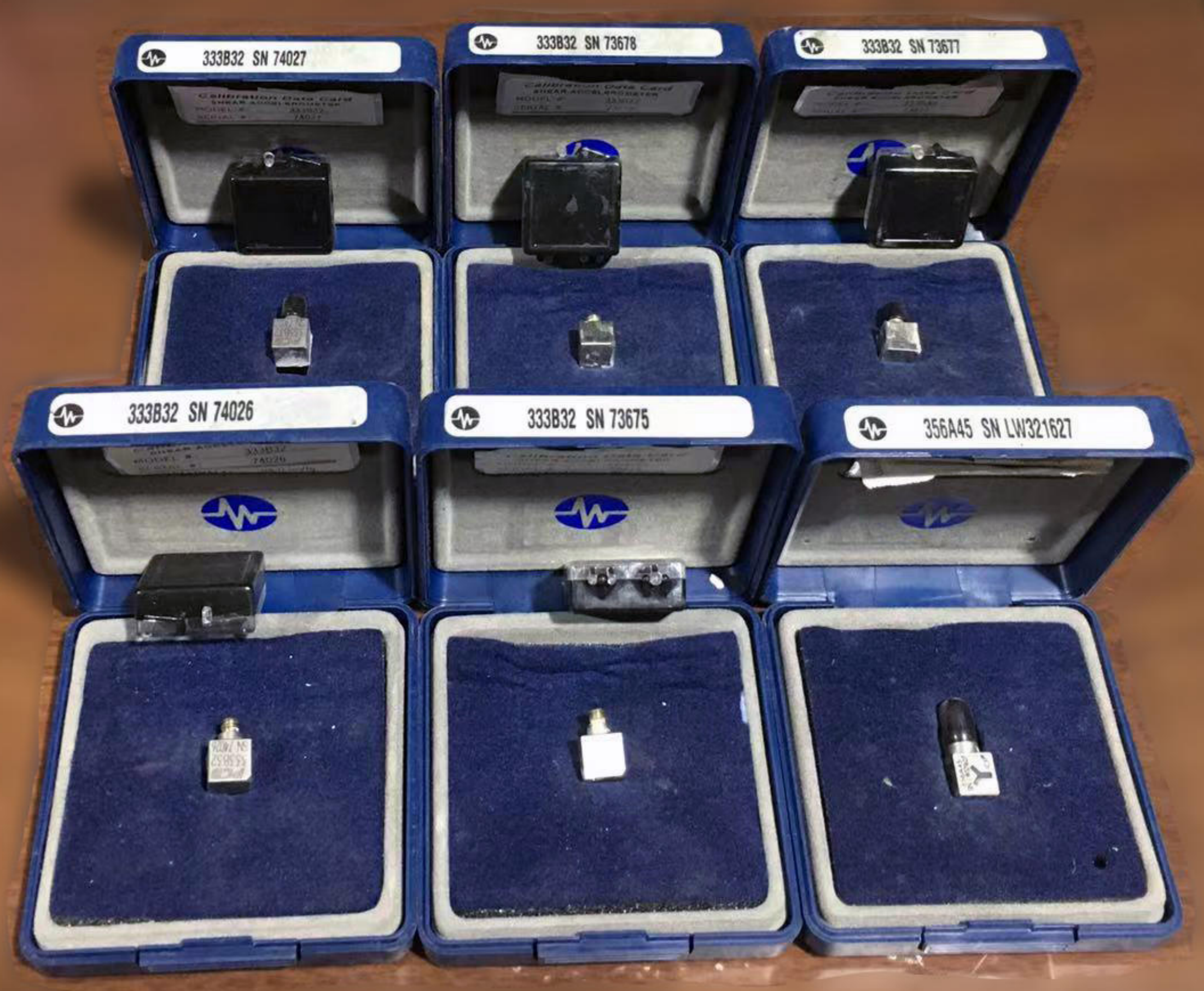}
      \caption{Sensors}
  \end{subfigure}
  \caption{Experimental Equipment}\label{fig5}
\end{figure}

To conduct experimental modal analysis, an impact hammer was used to simulate impact loading as the external excitation for the experimental model. The acceleration 
response of the model structure was collected using eight accelerometers, with a sampling frequency set at 3200 Hz and a data collection duration of 10 seconds. 
In accordance with ISO 7626-5 standards, the independent impact hammer tests were repeated 20 times. The vibration testing equipment used in the experiments is shown in Figure \ref{fig5}.
\begin{table}[ht]
    \centering
    \caption{Vibration Testing Equipment}
    \begin{tabular}{|c|c|c|c|}
    \hline
    Equipment & Model & Measurement Range&number \\ \hline
    Impact Hammer & Model PCB-086D20 & ±5000 lbf pk (±22240 N pk) &1\\ \hline
    Data Acquisition Card & Model NI 9231 & ——&2 \\ \hline
    Tri-axial Accelerometer & Model PCB-356A45 & ±50 g pk (±490 m/s² pk)&2 \\ \hline
    Uni-axial Accelerometer & Model PCB-333B32 & ±50 g pk (±490 m/s² pk)&6  \\ \hline
    \end{tabular}
    \label{tab:equipment}
    \end{table}
    
The locations 
of the two interchangeable excitation points and all sensor measurement points are shown in Figure \ref{fig4}. The excitation direction was along the negative z-axis. 
In addition 
to the three-axis sensors located at both ends of the bridge, the remaining uniaxial sensors recorded the vertical acceleration of the experimental model (positive z-axis). The 
positive directions for the three-axis accelerometers corresponded to the positive z-axis, x-axis, and y-axis as shown in the Figure \ref{fig5}.

\begin{figure}[!ht]
    \centering
    \begin{subfigure}{0.31\textwidth}
        \centering
        \includegraphics[width=\textwidth]{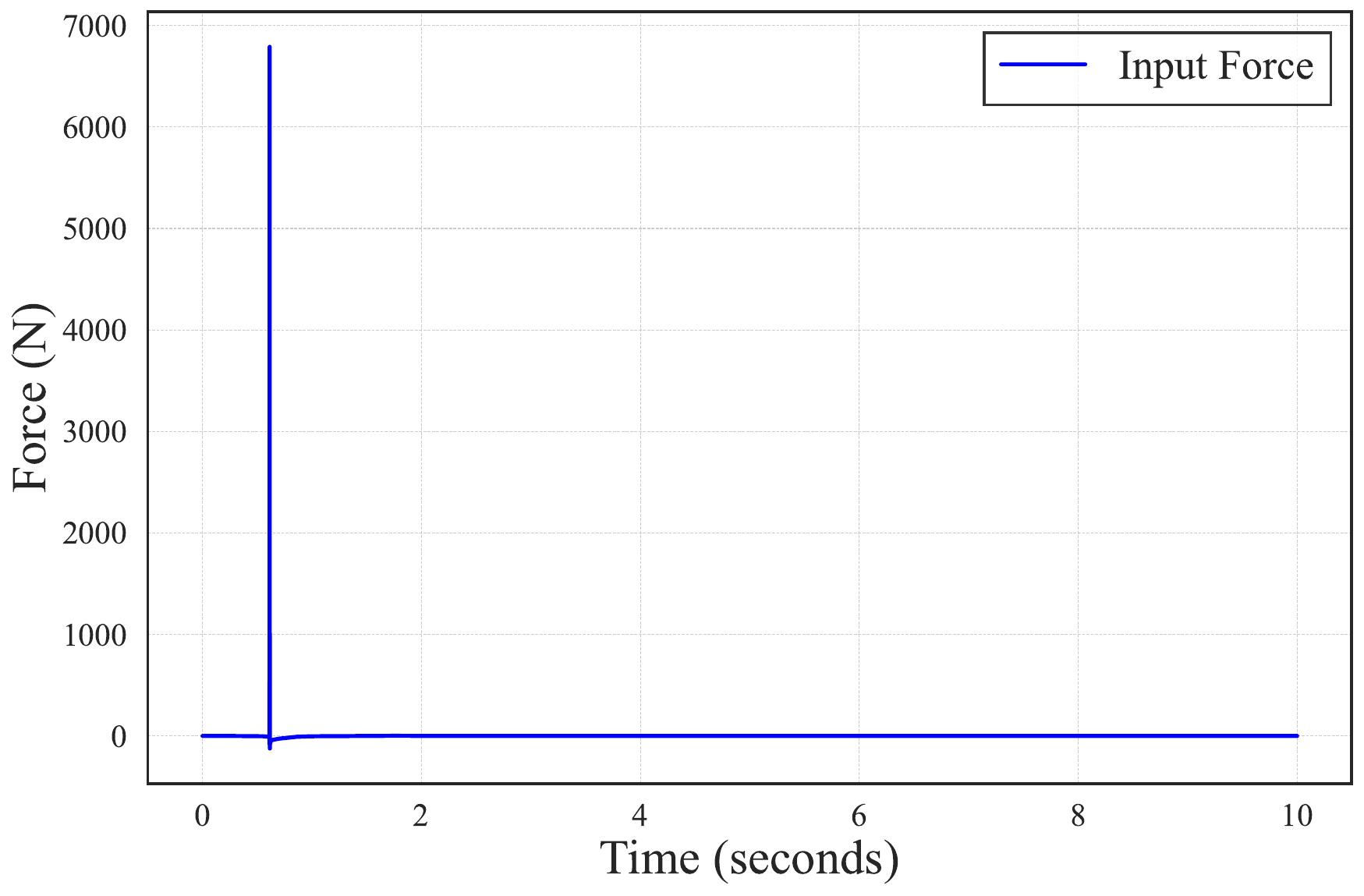}
        \caption{Input Force}
    \end{subfigure}
    \hspace{0.5em}
    \begin{subfigure}{0.31\textwidth}
        \centering
        \includegraphics[width=\textwidth]{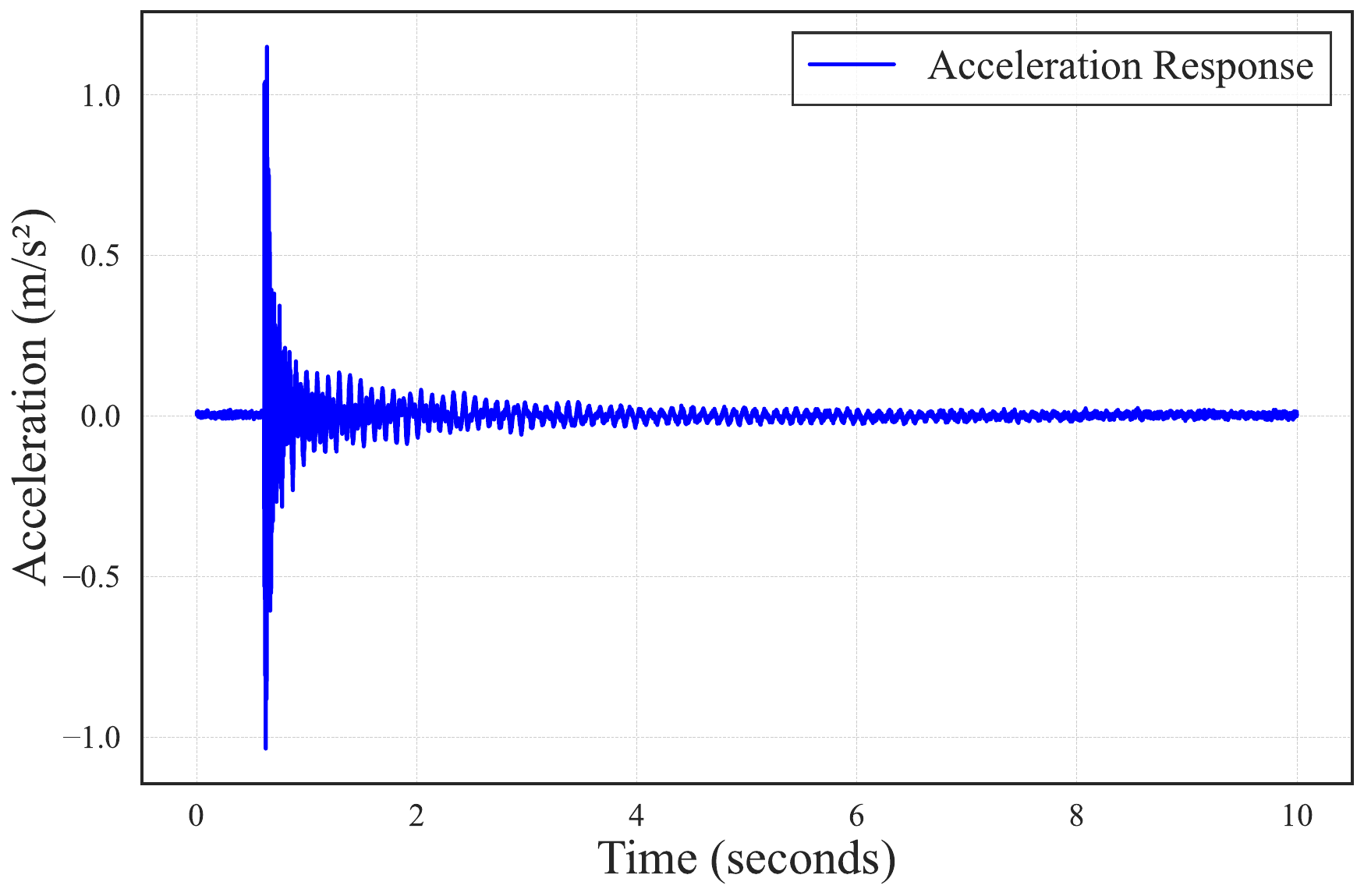}
        \caption{Output Acceleration}
    \end{subfigure}
    \hspace{0.5em}
    \begin{subfigure}{0.31\textwidth}
        \centering
        \includegraphics[width=\textwidth]{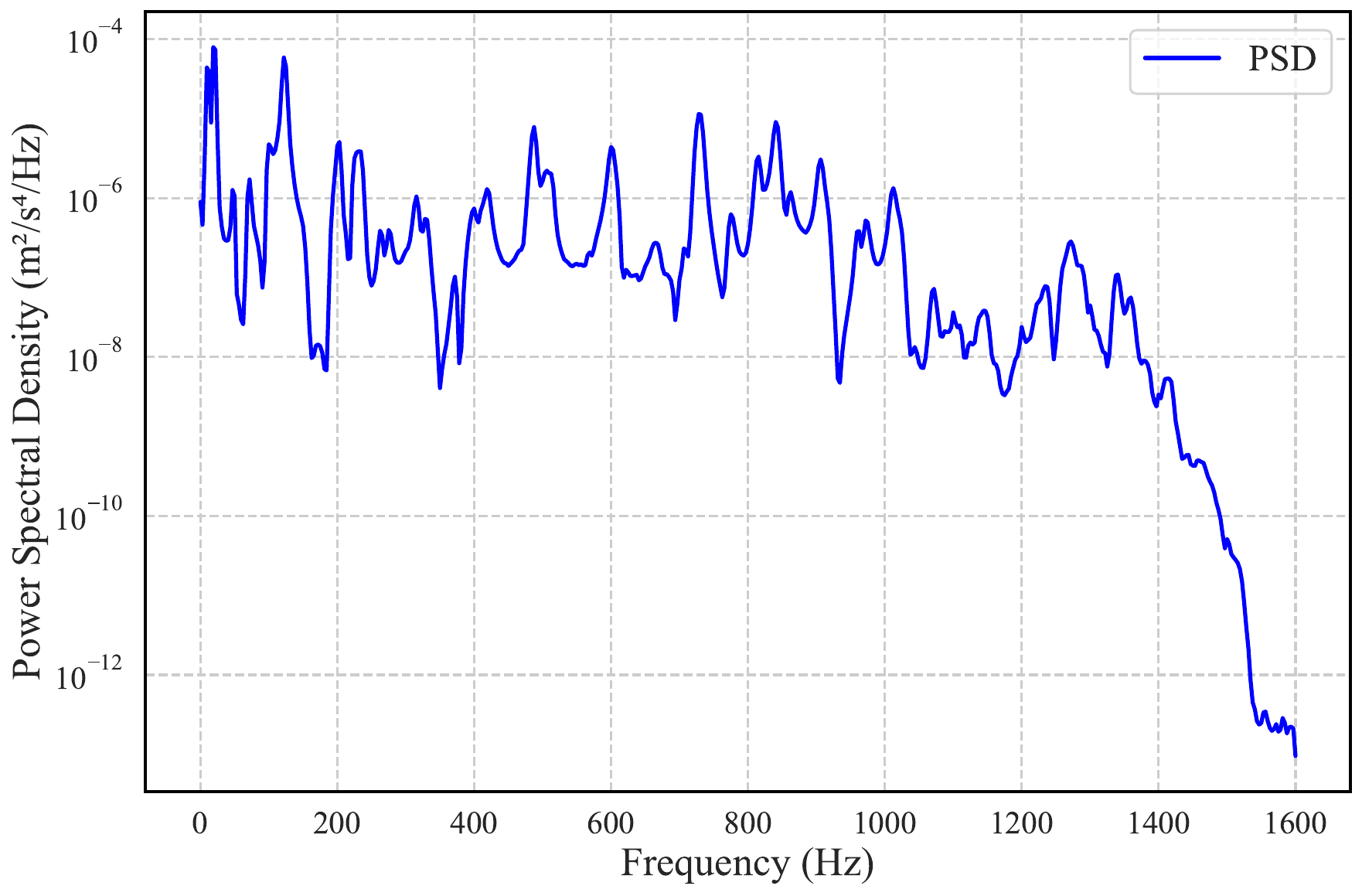}
        \caption{Power Spectral Density}
    \end{subfigure}
    \caption{Experimental Data Display}
    \label{fig:input_and_output}
\end{figure}

This modal analysis is based on data collected from a scaled experimental setup. A total of 20 sets of 10-second raw data segments were utilized, without applying any 
filtering or noise reduction techniques. The analysis incorporated acceleration data collected from eight sensors.
In contrast to numerical simulations, real-world structures inherently have infinite degrees of freedom. However, due to the nature of the excitation, limitations of the 
experiment, and the inherent characteristics of the structure itself, it is often difficult to identify higher-order frequencies and modes. Therefore, in this study, we 
focus on comparing the first nine modes and their corresponding frequencies.
The monitored data in the scaled experiment represent impulse responses with the input load. The acceleration of the first sensor and the acceleration power spectral 
density (PSD) are shown in Figure \ref{fig:input_and_output}.

\subsubsection{Experimental Modal Analysis}
This section aims to compare the performance of PCSSI, SSI-Cov, and SSI-Data in experimental modal analysis, focusing on modal frequency identification and mode 
shape characterization. To ensure a fair comparison, we used the same data matrix \( Y_f, Y_p \in \mathbb{R}^{50 \times 31000} \) for both SSI-Cov and SSI-Data, 
while for PCSSI, we used \( Y_f \in \mathbb{R}^{20 \times 31000} \) and \( Y_p \in \mathbb{R}^{50 \times 31000} \). Additionally, all three methods employed the 
same stabilization diagram algorithm with identical parameter settings. Due to the possibility of errors in modal analysis results, even after filtering through 
the stabilization diagram algorithm, we manually selected the first nine modal frequencies obtained by PCSSI for comparison based on their distribution characteristics, 
as shown in Table \ref{tab:PCSSI_data1}. In essence, the stabilization diagram can be considered a data-driven 
clustering algorithm, while manual selection serves as a form of human-guided clustering. The raw modal data extracted from the stabilization diagram using PCSSI, 
SSI-Cov, and SSI-Data can be found in \ref{raw_data}.

\begin{table}[ht]
\centering
\caption{PCSSI Artificially Selected Modal Frequencies}\label{tab:PCSSI_data1}
\begin{tabular}{|c|c|c|c|c|c|c|c|c|c|}
\hline
\# & Mode 1 & Mode 2 & Mode 3 & Mode 4 & Mode 5 & Mode 6 & Mode 7 & Mode 8 & Mode 9 \\
\hline
1  & 98.09  & 153.37  & 227.41  & 320.19  & 416.84  & 512.25  & 602.57  & 728.63  & 820.48  \\
2  & 98.33  & 153.46  & 227.87  & 319.93  & -  & 512.40  & 601.63  & 729.61  & 818.19  \\
3  & 98.20  & 153.19  & 229.49  & 319.65  & 415.96  & 513.48  & 601.07  & 729.42  & 813.67  \\
4  & 98.98  & 154.13  & 227.62  & 318.22  & 405.95  & 511.32  & 608.63  & 732.82  & 821.43  \\
5  & 99.26  & 154.17  & 228.15  & 318.55  & -  & 510.82  & 608.41  & 730.63  & 822.90  \\
6  & 98.45  & 153.57  & 227.54  & 319.41  & 408.52  & 511.01  & 609.74  & 731.02  & 820.57  \\
7  & 98.51  & 153.56  & 227.77  & 319.34  & 408.63  & 510.85  & 608.44  & 730.07  & 822.82  \\
8  & 103.46 & 157.50  & 228.52  & 318.15  & -  & 511.29  & 608.16  & 728.31  & 817.55  \\
9  & 99.32  & 155.42  & 227.97  & 318.14  & 407.39  & 511.58  & 607.65  & 730.63  & 818.56  \\
10 & 98.64  & 153.87  & 227.64  & 319.34  & 409.47  & 511.69  & 607.78  & 728.94  & 822.87  \\
11 & 100.19 & 156.18  & 228.48  & 318.18  & 407.22  & 511.78  & 607.78  & 730.10  & 818.44  \\
12 & 99.53  & 156.57  & 227.50  & 318.21  & -  & 511.16  & -  & 731.06  & 823.03  \\
13 & 99.78  & 154.80  & 228.29  & 318.71  & 412.79  & 511.62  & 609.13  & 729.58  & 822.49  \\
14 & 99.45  & 154.03  & 227.78  & 317.73  & -  & 510.81  & 609.18  & 731.09  & 825.82  \\
15 & 98.26  & 153.37  & 227.44  & 319.17  & -  & 512.50  & 606.55  & 728.81  & 820.11  \\
16 & 98.78  & 153.79  & 228.00  & 319.49  & -  & 511.22  & -  & 729.58  & 822.61  \\
17 & 98.46  & 153.76  & 227.47  & 319.54  & 408.20  & 511.03  & 607.99  & 729.39  & 821.86  \\
18 & 101.74 & 155.77  & 227.09  & 317.52  & -  & 505.98  & 603.42  & 720.48  & 811.85  \\
19 & 100.96 & 157.37  & 227.41  & 317.38  & 410.54  & 507.01  & 600.55  & 723.19  & 846.56  \\
20 & 101.51 & -  & 227.34  & 317.33  & 410.18  & 507.14  & -  & -  &  846.37 \\
\hline
\end{tabular}
\end{table}

By comparing the results in Table \ref{tab:PCSSI_data1}, Table \ref{tab:SSI_COV_data}, and Table \ref{tab:SSI_Data_data}, it is evident that both PCSSI and SSI-Data exhibit 
advantages in terms of accuracy and stability, with PCSSI demonstrating particularly strong performance in identifying low-order modal frequencies. This advantage becomes 
especially prominent when the number of experimental trials and available data observations are limited. In contrast, as shown in Table \ref{tab:SSI_COV_data}, the SSI-Cov 
method tends to underestimate modal frequencies and often misses certain low-order frequencies. 

\begin{figure}[!ht]
    \centering
    \begin{subfigure}{0.35\textwidth}
        \centering
        \includegraphics[width=\textwidth]{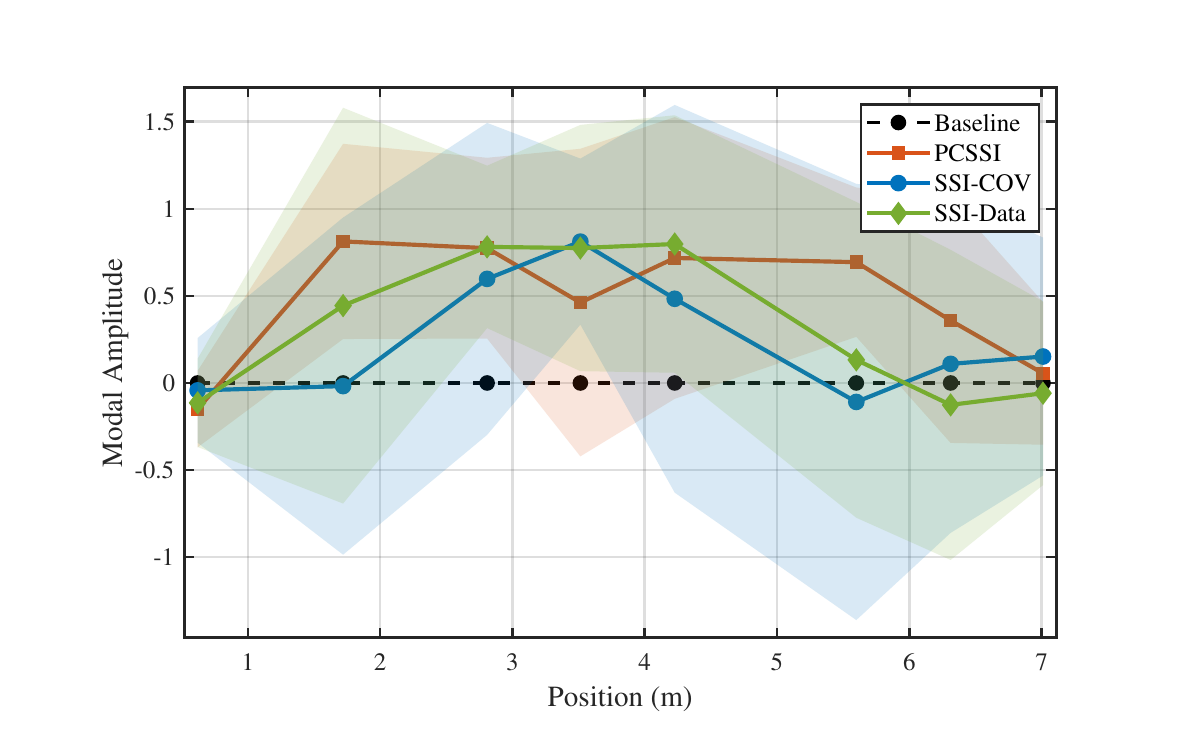}
        \caption{First-order}
    \end{subfigure}
    \hspace{-1.5em}
    \begin{subfigure}{0.35\textwidth}
        \centering
        \includegraphics[width=\textwidth]{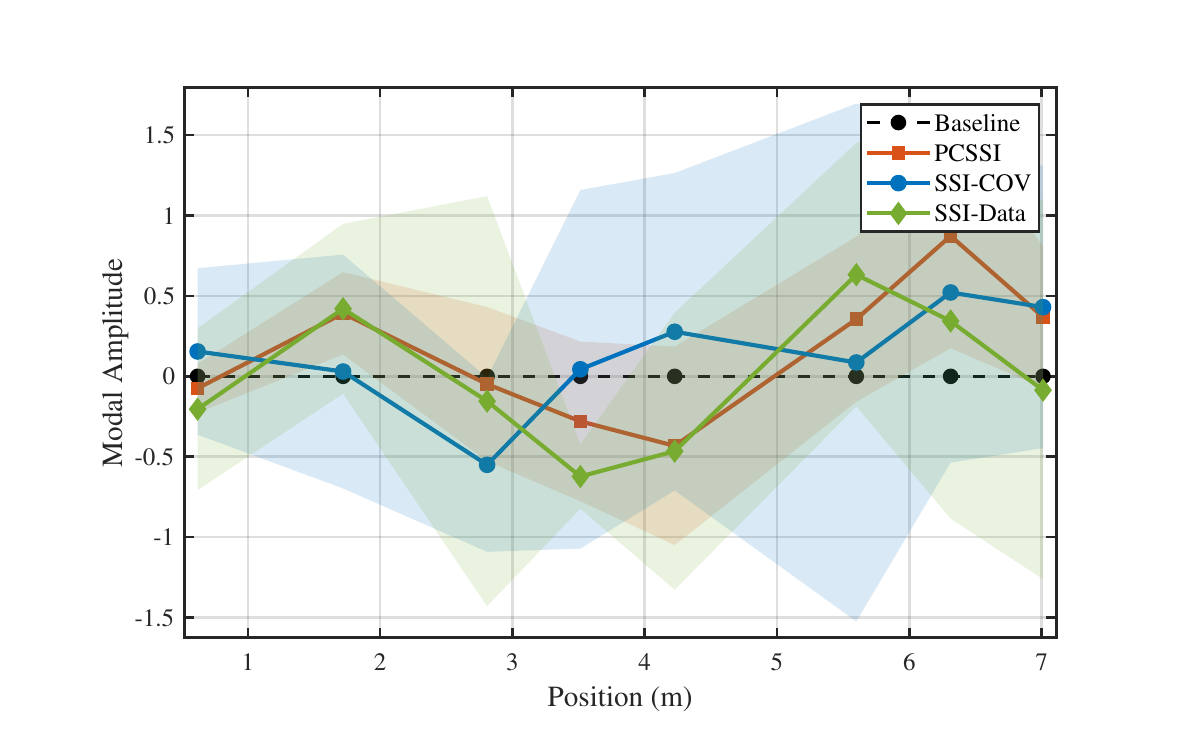}
        \caption{Second-order}
    \end{subfigure}
    \hspace{-1.5em}
    \begin{subfigure}{0.35\textwidth}
        \centering
        \includegraphics[width=\textwidth]{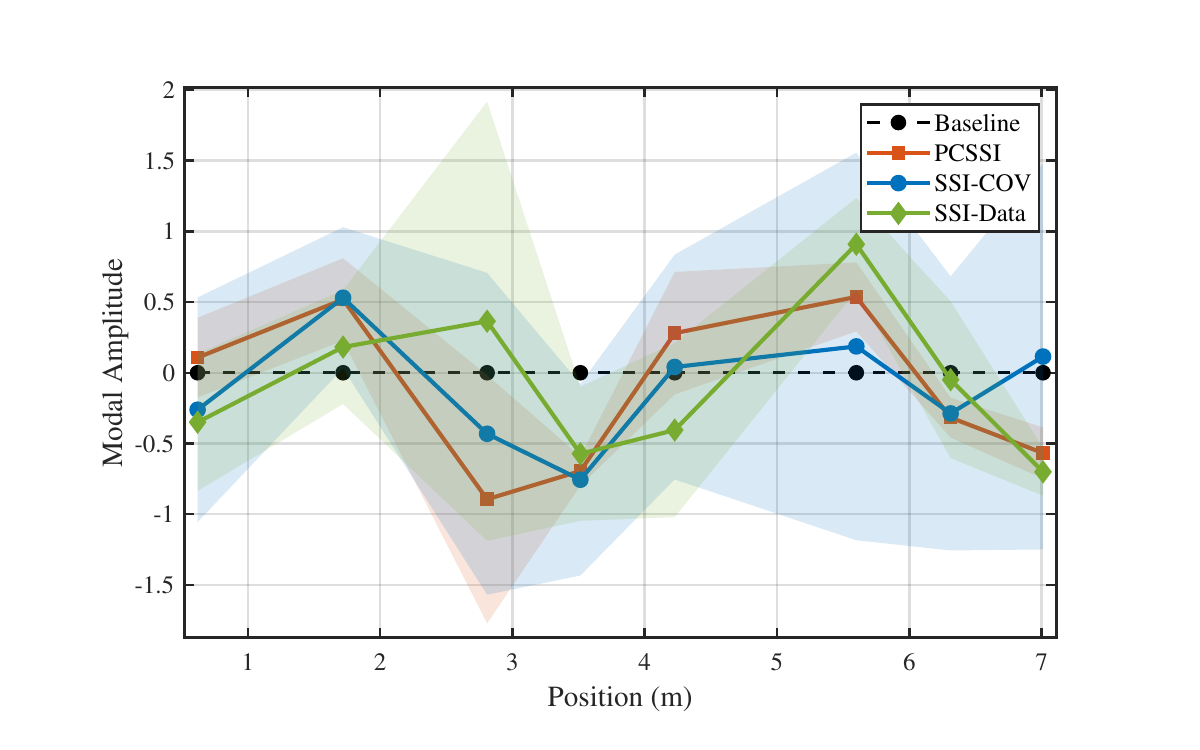}
        \caption{Third-order}
    \end{subfigure}
  
    \vspace{-0.5em}

    \begin{subfigure}{0.35\textwidth}
        \centering
        \includegraphics[width=\textwidth]{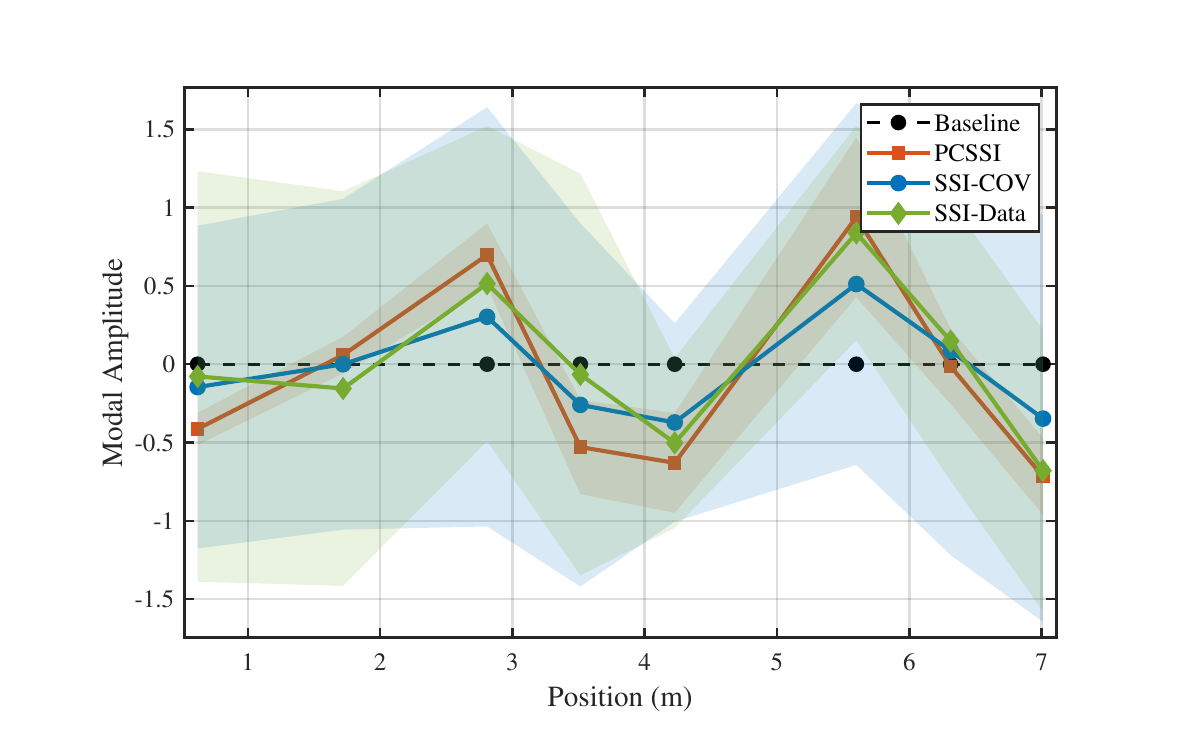}
        \caption{Fourth-order}
    \end{subfigure}
    \hspace{-1.5em}
    \begin{subfigure}{0.35\textwidth}
        \centering
        \includegraphics[width=\textwidth]{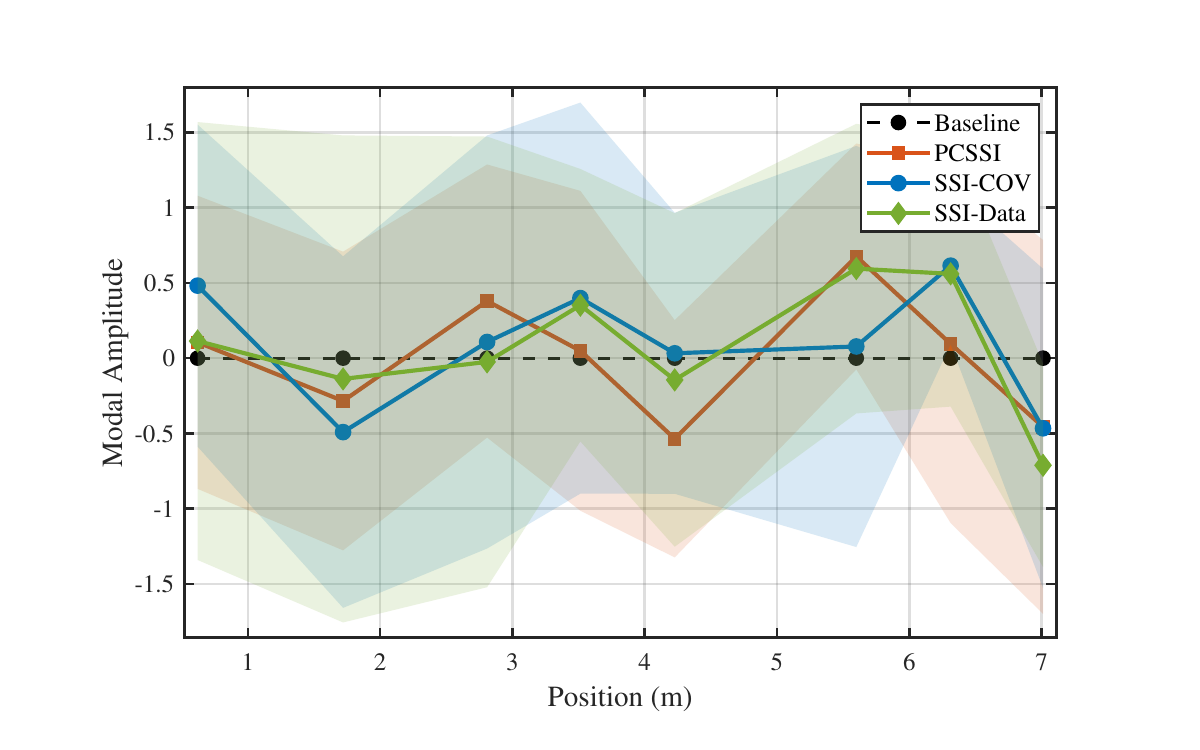}
        \caption{Fifth-order}
    \end{subfigure}
    \hspace{-1.5em}
    \begin{subfigure}{0.35\textwidth}
        \centering
        \includegraphics[width=\textwidth]{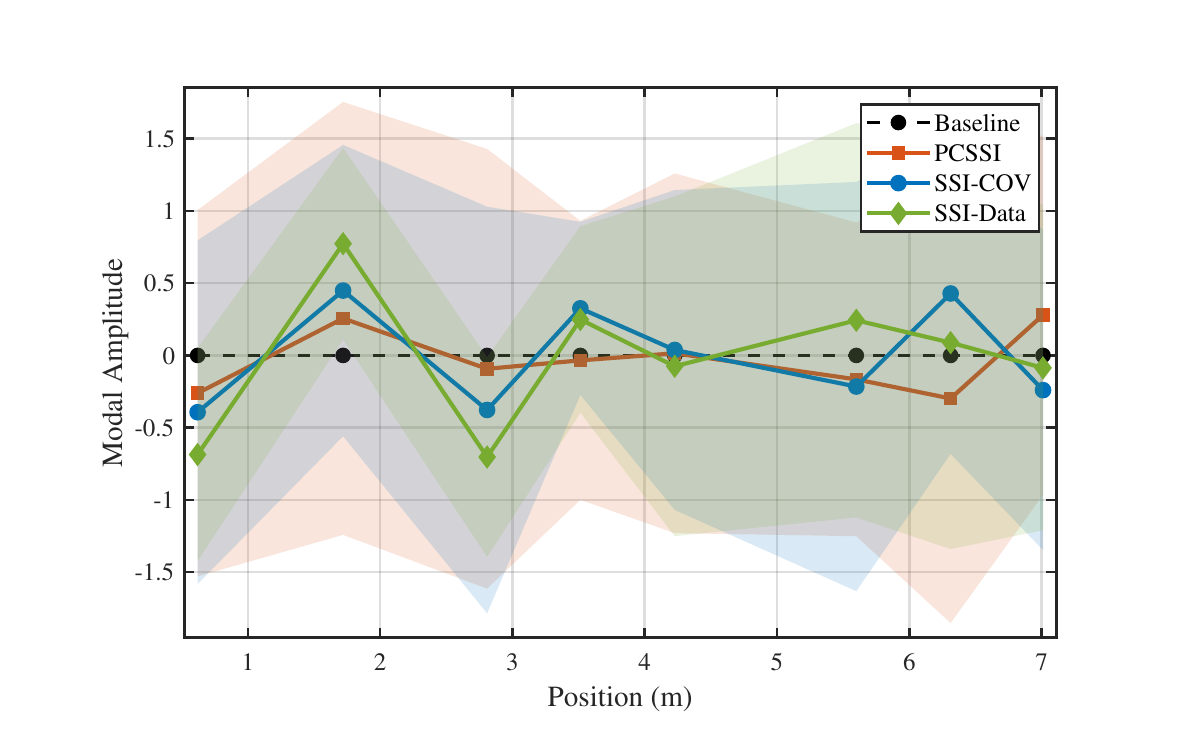}
        \caption{Sixth-order}
    \end{subfigure}
    \caption{Comparison of the first six mode shapes identified by PCSSI, SSI-Cov, and SSI-Data.}
    \label{fig:PCSSI_and_SSI-Cov_mode_shape_experimental}
\end{figure}

This makes it challenging to directly obtain an accurate mean of the mode shapes, 
requiring manual selection for reliable assessment. For example, in Table \ref{tab:SSI_COV_data}, the first-order frequency estimates for trials 11 to 14 exhibit significant deviations.
In addition to the comparison of modal frequencies, we also analyzed the first six experimental mode shapes, computing the mean and 95\% confidence interval for each mode. Although 
the scaled model used in the experiment is a curved bridge structure, we chose to project the mode shapes onto a flat plane for clarity and ease of visualization. The detailed results 
are presented in Figure \ref{fig:PCSSI_and_SSI-Cov_mode_shape_experimental}.  

Each subplot illustrates the specific mode shapes identified using the PCSSI, SSI-Cov, and SSI-Data methods, with the shaded regions around the mode shapes representing the confidence 
intervals. Notably, PCSSI consistently demonstrates greater stability in variance control across all mode shapes, with its variance either lower than or comparable to that of the 
SSI-Cov and SSI-Data methods.  
Overall, the results of PCSSI and SSI-Data are largely consistent, whereas SSI-Cov exhibits greater discrepancies in most cases. This discrepancy primarily arises from the tendency 
of SSI-Cov to underestimate modal frequencies, leading to errors in determining modal frequency orders. Consequently, these errors affect the calculation of the mean and confidence 
intervals, resulting in larger deviations in mode shape estimation and, in some cases, producing irregular or non-physical mode shapes.

\begin{figure}[!ht]
    \centering
    \begin{subfigure}{0.33\textwidth}
        \centering
        \includegraphics[width=\textwidth]{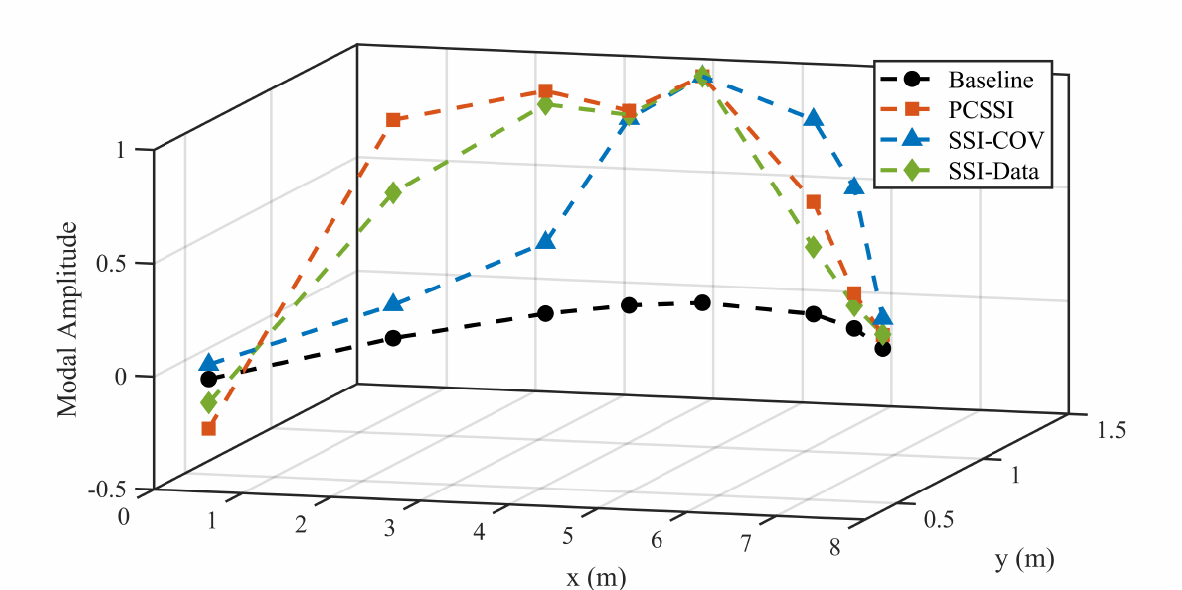}
        \caption{First-order}
    \end{subfigure}
    \hspace{-1.5em}
    \begin{subfigure}{0.33\textwidth}
        \centering
        \includegraphics[width=\textwidth]{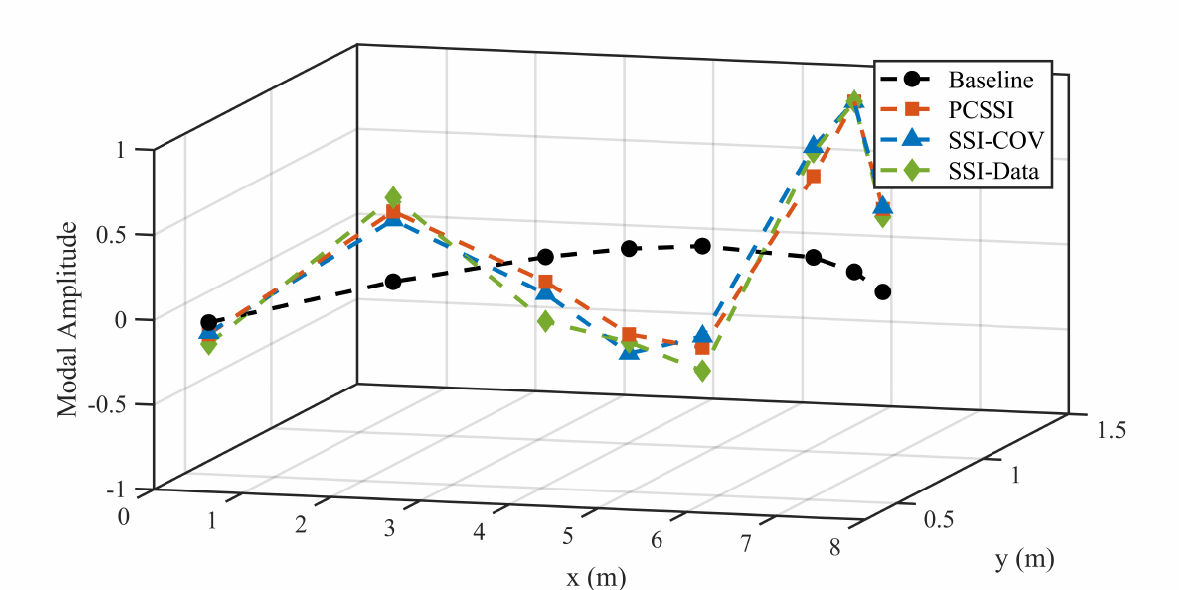}
        \caption{Second-order }
    \end{subfigure}
    \hspace{-1.5em}
    \begin{subfigure}{0.33\textwidth}
        \centering
        \includegraphics[width=\textwidth]{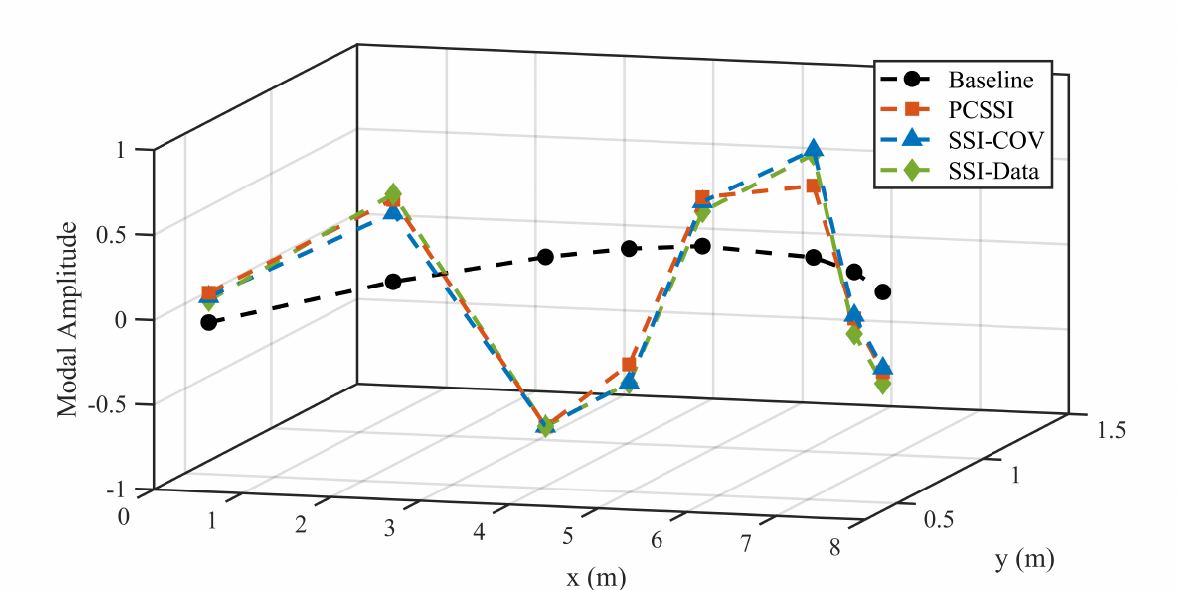}
        \caption{Third-order}
    \end{subfigure}
  
    \vspace{0.5em}

    \begin{subfigure}{0.33\textwidth}
        \centering
        \includegraphics[width=\textwidth]{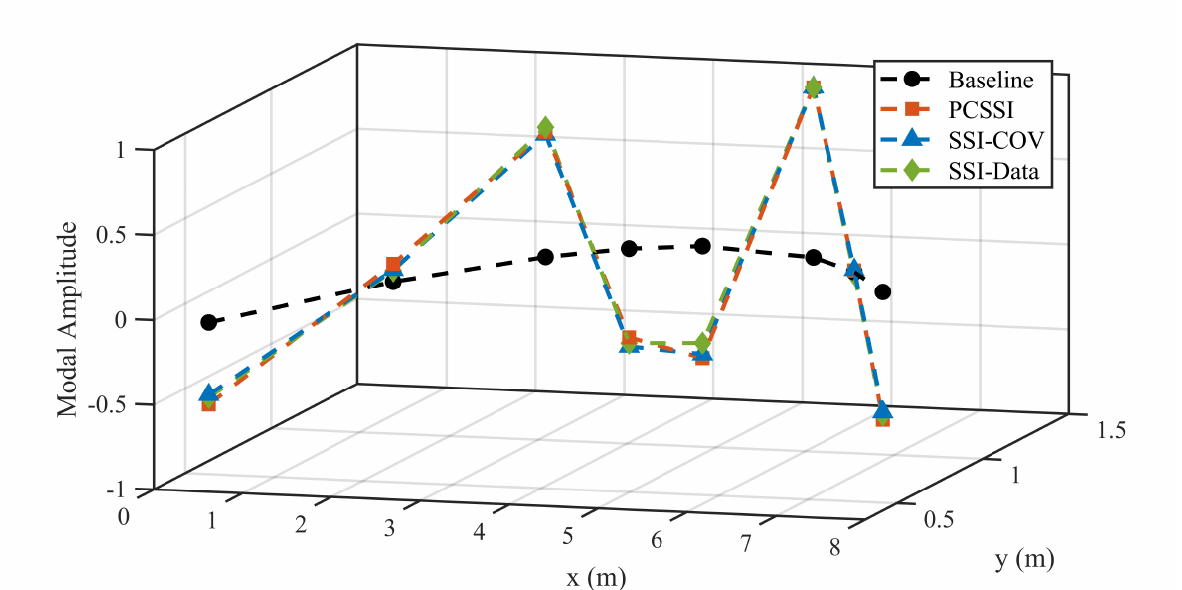}
        \caption{Fourth-order}
    \end{subfigure}
    \hspace{-1.5em}
    \begin{subfigure}{0.33\textwidth}
        \centering
        \includegraphics[width=\textwidth]{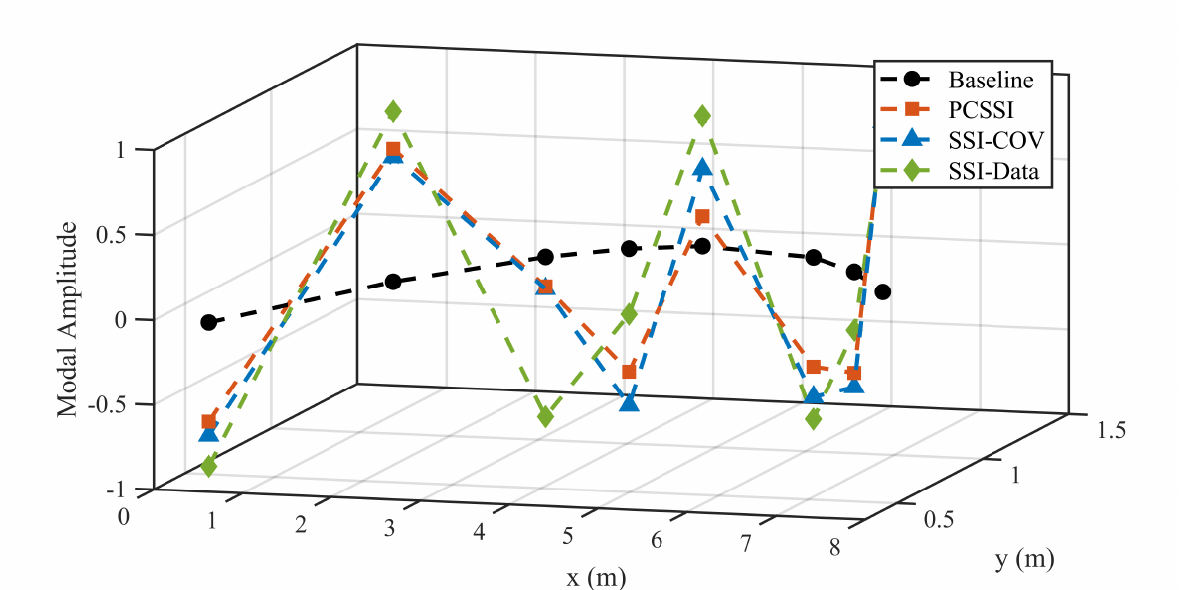}
        \caption{Fifth-order}
    \end{subfigure}
    \hspace{-1.5em}
    \begin{subfigure}{0.33\textwidth}
        \centering
        \includegraphics[width=\textwidth]{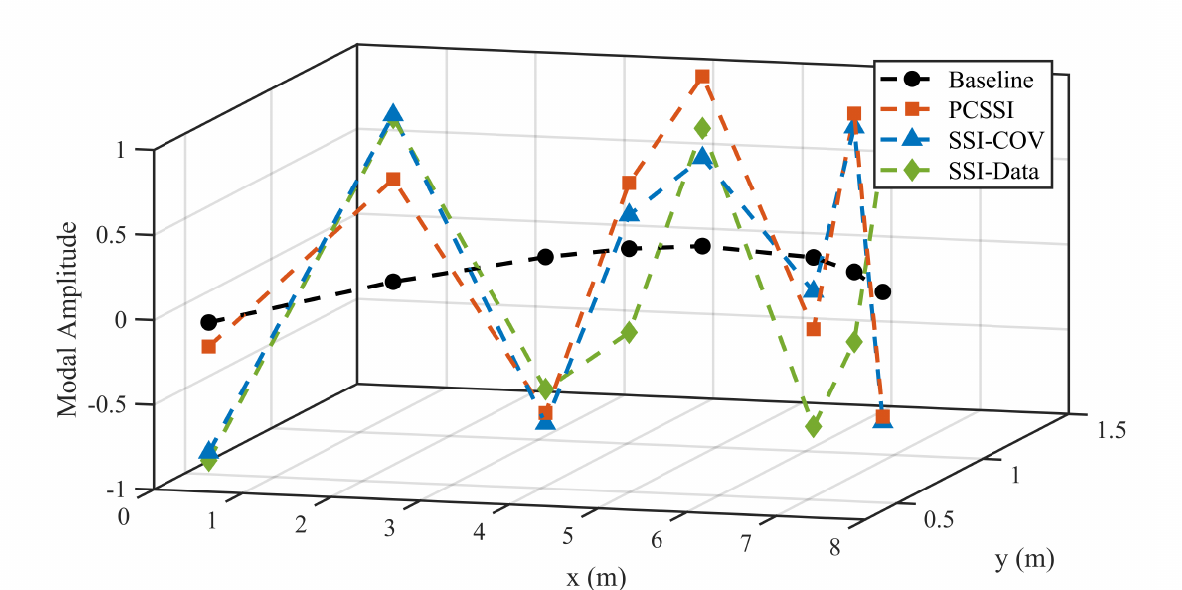}
        \caption{Sixth-order}
    \end{subfigure}
    \caption{Comparison of the first six mode shapes identified by PCSSI, SSI-Cov, and SSI-Data.}
    \label{fig:PCSSI_and_SSI-Cov_mode_shape_experimental1}
\end{figure}

To facilitate a clear comparison, we selected the fourth dataset and applied the PCSSI, SSI-Cov, and SSI-Data methods to perform a representative 3D visualization of the 
modal analysis (see Figure \ref{fig:PCSSI_and_SSI-Cov_mode_shape_experimental1}). From the results, it can be observed that for the first-order frequency, PCSSI and SSI-Data 
yield highly consistent results, while SSI-Cov exhibits a noticeable deviation. From a physical perspective, the outcomes from PCSSI and SSI-Data are considered more reliable.
In addition, apart from minor differences in the first-order mode shape, all three methods demonstrate stable and consistent performance in identifying subsequent mode shapes, 
showing minimal variation that closely aligns with their respective mean shapes. 

Among these, the PCSSI method shows the most pronounced advantages in terms of accuracy and stability, 
particularly under complex experimental conditions or limited data scenarios, effectively capturing the true characteristics of the structural modes. On the other hand, 
the SSI-Data method typically requires a larger amount of data to fully leverage its advantages and achieve comparable stability.
In comparison, the SSI-Cov method exhibits weaker performance regarding modal frequency identification and mode shape stability. When experimental data is limited or the 
signal-to-noise ratio is low, it tends to introduce larger errors, significantly complicating manual analysis and interpretation. Therefore, in practical applications, 
the SSI-Cov method usually necessitates supplemental analysis guided by physical insights to correctly identify the modal orders and frequencies, thus ensuring the accuracy 
of subsequent evaluations.

\section{Conclusions}\label{sec:conclusions}
This paper presents a unified framework for Stochastic Subspace Identification (SSI) methods, integrating instrumental variables to establish a connection between 
the SSI-Cov and SSI-Data approaches. Theoretical analysis demonstrates that, given an infinite amount of observational data, both methods converge to the same solution.
Building upon this foundation, this paper introduces the Principal Component Stochastic Subspace Identification (PCSSI) algorithm, which significantly enhances noise reduction 
capability under limited data conditions and reduces variance while maintaining computational efficiency comparable to classical SSI methods.   

The superior performance of PCSSI stems from its dual noise suppression mechanisms. First, SVD effectively extracts the principal signal subspace of \( Y_p \), filtering out 
most noise components. Second, by projecting the measurement matrix \( Y_f \) onto this principal signal subspace, PCSSI achieves excellent noise reduction and numerical 
stability, resulting in significantly lower variance in frequency and mode shape estimations compared to SSI-Cov and SSI-Data in most cases.
Numerical simulations on the Lysefjord bridge dataset, which contains 70\% non-stationary noise, demonstrate that PCSSI achieves minimal frequency deviation while 
maintaining high mode shape accuracy across 5000 Monte Carlo trials. 

Additionally, theoretical analysis of matrix condition numbers confirms that PCSSI and 
SSI-Data offer superior numerical stability compared to SSI-Cov. It is also shown that SSI-Cov may suffer from potential numerical instability when a sufficiently 
large amount of data is available. This claim is further validated through 5000 Monte Carlo simulations.  

Experimental validation on a 1:12 scaled overpass model further demonstrates that, compared to classical SSI methods, PCSSI accurately identifies modal frequencies 
while reconstructing mode shapes with strong stability and adherence to physical principles. These results confirm PCSSI's capability to process large-scale sensor 
data and operate effectively in complex noise environments, making it highly applicable to real-world engineering structures.  

Overall, PCSSI integrates the theoretical robustness of SSI-Data with the computational efficiency of SSI-Cov, offering an accurate and efficient solution for modal 
analysis in civil engineering. This method not only advances research in stochastic subspace identification but also provides a scalable approach for structural health 
monitoring and operational modal analysis.

\section{Acknowledgments}
This research has been supported by the China National Key R\&D Program (2022YFB26-02103), National Natural Science Foundation of China (General Program, 52378294), 
GuangDong Basic and Applied Basic Research Foundation (2024A1515013224), and Shenzhen Science and Technology Program (GXWD2023112914310001).


\appendix
\section{Orthogonal Projection}\label{projection}
\subsection{Orthogonal Projection onto the Row Space}
This section aims to use SVD to prove two expressions related to the projection matrix, as detailed below.
The orthogonal projection of a matrix \( A \in \mathbb{R}^{n \times m} \) onto the 
row space of a matrix \( B \in \mathbb{R}^{m \times n} \) is mathematically defined as:  
\begin{equation}
    A/B = A B^\top (B B^\top)^\dagger B = A \Pi_B, \quad \Pi_B \triangleq B^\top (B B^\top)^\dagger B,
\end{equation}
where \( \dagger \) denotes the Moore-Penrose pseudoinverse, and \( \Pi_B \) is the 
projection matrix. This matrix can be computed numerically robustly and efficiently 
using the LQ decomposition, which is the transpose of the QR decomposition.
This approach is commonly used in the classical SSI-Data method. However, to 
illustrate the relationship between the SSI-Cov and SSI-Data methods in the limit 
of infinite data, this paper adopts the SVD approach for a conceptual explanation. 
 The detailed analysis is as follows: First, let the SVD of \( B \) be 
expressed as:
\begin{equation}
    B = U_k S_k V_k^\top,
\end{equation}
where \( U_k \in \mathbb{R}^{m \times k} \) is the matrix of left singular vectors, \( S_k \in \mathbb{R}^{k \times k} \) is the diagonal matrix of singular values,  
\( V_k \in \mathbb{R}^{n \times k} \) is the matrix of right singular vectors, and \( k = \text{rank}(B) \).  
Using the SVD, the matrix \( B B^\top \) and its pseudoinverse \((B B^\top)^\dagger\) can be represented as:  
\begin{equation}
    B B^\top = U_k S_k^2 U_k^\top, \quad (B B^\top)^\dagger = U_k S_k^{-2} U_k^\top,
\end{equation}
where \( S_k^2 \) contains the squared singular values of \( S_k \), and \( S_k^{-2} \) contains their reciprocals.  
Substituting the SVD into the expression \( \Pi_B = B^\top (B B^\top)^\dagger B \), we obtain:  
\begin{equation}
    \Pi_B = (V_k S_k U_k^\top)(U_k S_k^{-2} U_k^\top)(U_k S_k V_k^\top).
\end{equation}
Using the orthogonality properties of the SVD (\( U_k^\top U_k = I_k \) and \( U_k U_k^\top \) being the projection matrix of \( B B^\top \)), we simplify step-by-step:  
\begin{equation}
    \Pi_B = V_k S_k S_k^{-2} S_k V_k^\top = V_k V_k^\top.
\end{equation}
This result demonstrates that the structure of the orthogonal projection matrix \( \Pi_B \) is determined by the right singular vectors \( V_k \).

\subsection{The Effect of a Matrix Multiplied by Its Pseudoinverse}
Let the SVD decomposition of the matrix \( \Psi^\top \) be expressed as:  
\begin{equation}
    \Psi = U_1 \Sigma_1 V_1^\top, \quad \Psi^\top = V_1 \Sigma_1 U_1^\top, 
\end{equation}
where \( U_1 \) is an \( m \times r \) matrix containing the left singular vectors, \( \Sigma_1 \) is an \( r \times r \) diagonal matrix containing the nonzero singular values,  
\( V_1 \) is an \( n \times r \) matrix containing the right singular vectors,  
and \( r = \text{rank}(\Psi^\top) \).  
The pseudoinverse of the matrix \( (\Psi^\top)^\dagger \) is given by:  
\begin{equation}
    (\Psi^\top)^\dagger = U_1 \Sigma_1^{-1} V_1^\top.
\end{equation}
Now, consider the product \( \Psi^\top (\Psi^\top)^\dagger \), and derive it using an approach similar to the previous section:  
\begin{align}
    \Psi^\top (\Psi^\top)^\dagger &= (V_1 \Sigma_1 U_1^\top)(U_1 \Sigma_1^{-1} V_1^\top) \\  
    &= V_1 \Sigma_1 \Sigma_1^{-1} V_1^\top \\  
    &= V_1 V_1^\top.
\end{align}
Thus, the matrix \( \Psi^\top (\Psi^\top)^\dagger \) is equal to \( V_1 V_1^\top \), which is a projection matrix that projects onto the 
column space of \( \Psi^\top \). In other words, this matrix projects any vector onto \( \text{row}(\Psi) \) (the row space of \( \Psi \), 
which is the same as the column space of \( \Psi^\top \)).  

\section{Performance Differences Between PCSSI and SSI Methods}\label{numerical_stability}

Although, as analyzed in Section \ref{sec:SSI}, theoretically, PCSSI, SSI-CoV, and SSI-Data can completely eliminate noise and converge to the same solution under 
infinite data conditions, practical applications are constrained by factors such as sampling time and computational resources. As a result, modal analysis is typically 
performed with finite data.
Considering the role of instrumental variables and assuming that the noise has a mean of zero, we obtain:
\begin{equation}\label{eq_{19}} 
    \frac{1}{j} Y_f\Psi^\top = \frac{1}{j} (O_i \hat{X}_i \Psi^\top) + O_j(\varepsilon) \implies \mathbb{E} \frac{1}{j} [Y_f\Psi^\top] = O_i \hat{X}_i \Psi^\top 
\end{equation}
Thus, even with finite data, the mean of the SSI method should remain consistent. Consequently, the performance differences primarily depend on variance, 
computational efficiency, and numerical stability.
This section analyzes the performance differences between PCSSI, SSI-CoV, and SSI-Data under limited data conditions from these three perspectives and finally 
validates our reasoning through numerical examples. 

\subsection{Numerical Computation Efficiency Analysis}

From the perspective of variance, it is theoretically possible to analyze its impact by computing the covariance matrix. However, due to the nonlinear nature of eigenvalue 
decomposition, theoretical analysis is often complex or requires simplifications\cite{reynders2008uncertainty,reynders2016uncertainty,reynders2021uncertainty}.  Therefore, rather than performing a quantitative theoretical analysis of variance, 
this study employs Monte Carlo simulations for empirical validation (see \ref{numerical2} for details).

Qualitatively, both PCSSI and SSI-Data utilize projection matrices as instrumental variables. However, PCSSI distinguishes itself by applying SVD prior to projection,
 which removes components with small singular values and isolates the primary signal subspace. This process inherently enhances noise reduction, leading to an expected 
 reduction in variance compared to SSI-Data. While a theoretical variance comparison between PCSSI and SSI-CoV is more challenging, empirical results from 
 practical applications generally indicate that PCSSI exhibits lower variance than SSI-CoV.

In terms of computational efficiency, for an \( m \times n \) matrix (assuming \( m \geq n \)), the computational complexities of QR decomposition and SVD using classical 
numerical algorithms are summarized in Table \ref{tab:qr_svd_complexity}. Since SSI-CoV does not require the use of projection matrices, it generally achieves the fastest 
computational performance under identical dimensions. For both PCSSI and SSI-Data, QR decomposition typically has lower computational complexity than SVD, making it more 
suitable for fast computations. However, SVD provides superior noise reduction at the expense of higher computational cost, making it more effective for handling 
ill-conditioned matrices. 

\begin{table}[h]
    \centering
    \caption{Computational Complexity of QR and SVD Decompositions}
    \begin{tabular}{|c|c|c|}
        \hline
        Decomposition Method & (\( m \times n \) Matrix, \( m \geq n \)) & (\( n \times n \) Square Matrix) \\
        \hline
        QR Decomposition (Householder) & \( O(mn^2) \) & \( O(n^3) \) \\
        \hline
        SVD (Golub-Kahan) & \( O(mn^2) \) & \( O(n^3) \) \\
        \hline
    \end{tabular}
    \label{tab:qr_svd_complexity}
\end{table}

For SSI-Data and PCSSI, different sizes of Hankel matrices are used. Specifically, the Hankel matrix for SSI-Data is of size \( \mathbb{R}^{2im \times j} \), whereas for 
PCSSI it is \( \mathbb{R}^{im \times j} \). This results in computational complexities of \( O(2imj^2) \) and \( O(imj^2) \), respectively. Thus, under identical computational 
conditions, PCSSI exhibits computational efficiency comparable to that of SSI-Data.

In terms of numerical stability, SSI-CoV requires computing \( Y_fY_p^\top \), which can amplify the condition number of the matrix, introducing potential numerical 
instability risks. In contrast, PCSSI and SSI-Data extract the primary signal components from \( Y_p \) using projection matrices, which should enhance numerical stability.

\subsection{Numerical Computation Stability Analysis}

This section analyzes the numerical computation stability of the three methods by examining the condition numbers of the relevant product matrices, 
defined as the ratio of the largest to the smallest singular value. This analysis highlights the numerical stability differences among SSI-Cov, SSI-Data, and PCSSI. 
From Equation \ref{eq_{19}}, we derive the following fundamental expression:
\begin{equation}
    \frac{1}{j} Y_f\Psi^\top = \frac{1}{j}(O_i \hat{X}_i \Psi^\top) + O_j(\varepsilon)  
\end{equation}
Theoretically, minimizing the error term \( O_j(\varepsilon) \) is desirable. However, in practical numerical computations, this ideal scenario may 
introduce potential stability issues. Since \( \operatorname{rank}(O_i \hat{X}_i \Psi^\top) = n \), the smallest singular value is typically dominated 
by the noise term \( O_j(\varepsilon) \). As the data length \( j \) increases, the influence of noise diminishes, but the condition number of the
 matrix may grow significantly, thereby worsening numerical stability.

In particular, improper selection of the instrumental variable \( \Psi \) can result in a highly unbalanced singular value distribution, leading to an 
excessively large norm of the matrix product. Consequently, numerical errors associated with extremely small singular values may be significantly 
amplified, reducing computational stability.
The SSI-Cov method constructs the matrix
\begin{equation}
    M_{\text{Cov}} \doteq \frac{1}{j} Y_f Y_p^\top 
\end{equation}
with its condition number defined as
\begin{equation}
    \kappa(M_{\text{Cov}}) = \frac{\sigma_{\max}(M_{\text{Cov}})}{\sigma_{\min}(M_{\text{Cov}})}  
\end{equation}
where \( \kappa(\cdot) \) denotes the condition number of a matrix. As a result, while SSI-Cov theoretically 
benefits from increasing \( j \), in practical applications, excessive \( j \) can lead to numerical instability, necessitating a trade-off between numerical stability and noise 
suppression to achieve optimal performance.  

The SSI-Data method constructs an orthogonal projection matrix \( \Pi_{Y_p} \) by performing a QR decomposition on the past output matrix \( Y_p \). Theoretically, this matrix can 
also be expressed as:
\begin{equation}
    M_{\text{Data}} \doteq \frac{1}{j} Y_f\Pi_{Y_p} = \frac{1}{j} Y_f V_p V_p^\top 
\end{equation}
Since \( V_p V_p^\top \) is an orthogonal projection matrix, it follows that:
\begin{equation}
    \kappa(M_{\text{Data}}) = \kappa\left(\frac{1}{j}Y_f \Pi_{Y_p}\right) = \kappa\left(\frac{1}{j}Y_f\right). 
\end{equation}
In other words, from the perspective of numerical stability, the SSI-Data method is equivalent to addressing the condition number of \( Y_f \), thereby avoiding the 
additional instability introduced in the SSI-Cov method due to the interaction between \( Y_p \) and noise. In general, the following inequality holds:
\begin{equation}
    \kappa\left(\frac{1}{j}Y_f\right) < \kappa\left(\frac{1}{j}Y_fY_p^\top\right),  
\end{equation}
indicating that SSI-Data exhibits significantly better numerical stability compared to the SSI-Cov method.
Building upon the SSI-Data method, PCSSI further applies a truncated SVD to \( Y_p \) and constructs an orthogonal projection matrix using the 
truncated right singular matrix \( V_s \), resulting in the following projected matrix:
\begin{equation}
    M_{\text{PCSSI}} \doteq \frac{1}{j} Y_f V_s V_s^\top
\end{equation}
Since \( V_s V_s^\top \) is an orthogonal projection matrix, its condition number remains the same as that of the SSI-Data method and 
primarily depends on the singular value distribution of \( \frac{1}{j} Y_f V_s \):
\begin{equation}
    \kappa(M_{\text{PCSSI}}) = \kappa\left(\frac{1}{j} Y_f V_s V_s^\top \right) = \kappa\left(\frac{1}{j} Y_f\right). 
\end{equation}
In general, the condition number of \( M_{\text{PCSSI}} \) is significantly lower than that of \( M_{\text{Cov}} \) and is comparable to or slightly better 
than that of \( M_{\text{Data}} \), depending on the dataset and truncation level. By leveraging truncated projection, PCSSI balances noise suppression and effective 
signal extraction. Ideally, this approach enhances robustness against noise and improves numerical stability compared to SSI-Data. 

However, excessive truncation 
may cause the projected principal signal subspace to deviate from the true signal subspace, leading to the loss of essential signal components. This loss can make the 
remaining useful information overly sparse, increasing its susceptibility to numerical error amplification in subsequent computations.
Overall, the numerical stability of \( \frac{1}{j} Y_f \Psi^\top \) heavily depends on the design of the instrumental variable \( \Psi \). A well-chosen \( \Psi \) not 
only effectively eliminates noise but also ensures that the norm of the product matrix remains within a reasonable range, thereby maintaining the stability of subsequent 
matrix decompositions and inverse operations. Conversely, an improperly designed \( \Psi \) may lead to extreme deviations in the matrix norm , 
which can, in turn, result in high condition numbers and numerical instability.

\subsection{Numerical Verification of Theoretical Convergence Under Sufficiently Large Monitoring Data}\label{numerical2}

\begin{figure}[!ht]
    \centering
    \begin{subfigure}{0.29\textwidth}
        \centering
        \includegraphics[width=\textwidth]{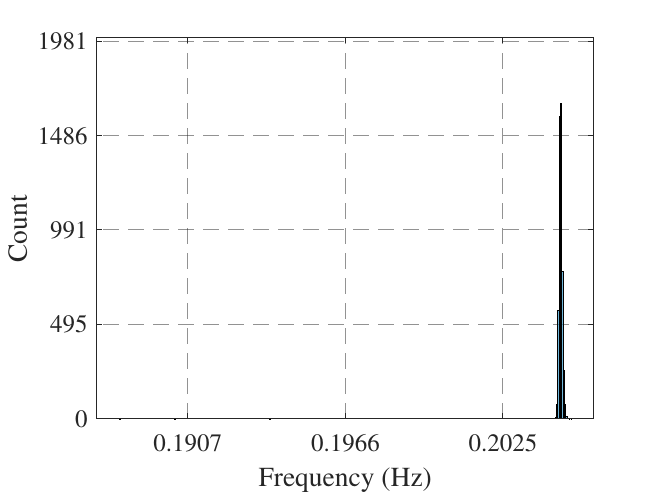}
        \caption{First-order  of PCSSI}
    \end{subfigure}
    \hspace{2em} 
    \begin{subfigure}{0.29\textwidth}
        \centering
        \includegraphics[width=\textwidth]{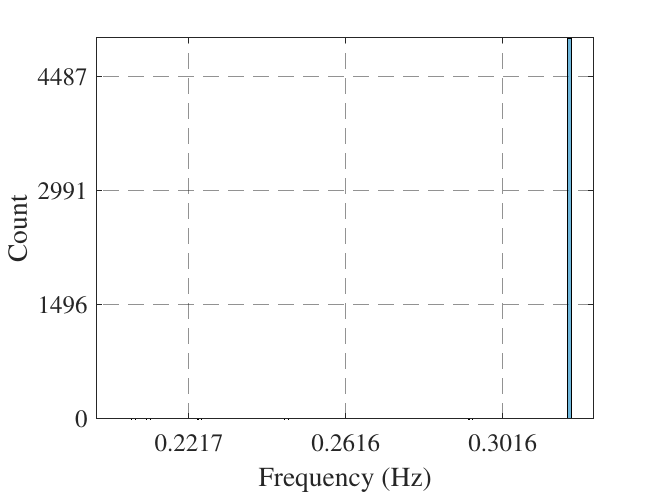}
        \caption{Second-order of PCSSI}
    \end{subfigure}
    \hspace{2em} 
    \begin{subfigure}{0.29\textwidth}
        \centering
        \includegraphics[width=\textwidth]{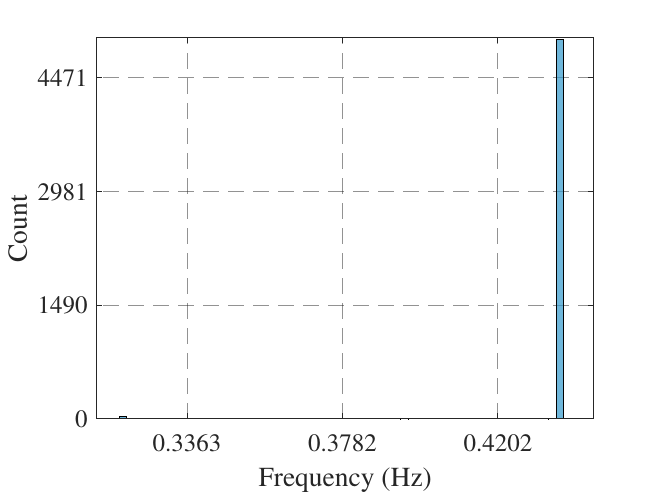}
        \caption{Third-order  of PCSSI}
    \end{subfigure}

    \vspace{0.5em} 

    \begin{subfigure}{0.29\textwidth}
        \centering
        \includegraphics[width=\textwidth]{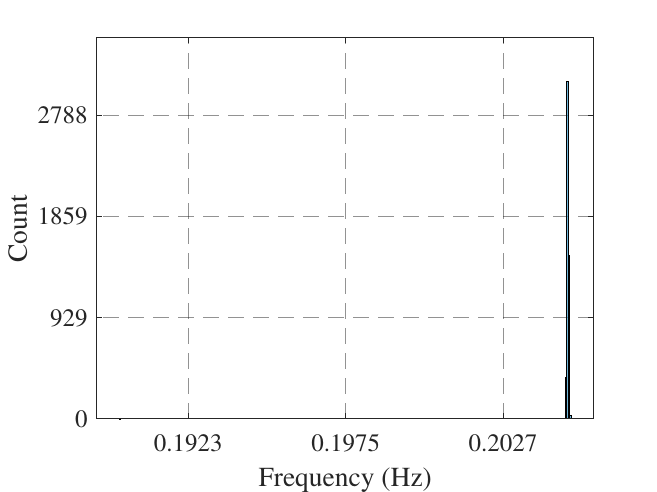}
        \caption{First-order of SSI-Cov}
    \end{subfigure}
    \hspace{2em} 
    \begin{subfigure}{0.29\textwidth}
        \centering
        \includegraphics[width=\textwidth]{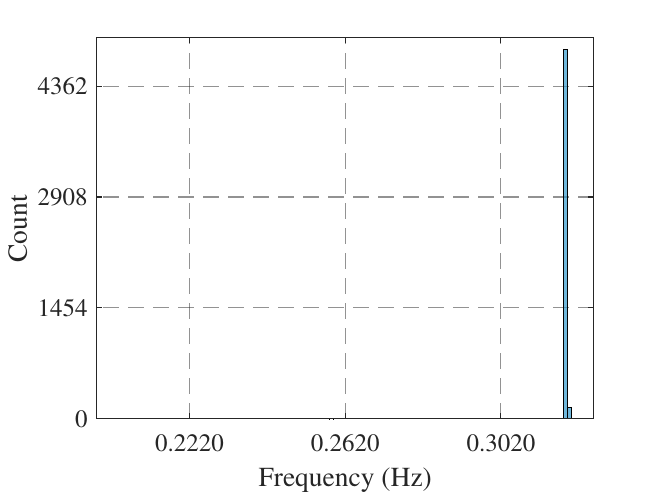}
        \caption{Second-order  of SSI-Cov}
    \end{subfigure}
    \hspace{2em} 
    \begin{subfigure}{0.29\textwidth}
        \centering
        \includegraphics[width=\textwidth]{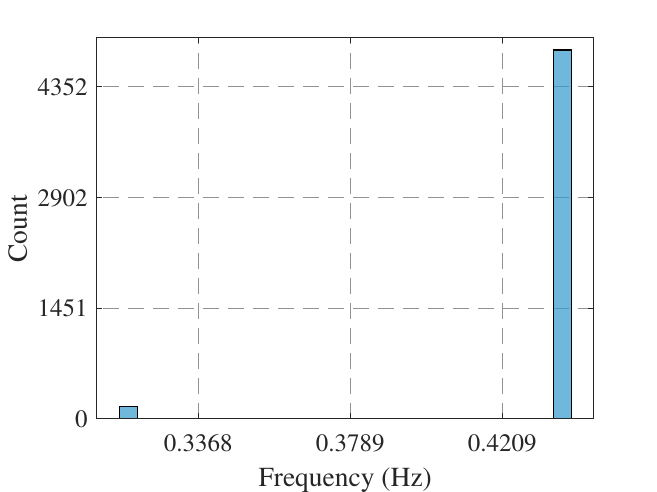}
        \caption{Third-order of SSI-Cov}
    \end{subfigure}

    \vspace{0.5em} 

    \begin{subfigure}{0.29\textwidth}
        \centering
        \includegraphics[width=\textwidth]{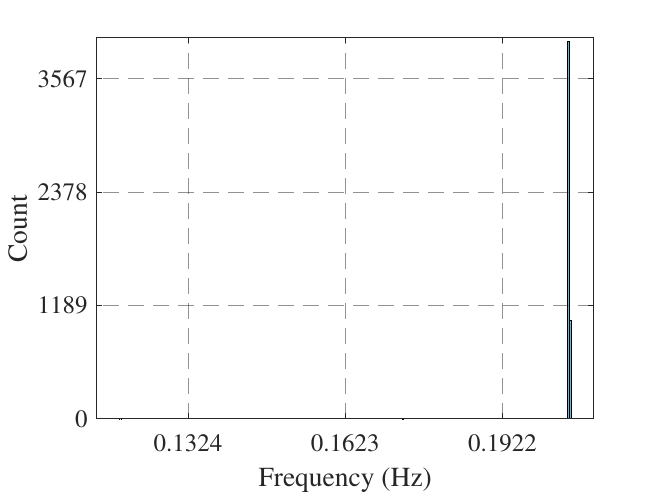}
        \caption{First-order of SSI-Data}
    \end{subfigure}
    \hspace{2em} 
    \begin{subfigure}{0.29\textwidth}
        \centering
        \includegraphics[width=\textwidth]{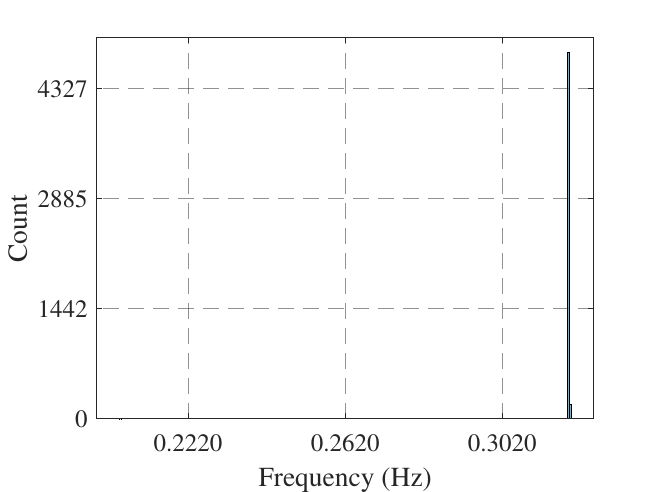}
        \caption{Second-order  of SSI-Data}
    \end{subfigure}
    \hspace{2em} 
    \begin{subfigure}{0.29\textwidth}
        \centering
        \includegraphics[width=\textwidth]{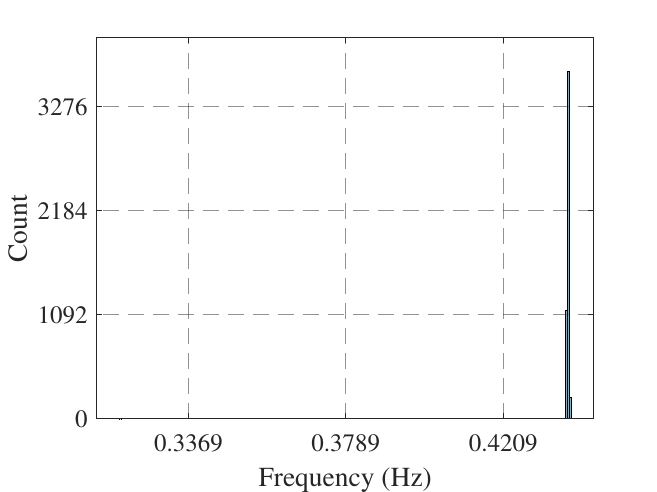}
        \caption{Third-order of SSI-Data}
    \end{subfigure}
    
    \caption{Comparison of the first three frequency distributions for PCSSI, SSI-Cov, and SSI-Data.}
    \label{fig:PCSSI_and_SSI-Cov5}
\end{figure}

\begin{figure}[!ht]
    \centering
    \begin{subfigure}{0.29\textwidth}
        \centering
        \includegraphics[width=\textwidth]{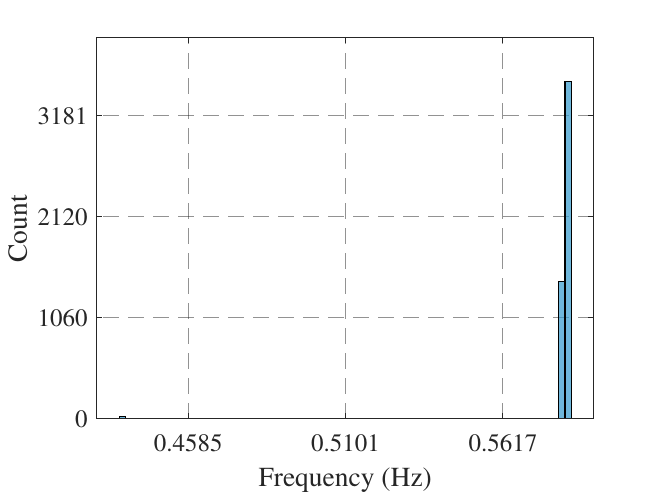}
        \caption{Fourth-order of PCSSI}
    \end{subfigure}
    \hspace{2em}
    \begin{subfigure}{0.29\textwidth}
        \centering
        \includegraphics[width=\textwidth]{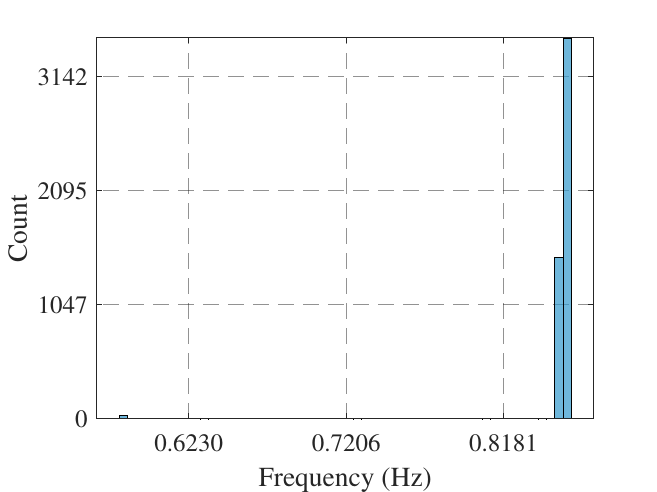}
        \caption{Fifth-order of PCSSI}
    \end{subfigure}
    \hspace{2em}
    \begin{subfigure}{0.29\textwidth}
        \centering
        \includegraphics[width=\textwidth]{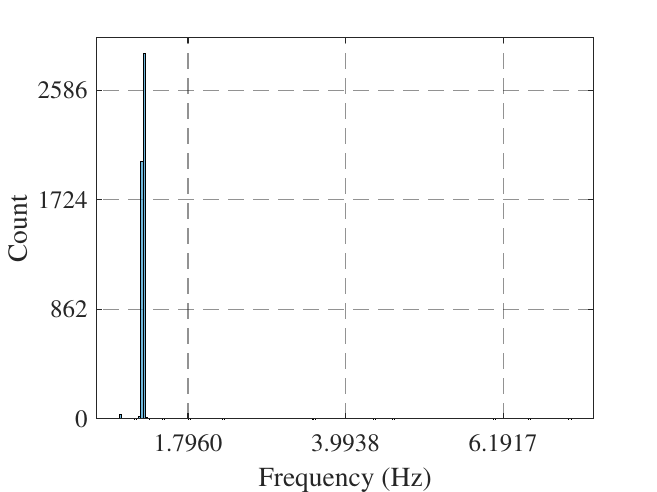}
        \caption{Sixth-order of PCSSI}
    \end{subfigure}
  
    \vspace{0.5em}

    \begin{subfigure}{0.29\textwidth}
        \centering
        \includegraphics[width=\textwidth]{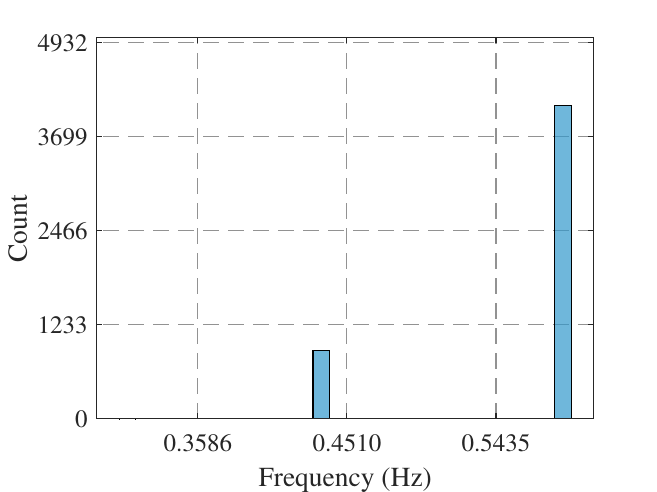}
        \caption{Fourth-order of SSI-Cov}
    \end{subfigure}
    \hspace{2em}
    \begin{subfigure}{0.29\textwidth}
        \centering
        \includegraphics[width=\textwidth]{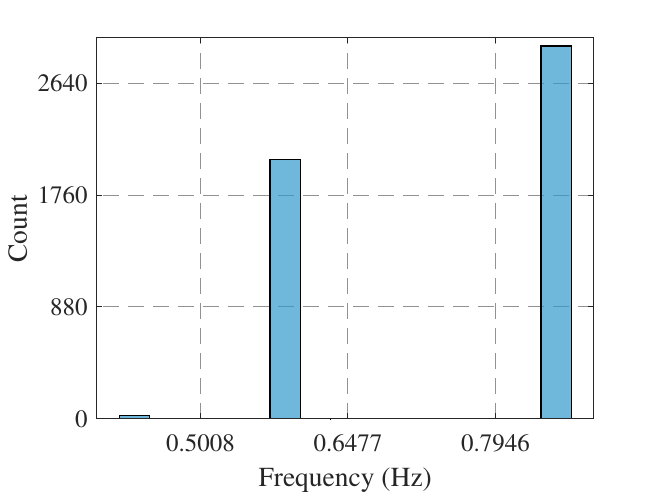}
        \caption{Fifth-order of SSI-Cov}
    \end{subfigure}
    \hspace{2em}
    \begin{subfigure}{0.29\textwidth}
        \centering
        \includegraphics[width=\textwidth]{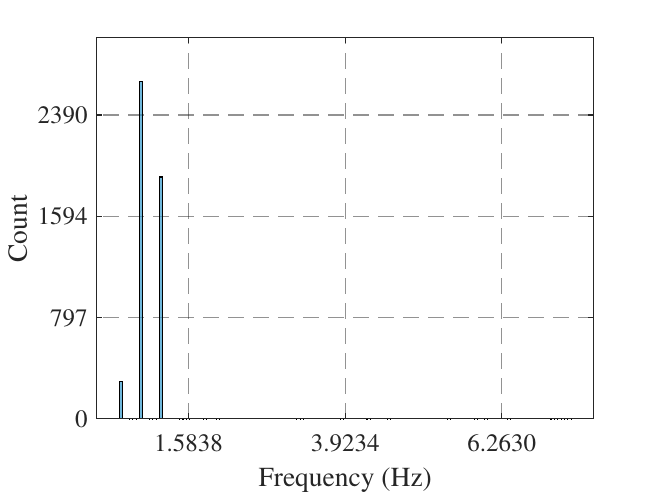}
        \caption{Sixth-order of SSI-Cov}
    \end{subfigure}

    \vspace{0.5em}

    \begin{subfigure}{0.29\textwidth}
        \centering
        \includegraphics[width=\textwidth]{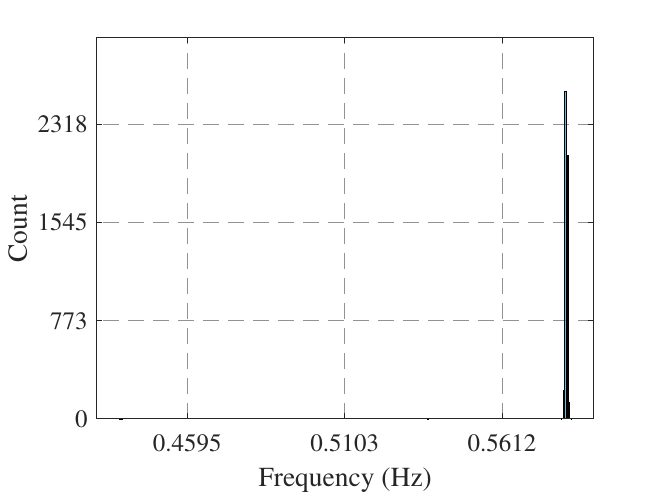}
        \caption{Fourth-order of SSI-Data}
    \end{subfigure}
    \hspace{2em}
    \begin{subfigure}{0.29\textwidth}
        \centering
        \includegraphics[width=\textwidth]{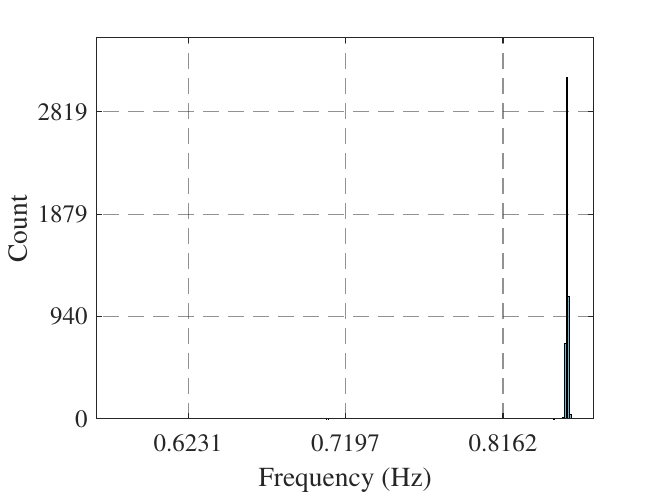}
        \caption{Fifth-order of SSI-Data}
    \end{subfigure}
    \hspace{2em}
    \begin{subfigure}{0.29\textwidth}
        \centering
        \includegraphics[width=\textwidth]{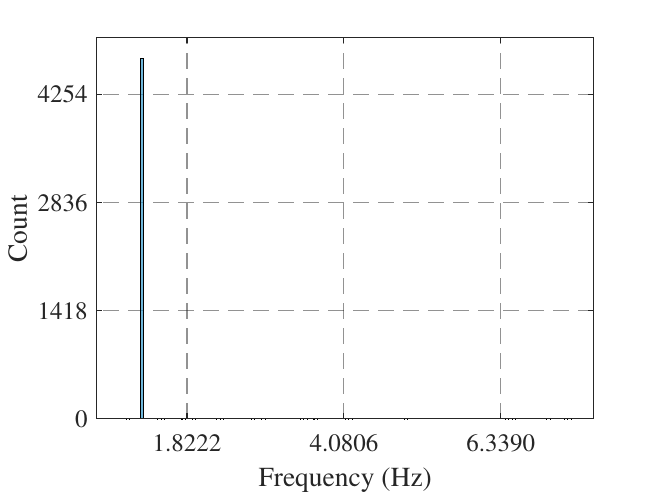}
        \caption{Sixth-order of SSI-Data}
    \end{subfigure}
    \caption{Comparison of the 4th, 5th, and 6th frequencies for PCSSI, SSI-Cov, and SSI-Data.}
    \label{fig:PCSSI_and_SSI-Cov6}
\end{figure}

\begin{figure}[!ht]
    \centering
    \begin{subfigure}{0.31\textwidth}
        \centering
        \includegraphics[width=\textwidth]{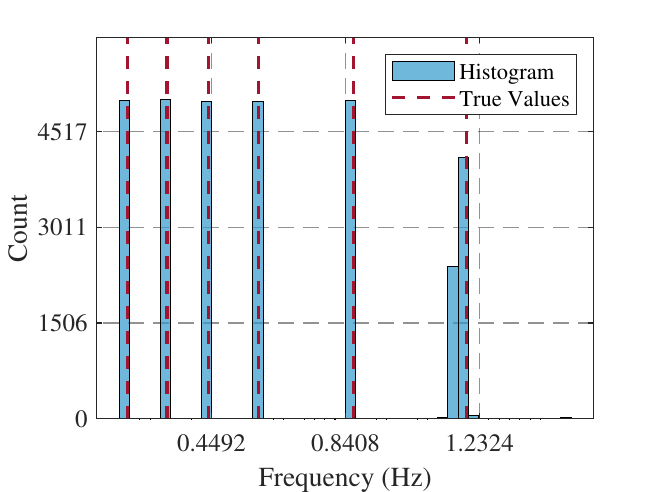}
        \caption{PCSSI}
    \end{subfigure}
    \hspace{0.5em}
    \begin{subfigure}{0.31\textwidth}
        \centering
        \includegraphics[width=\textwidth]{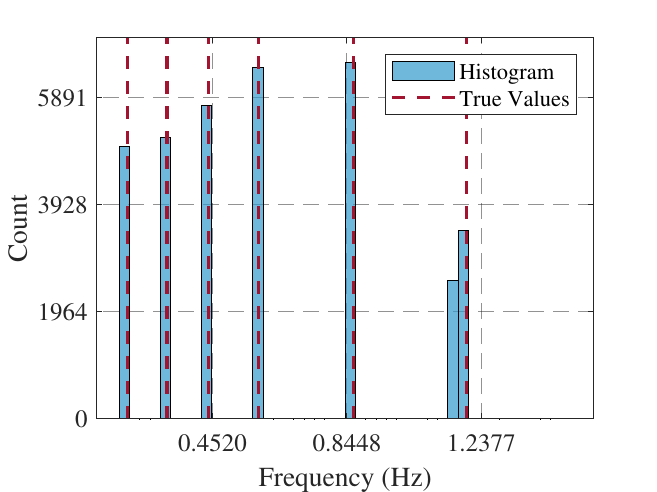}
        \caption{SSI-Cov}
    \end{subfigure}
    \hspace{0.5em}
    \begin{subfigure}{0.31\textwidth}
        \centering
        \includegraphics[width=\textwidth]{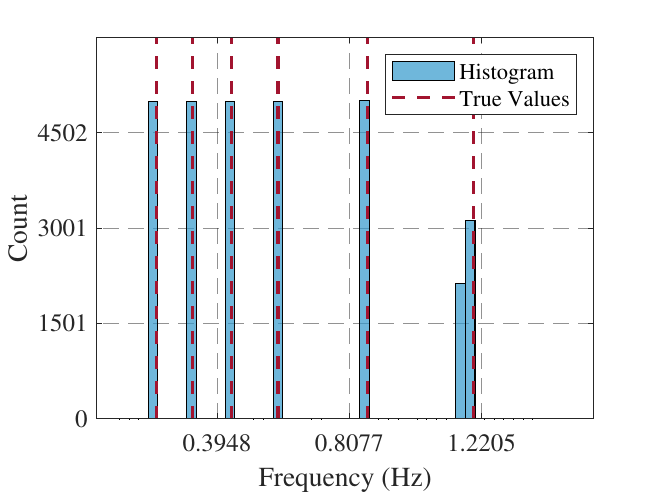}
        \caption{SSI-Data}
    \end{subfigure}
    \caption{Comparison of the identified truncated frequencies for PCSSI, SSI-Cov, and SSI-Data.}\label{fig:PCSSI_and_SSI-Cov4}
  \end{figure}

\begin{figure}[!ht]
    \centering
    \begin{subfigure}{0.29\textwidth}
        \centering
        \includegraphics[width=\textwidth]{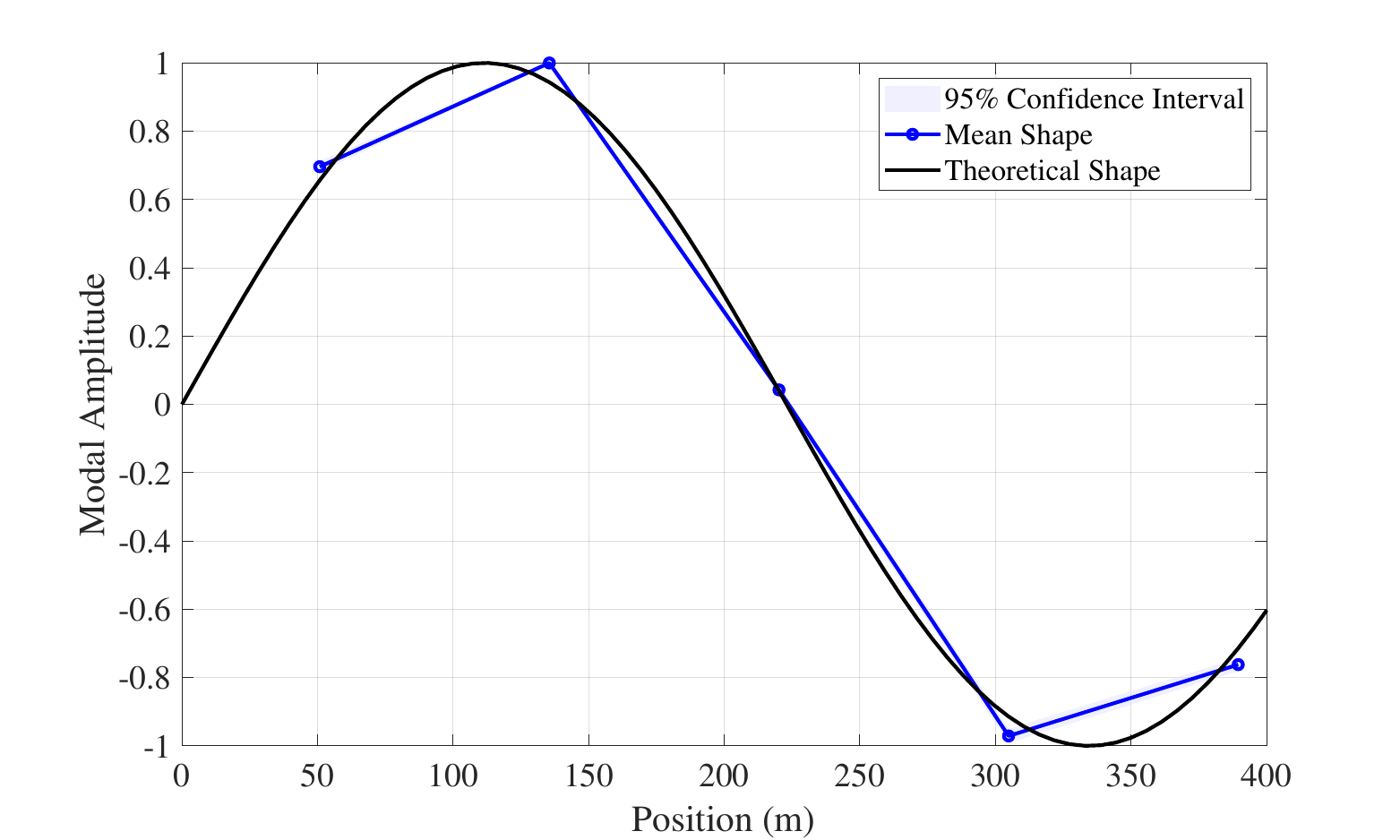}
        \caption{First-order of PCSSI}
    \end{subfigure}
    \hspace{2em}
    \begin{subfigure}{0.29\textwidth}
        \centering
        \includegraphics[width=\textwidth]{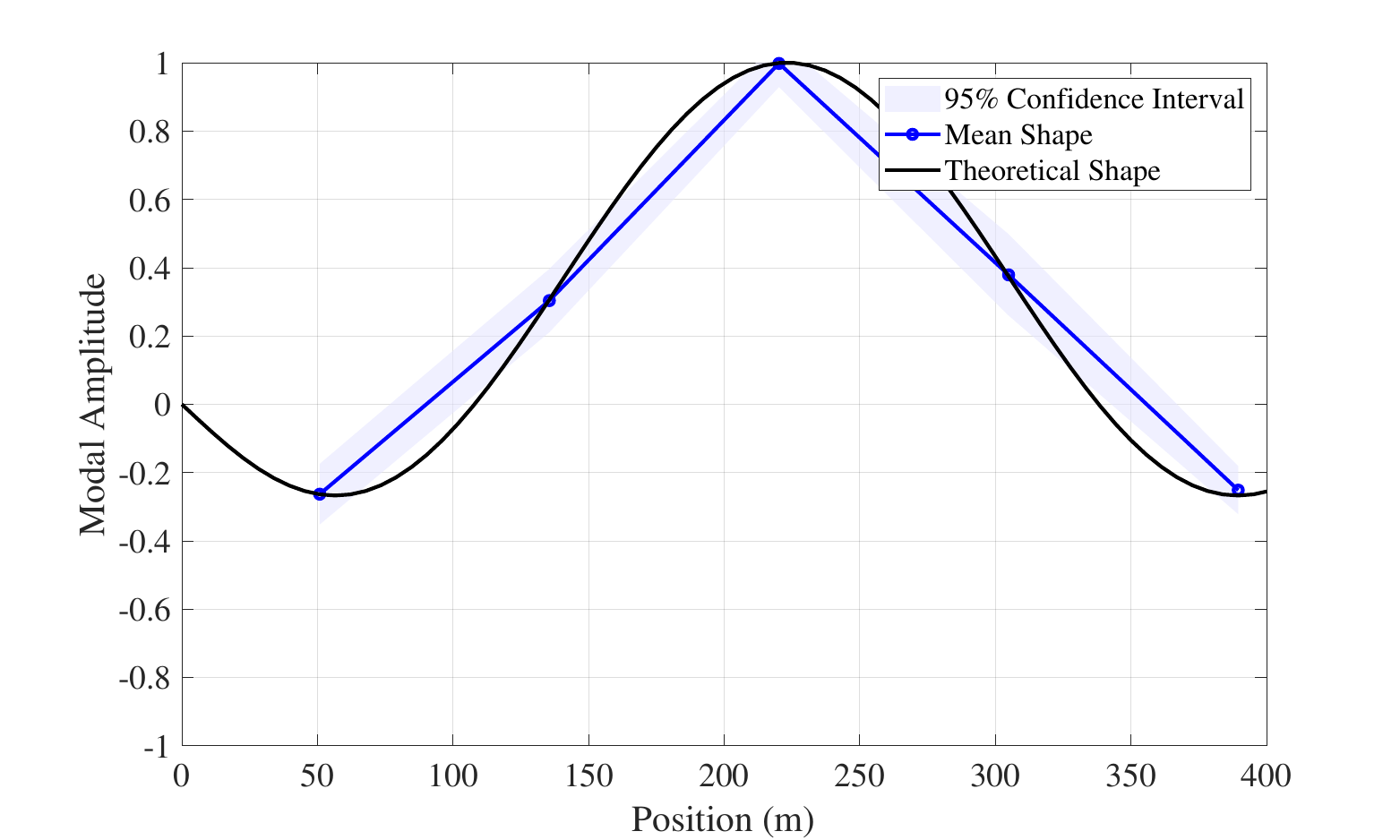}
        \caption{Second-order of PCSSI}
    \end{subfigure}
    \hspace{2em}
    \begin{subfigure}{0.29\textwidth}
        \centering
        \includegraphics[width=\textwidth]{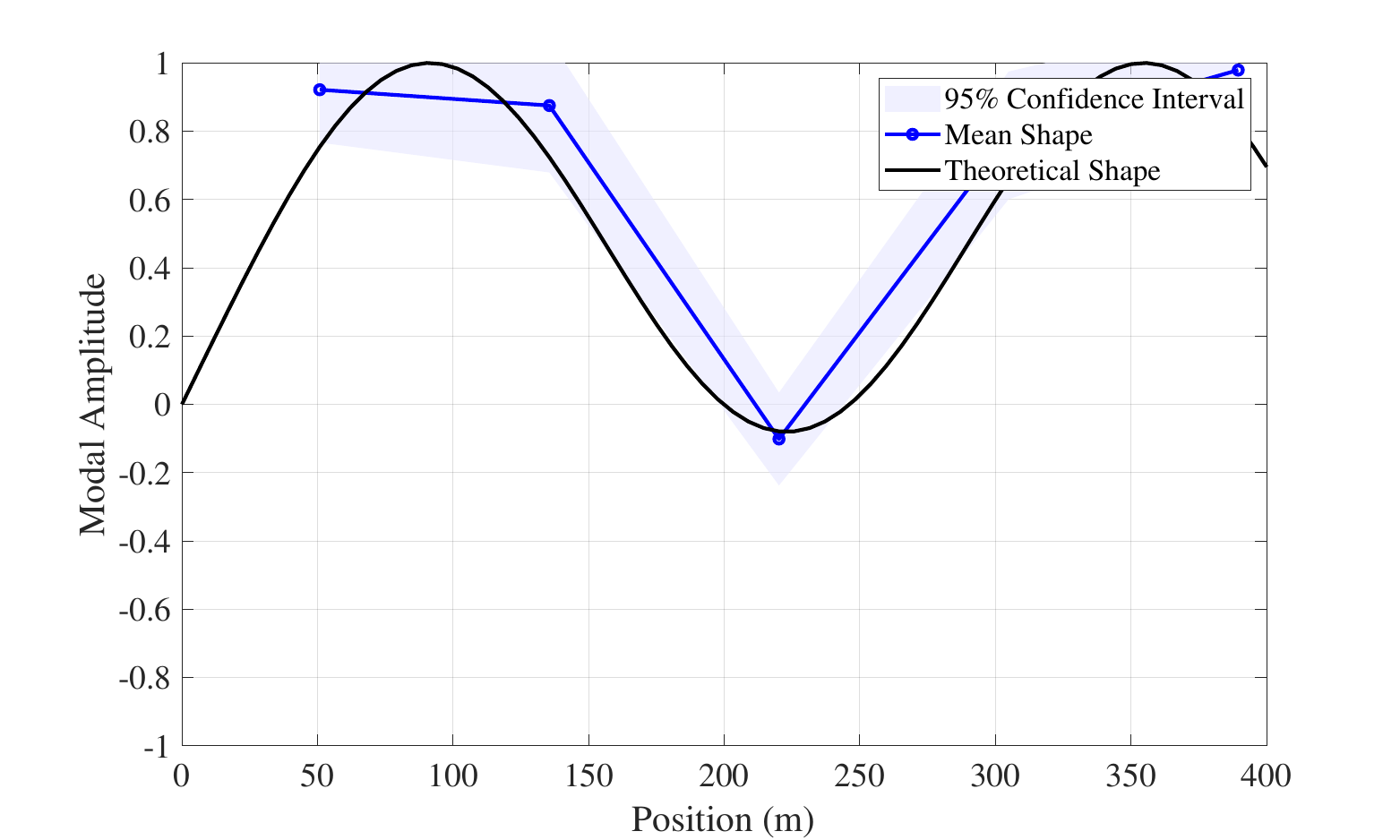}
        \caption{Third-order of PCSSI}
    \end{subfigure}
  
    \vspace{0.5em}

    \begin{subfigure}{0.29\textwidth}
        \centering
        \includegraphics[width=\textwidth]{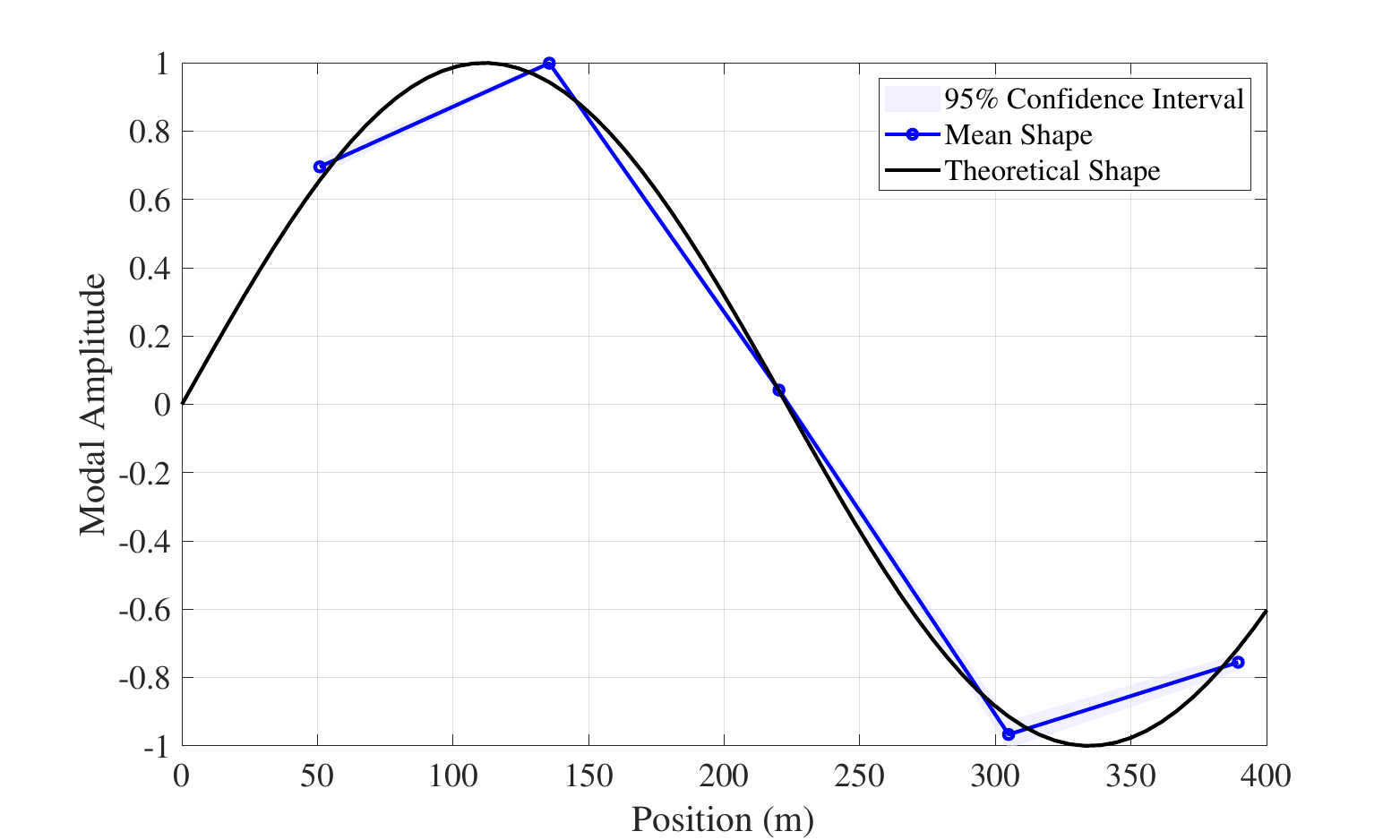}
        \caption{First-order of SSI-Cov}
    \end{subfigure}
    \hspace{2em}
    \begin{subfigure}{0.29\textwidth}
        \centering
        \includegraphics[width=\textwidth]{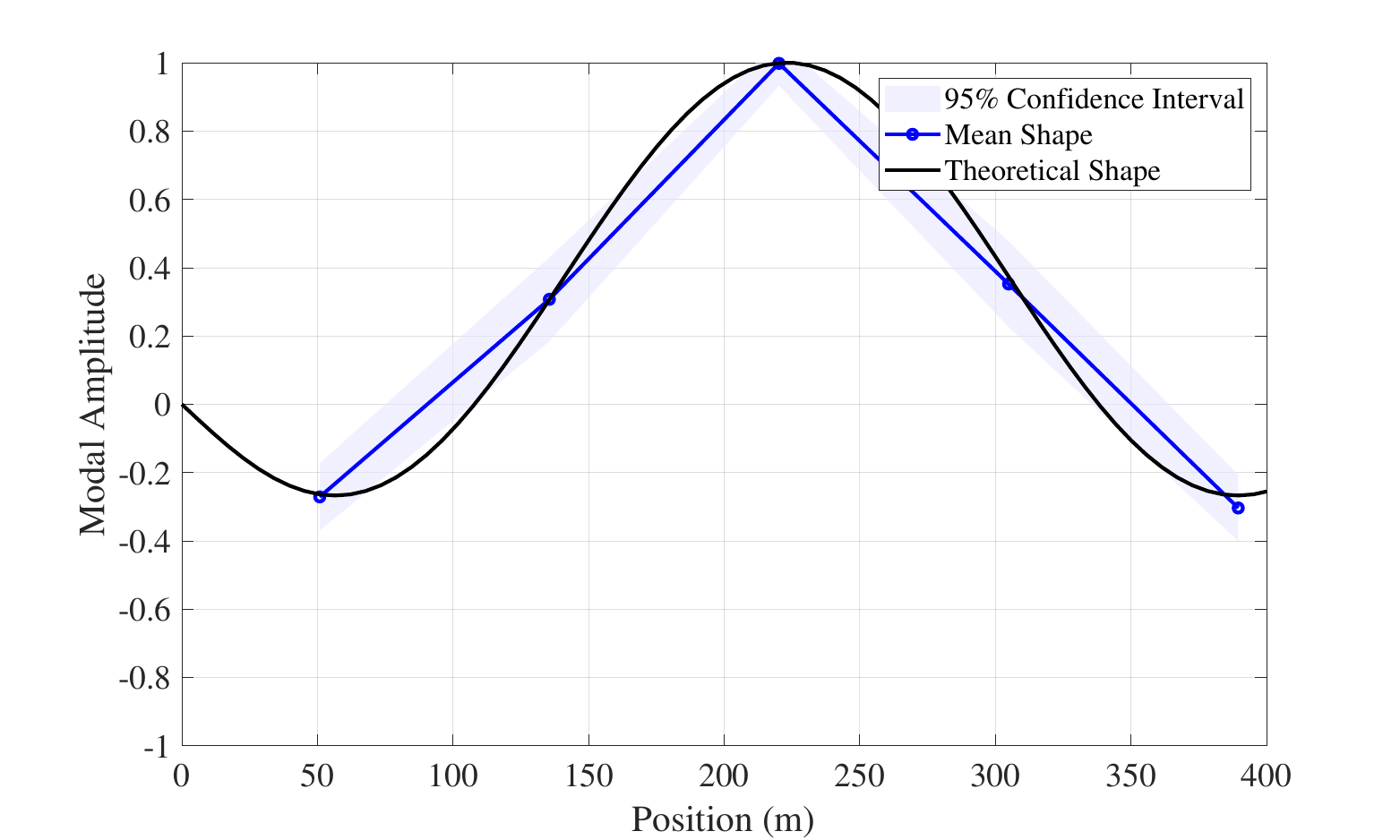}
        \caption{Second-order of SSI-Cov}
    \end{subfigure}
    \hspace{2em}
    \begin{subfigure}{0.29\textwidth}
        \centering
        \includegraphics[width=\textwidth]{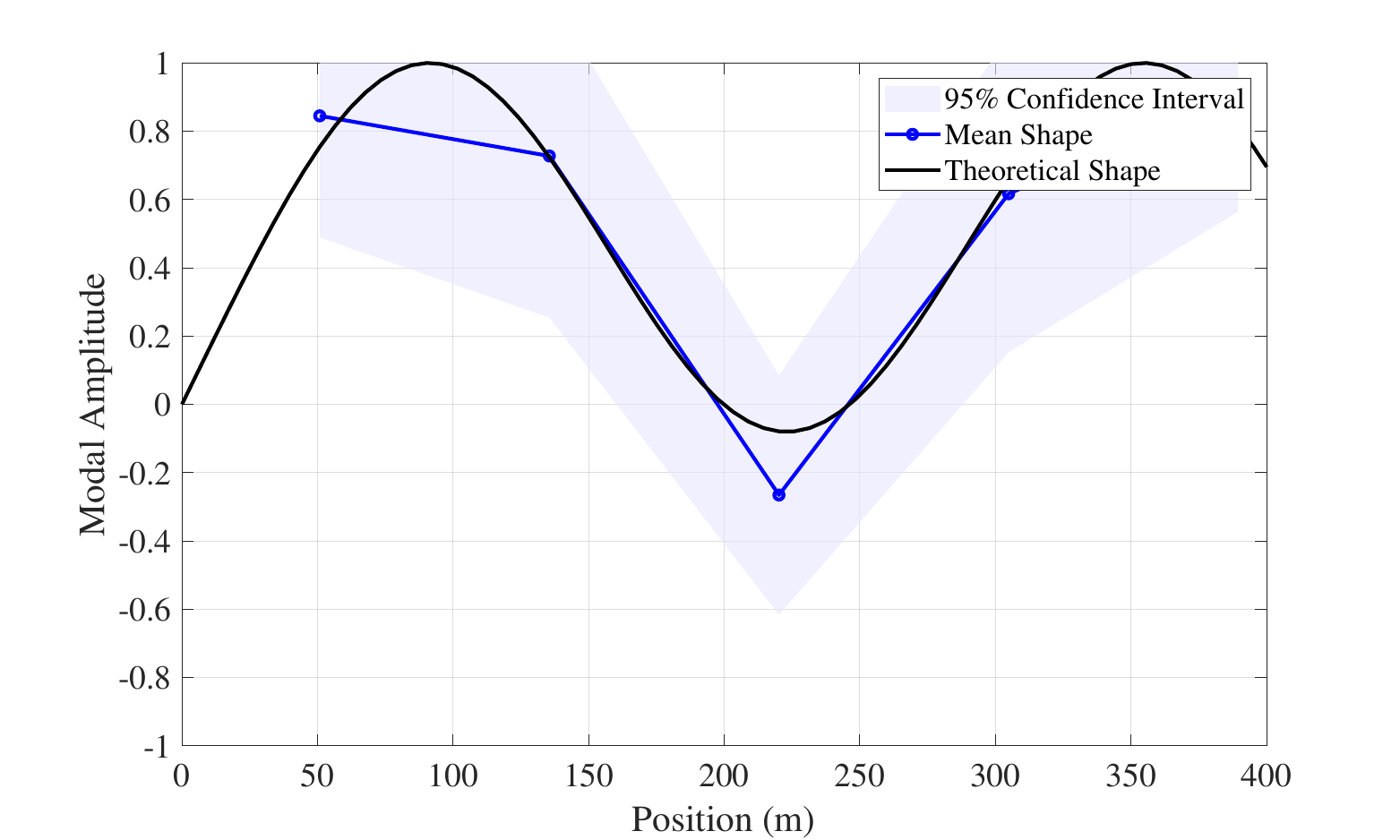}
        \caption{Third-order of SSI-Cov}
    \end{subfigure}

    \vspace{0.5em}

    \begin{subfigure}{0.29\textwidth}
        \centering
        \includegraphics[width=\textwidth]{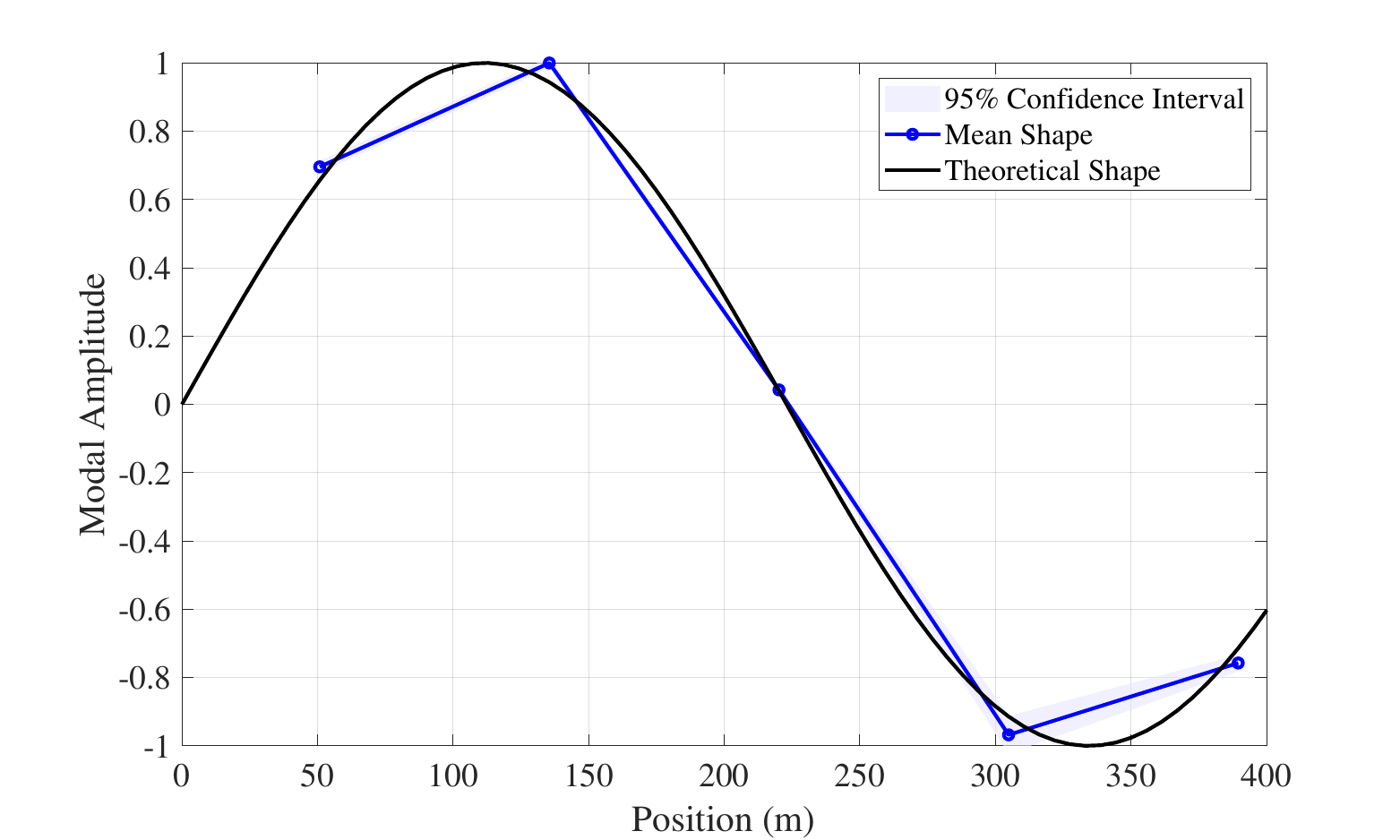}
        \caption{First-order of SSI-Data}
    \end{subfigure}
    \hspace{2em}
    \begin{subfigure}{0.29\textwidth}
        \centering
        \includegraphics[width=\textwidth]{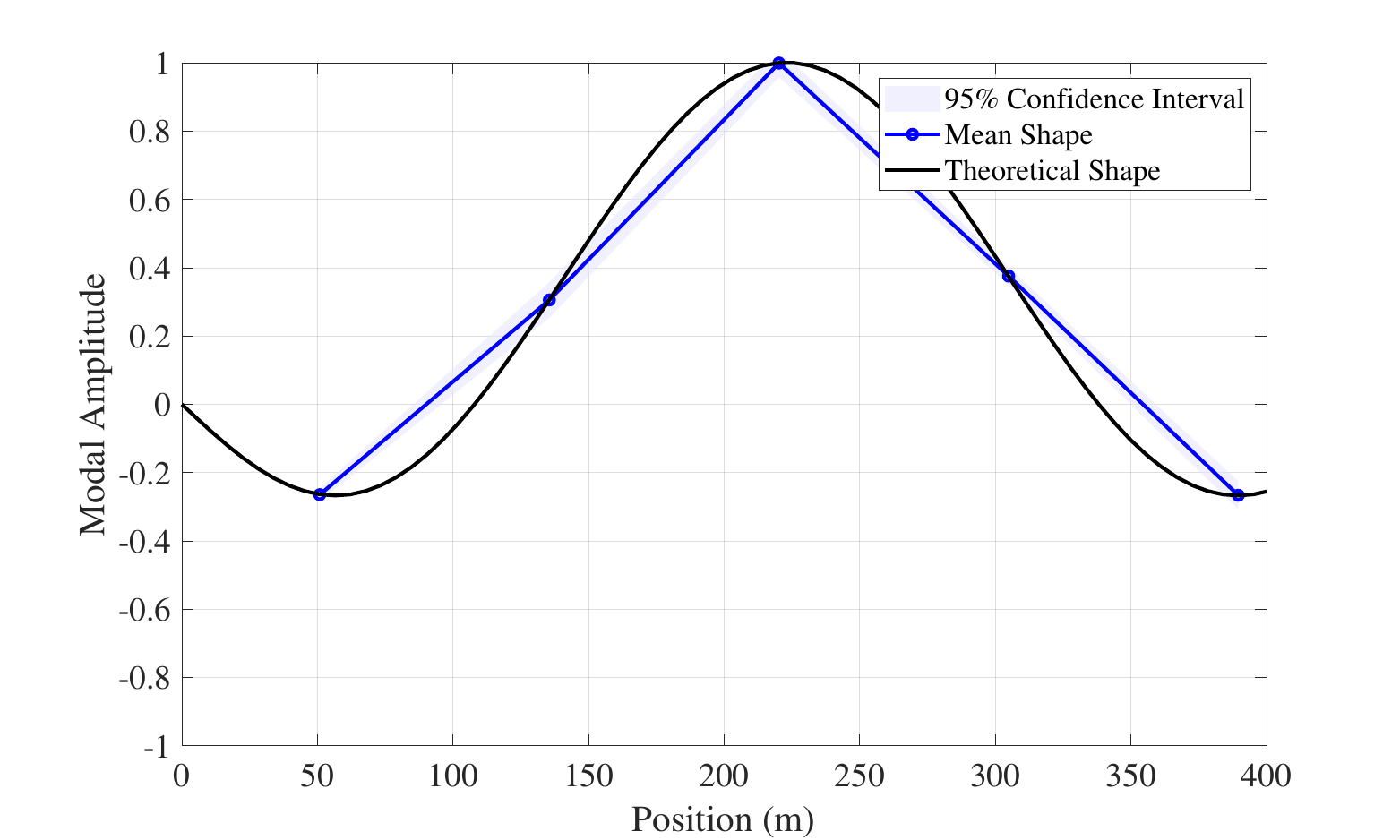}
        \caption{Second-order of SSI-Data}
    \end{subfigure}
    \hspace{2em}
    \begin{subfigure}{0.29\textwidth}
        \centering
        \includegraphics[width=\textwidth]{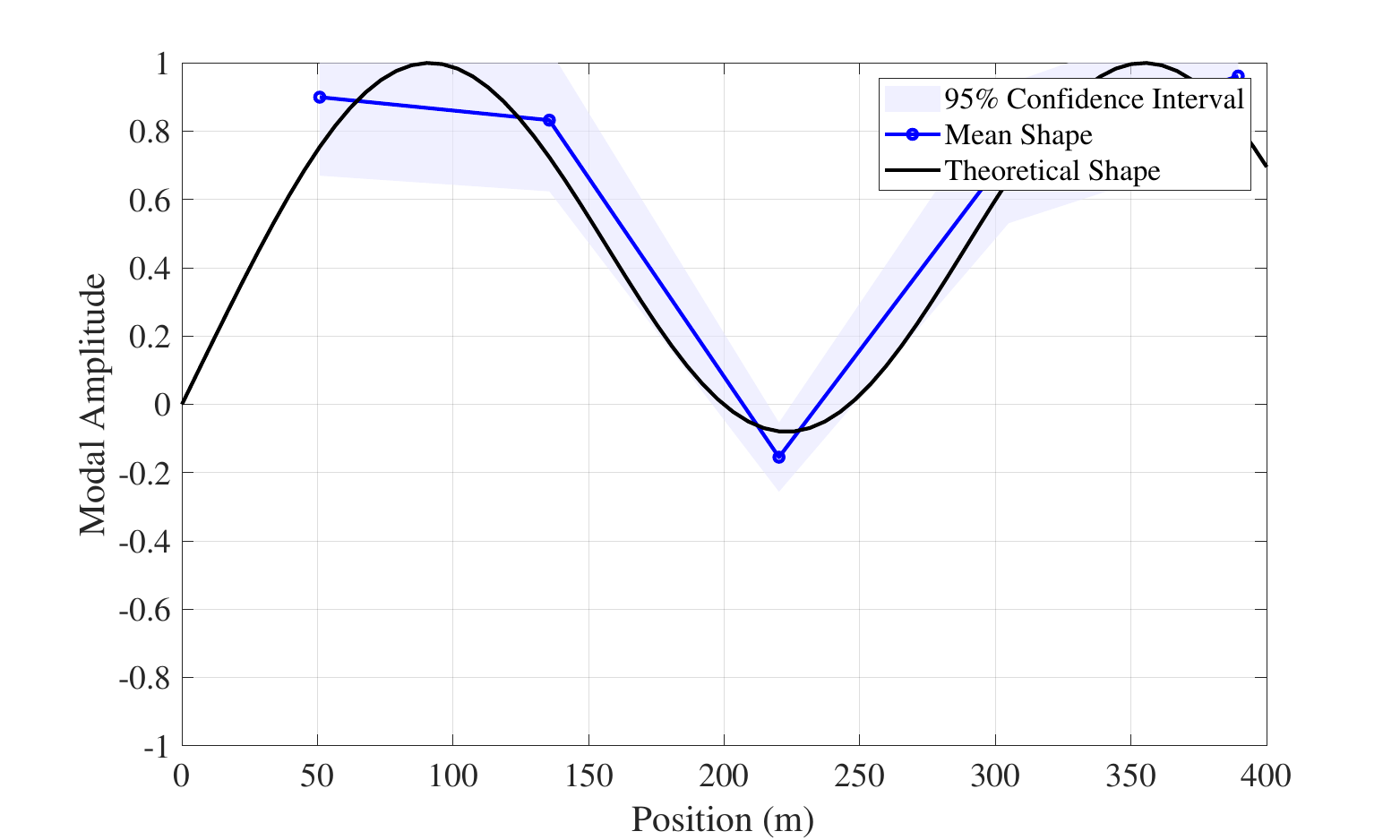}
        \caption{Third-order of SSI-Data}
    \end{subfigure}

    \caption{Comparison of the first three identified mode shapes for PCSSI, SSI-Cov, and SSI-Data.}
    \label{fig:PCSSI_and_SSI-Cov_mode_shape3}
\end{figure}

\begin{figure}[!ht]
    \centering
    \begin{subfigure}{0.29\textwidth}
        \centering
        \includegraphics[width=\textwidth]{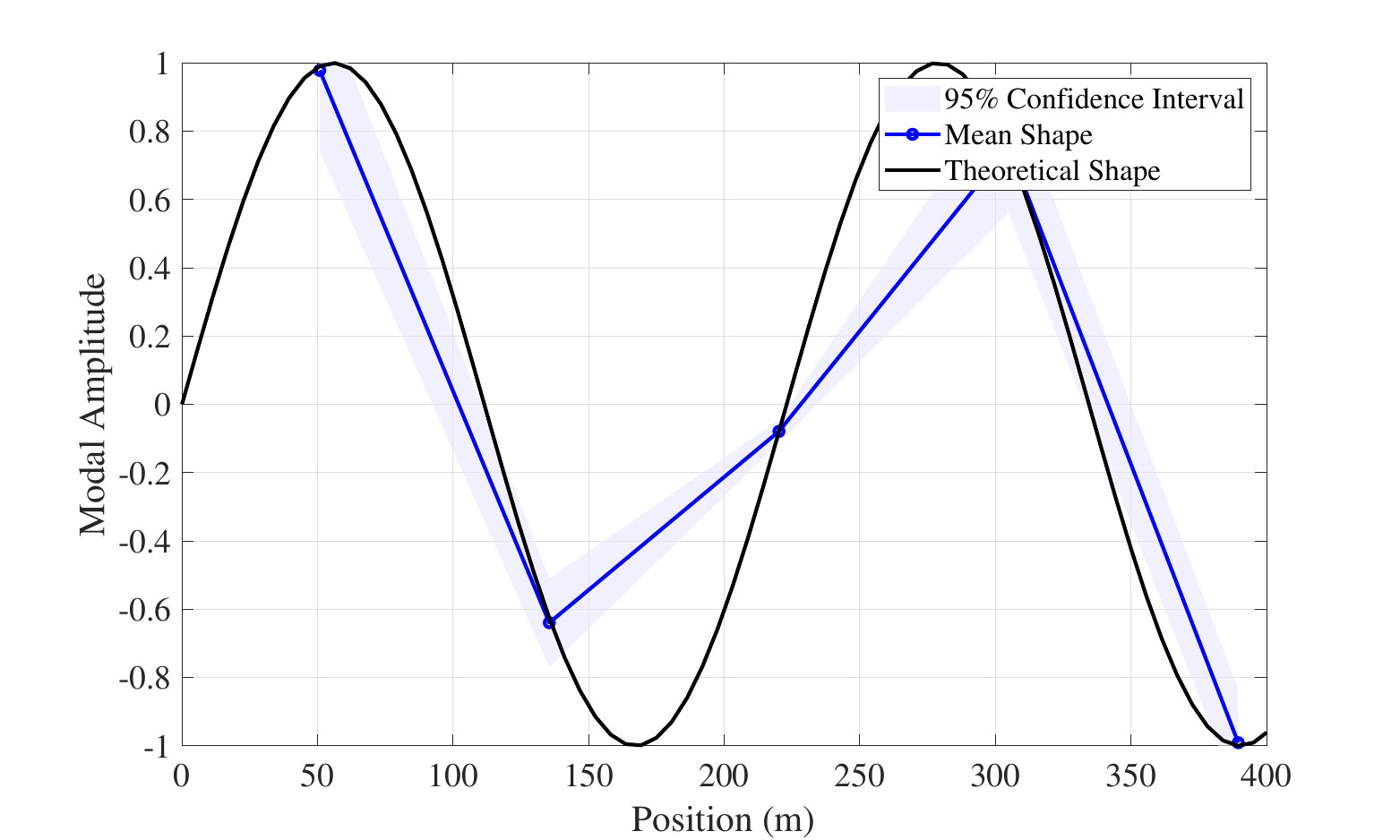}
        \caption{Fourth-order of PCSSI}
    \end{subfigure}
    \hspace{2em}
    \begin{subfigure}{0.29\textwidth}
        \centering
        \includegraphics[width=\textwidth]{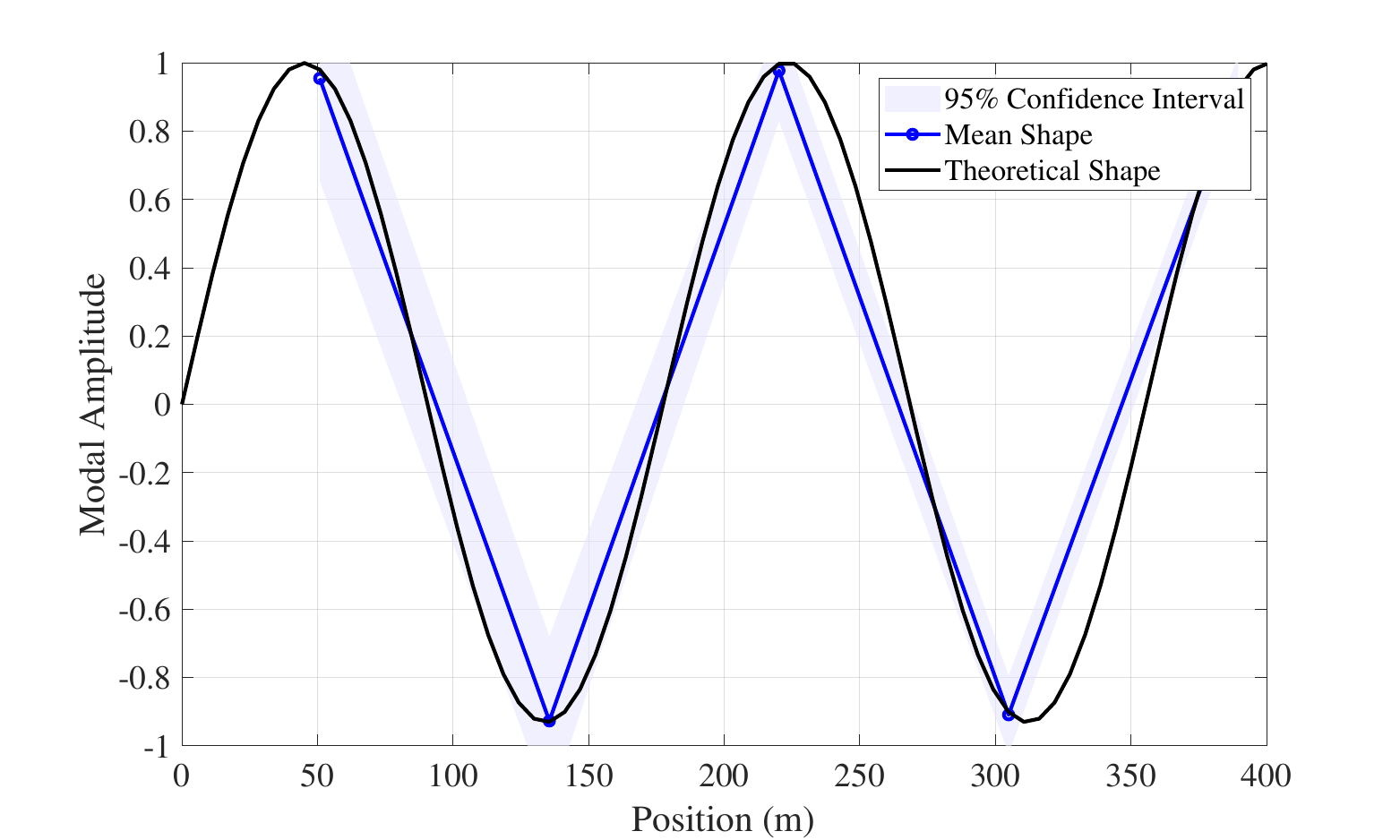}
        \caption{Fifth-order of PCSSI}
    \end{subfigure}
    \hspace{2em}
    \begin{subfigure}{0.29\textwidth}
        \centering
        \includegraphics[width=\textwidth]{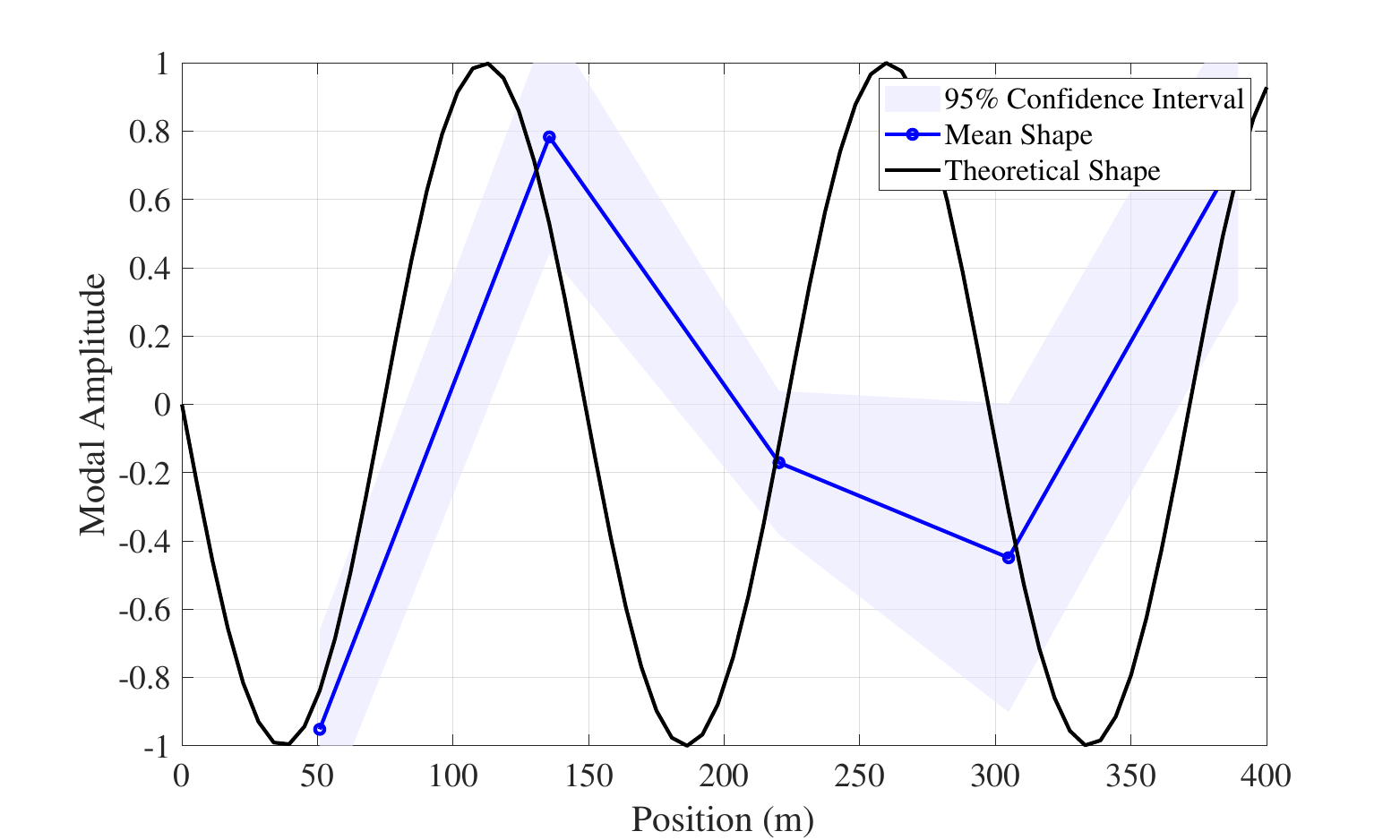}
        \caption{Sixth-order of PCSSI}
    \end{subfigure}
  
    \vspace{0.5em}

    \begin{subfigure}{0.29\textwidth}
        \centering
        \includegraphics[width=\textwidth]{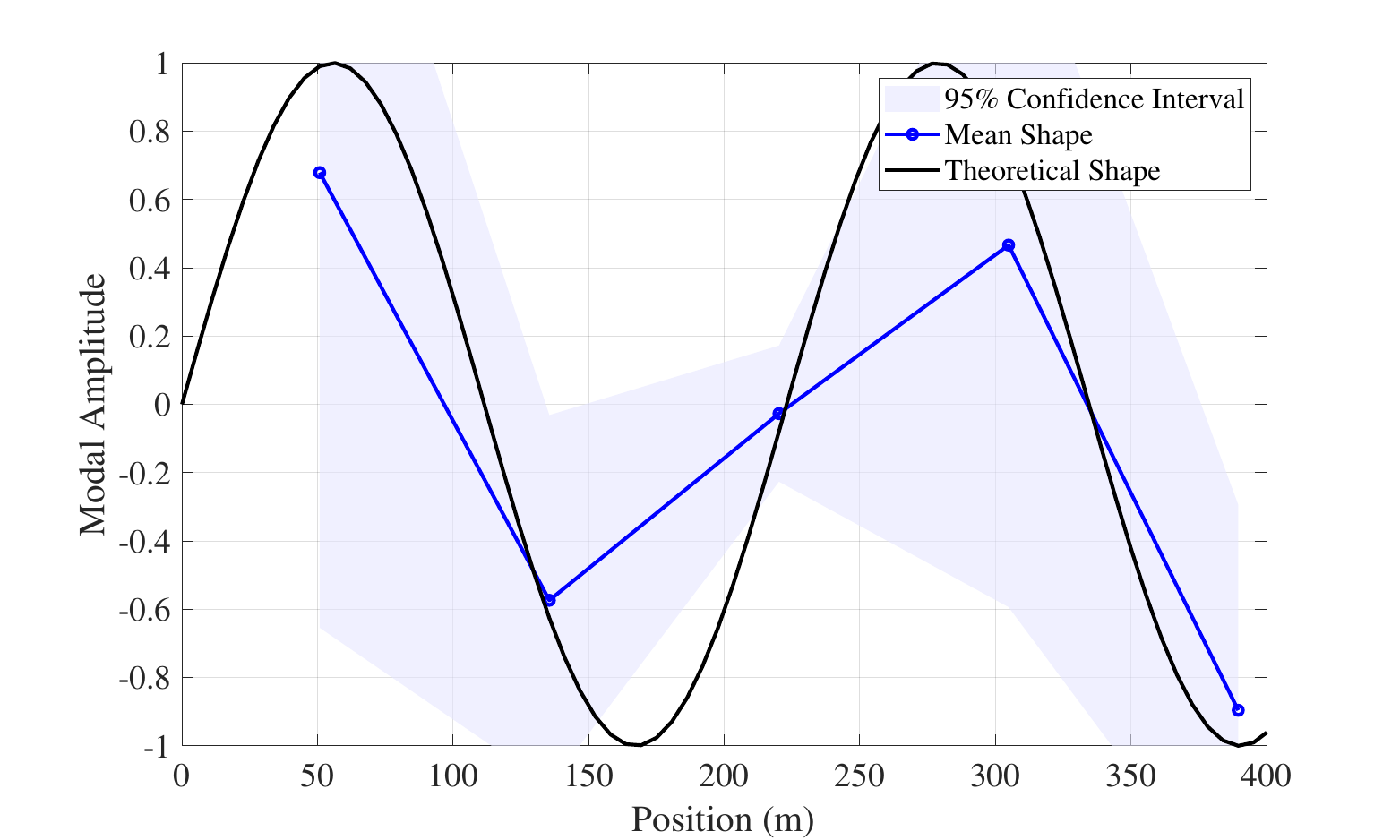}
        \caption{Fourth-order  of SSI-Cov}
    \end{subfigure}
    \hspace{2em}
    \begin{subfigure}{0.29\textwidth}
        \centering
        \includegraphics[width=\textwidth]{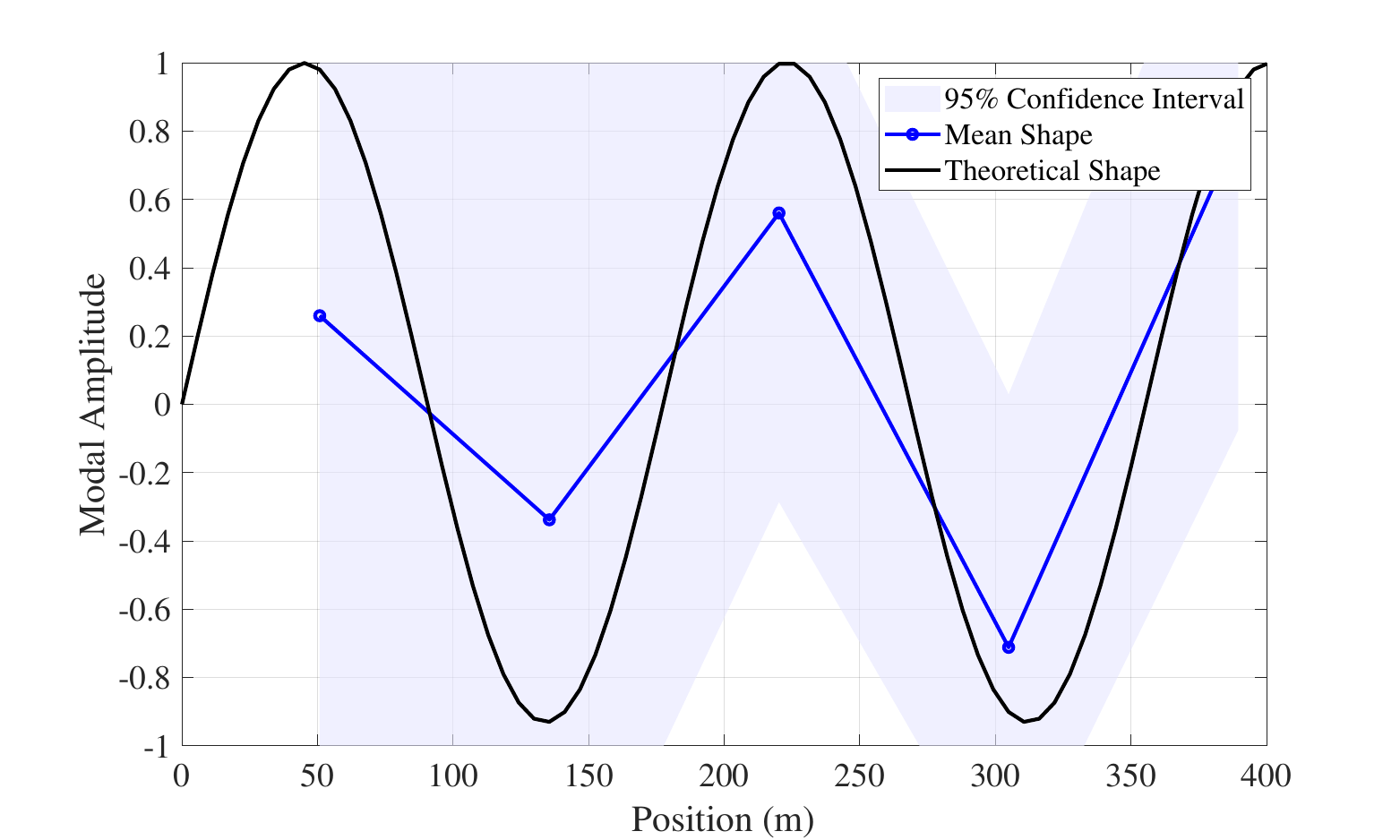}
        \caption{Fifth-order of SSI-Cov}
    \end{subfigure}
    \hspace{2em}
    \begin{subfigure}{0.29\textwidth}
        \centering
        \includegraphics[width=\textwidth]{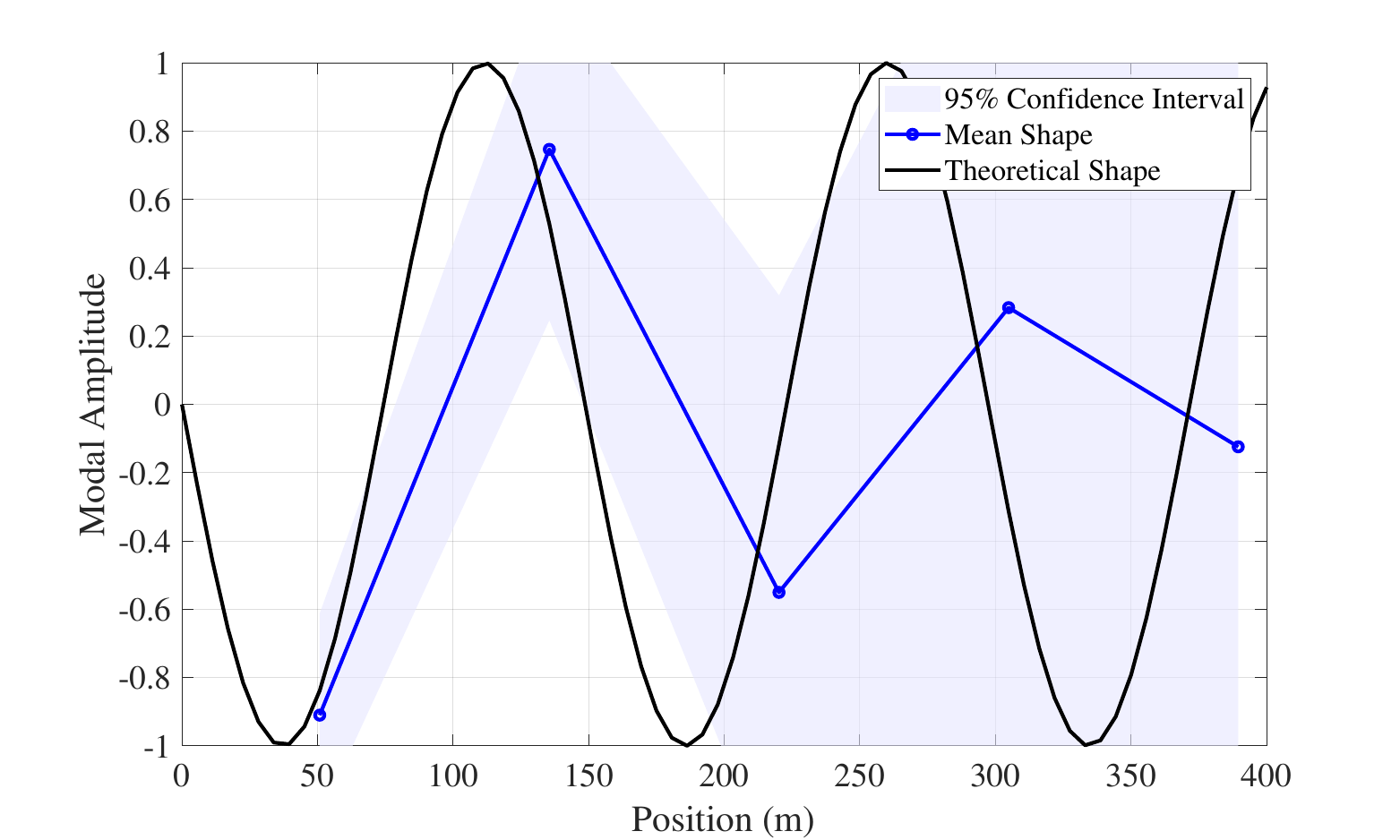}
        \caption{Sixth-order of SSI-Cov}
    \end{subfigure}

    \vspace{0.5em}

    \begin{subfigure}{0.29\textwidth}
        \centering
        \includegraphics[width=\textwidth]{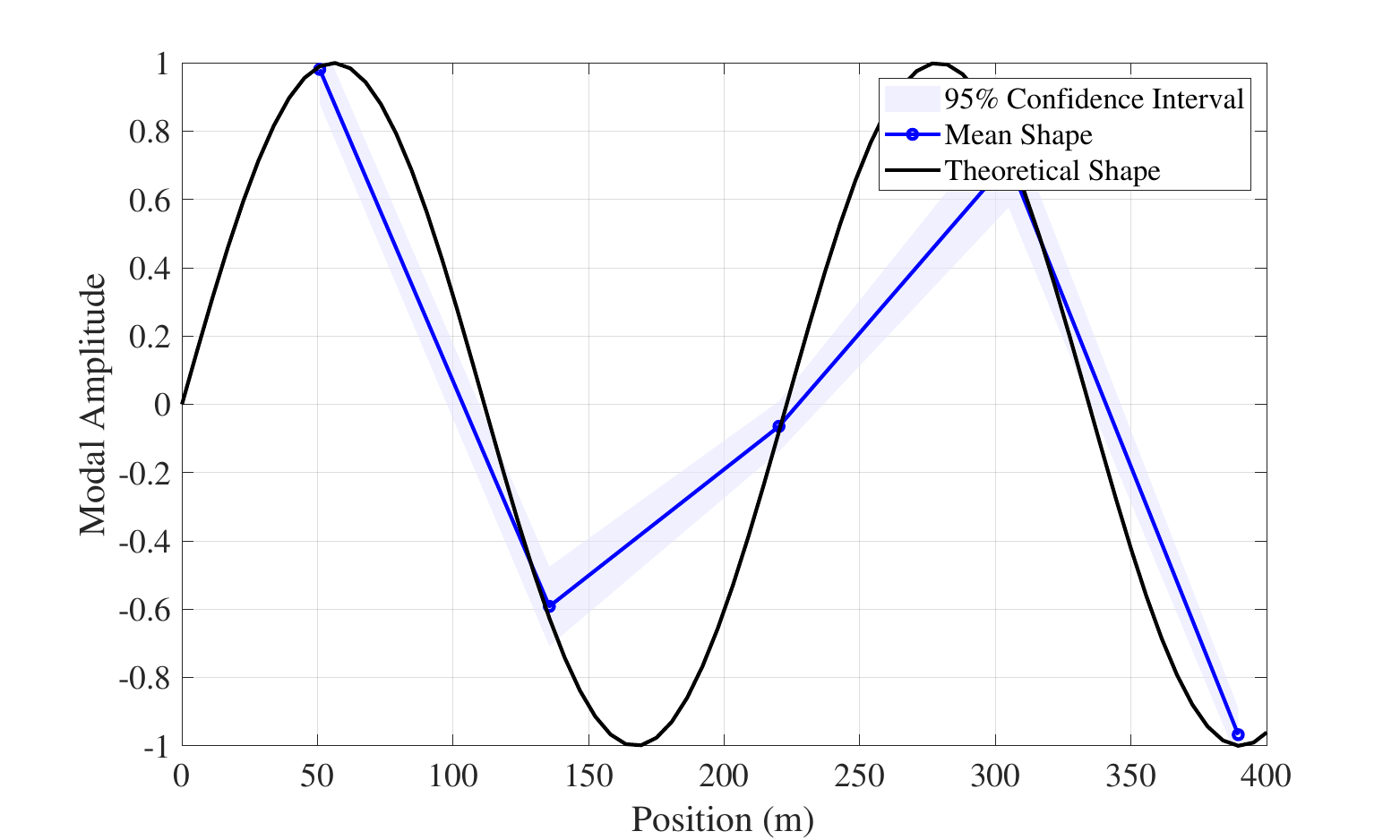}
        \caption{Fourth-order  of SSI-Data}
    \end{subfigure}
    \hspace{2em}
    \begin{subfigure}{0.29\textwidth}
        \centering
        \includegraphics[width=\textwidth]{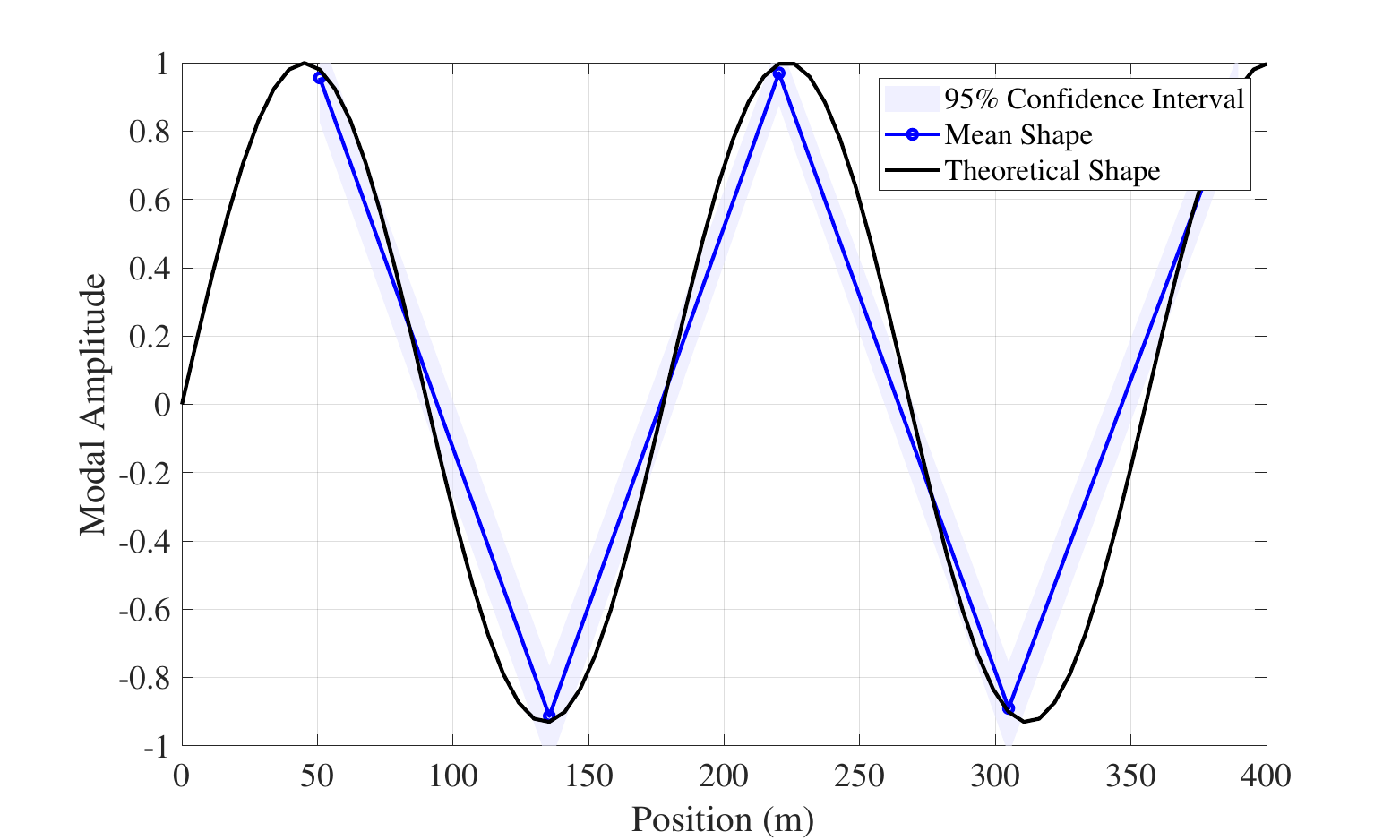}
        \caption{Fifth-order of SSI-Data}
    \end{subfigure}
    \hspace{2em}
    \begin{subfigure}{0.29\textwidth}
        \centering
        \includegraphics[width=\textwidth]{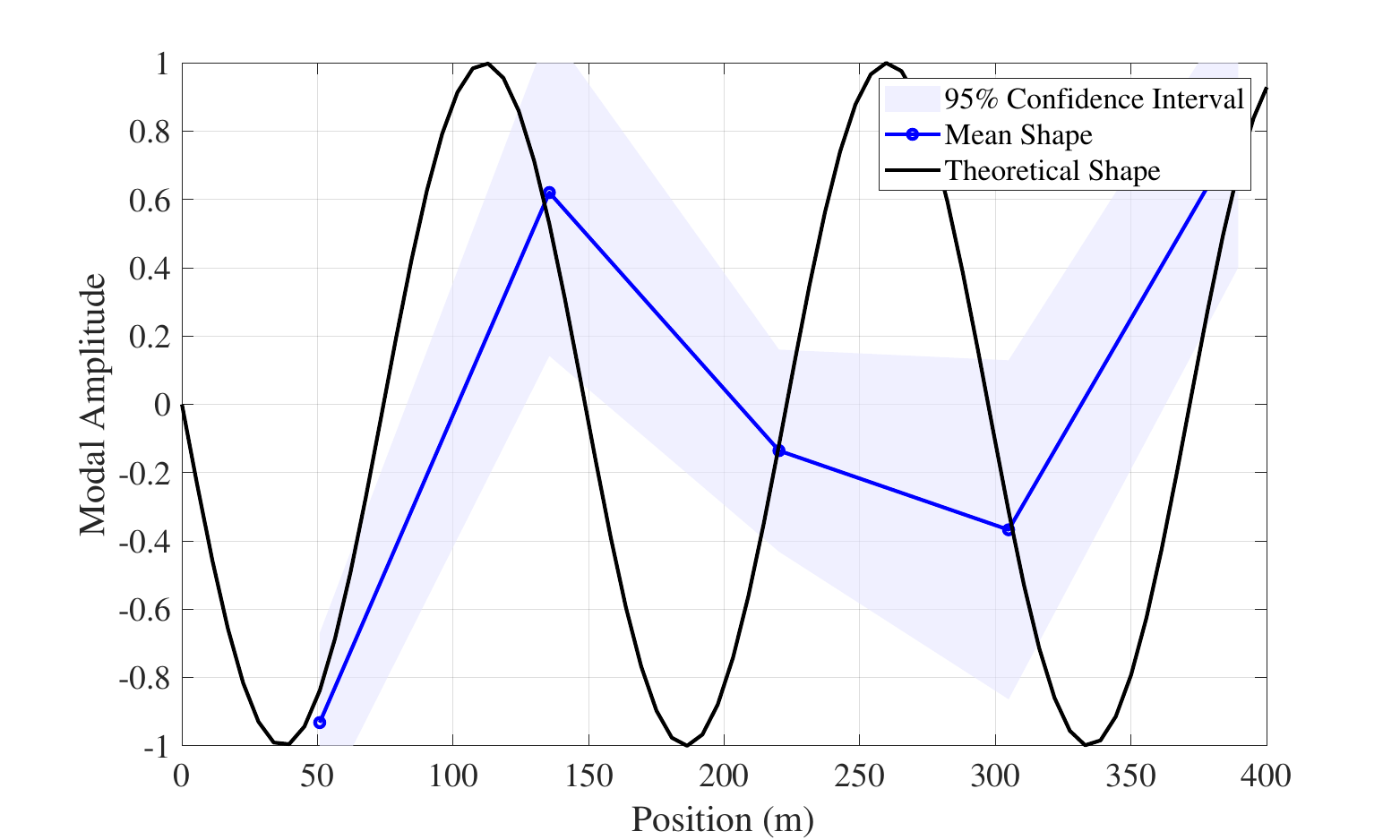}
        \caption{Sixth-order of SSI-Data}
    \end{subfigure}
    \caption{Comparison of the fourth to sixth identified mode shapes for PCSSI, SSI-Cov, and SSI-Data.}
    \label{fig:PCSSI_and_SSI-Cov_mode_shape4}
\end{figure}

To verify the theoretical analysis presented earlier and throughout the main text, we conducted modal analyses using all three methods for \( Y_f \in \mathbb{R}^{50 \times 50000} \). 
We set \( j = 50000 \) for the computation, as this data volume is sufficiently large to approximate the behavior expected in an infinite data scenario.
Furthermore, to ensure statistical robustness, we performed 5000 Monte Carlo simulations under these conditions. The frequency analysis results are presented in 
Figures \ref{fig:PCSSI_and_SSI-Cov5}, \ref{fig:PCSSI_and_SSI-Cov6}, and \ref{fig:PCSSI_and_SSI-Cov4}, while the corresponding mode shape analysis results are shown in 
Figures \ref{fig:PCSSI_and_SSI-Cov_mode_shape3} and \ref{fig:PCSSI_and_SSI-Cov_mode_shape4}.

To compare the performance of PCSSI and SSI-Data, both methods were computed using the same data matrix \( Y_p,Y_f \in \mathbb{R}^{50 \times 50000} \). Theoretically, their results 
should be identical, meaning that any observed differences can primarily be attributed to numerical errors and computational stability. As shown in the figures, compared to 
the \( Y_f \in \mathbb{R}^{18 \times 5000} \) used in the main text, SSI-Data exhibits a significant improvement in modal analysis performance, with its identified 
frequencies and mode shapes closely aligning with those of PCSSI and the true values. In practical terms, for large datasets, SSI-Data appears to demonstrate superior 
numerical stability compared to PCSSI, exhibiting lower variance.  
However, PCSSI offers greater flexibility by allowing different choices for the row and column dimensions of \( Y_f \), making it generally more advantageous than SSI-Data, 
particularly when working with smaller datasets.

Additionally, all three methods show substantial improvements in identifying the sixth modal frequency, eliminating previous cases where it was almost undetectable. 
However, the performance of SSI-Cov remains largely unimproved. This confirms our previous analysis: although increasing \( j \) effectively suppresses noise components, 
the condition number of \( Y_f Y_p^\top \) increases rapidly, leading to severe numerical instability. As a result, while the theoretical properties of SSI-Cov suggest 
that a larger \( j \) should enhance performance, in practical applications, increasing \( j \) does not necessarily yield better results. Instead, an optimal balance 
must be achieved between numerical stability and noise suppression to obtain reliable performance.  
In contrast, SSI-Data does not suffer from numerical stability issues due to its use of QR decomposition for projection calculations. Therefore, it can continually improve 
with increasing \( j \), leading to progressively better results.

Furthermore, through extensive numerical simulations, we observed that the SSI-Data method tends to underestimate the number of identified modal frequencies, occasionally 
missing some frequencies that actually exist. This issue is particularly pronounced in high-noise environments, leading to a conservative estimation bias. However, as \( j \) 
increases, this tendency gradually diminishes, and the occurrence of missing frequencies is significantly reduced.  
In contrast, the SSI-Cov method tends to overestimate the number of modal frequencies, often producing multiple closely spaced frequency values within a single estimation. Unlike 
SSI-Data, this characteristic does not improve with increasing \( j \); in fact, it may become more pronounced, resulting in a multi-peak distribution in the identified frequencies.  

PCSSI lies between SSI-Cov and SSI-Data in terms of estimation behavior. It generally does not miss frequencies but may occasionally produce slight overestimations. As a result, 
PCSSI's frequency estimates typically exhibit a single-peak distribution, though in rare cases, a weak secondary peak may appear. However, as \( j \) increases, this phenomenon 
gradually diminishes, enhancing the robustness of PCSSI.

Overall, as shown in the main text, when \( j \) is relatively small, numerical instability is not severe, making PCSSI the most effective method, followed by SSI-Cov, 
with SSI-Data performing the worst. When \( j \) is moderately large, PCSSI remains the best-performing method, while the performance of SSI-Cov and SSI-Data becomes 
comparable. However, when \( j \) is sufficiently large, PCSSI and SSI-Data exhibit similar performance, whereas the performance of SSI-Cov may deteriorate.  

These findings indicate that while increasing the data volume enhances the performance of all three methods, their responses to increasing \( j \) vary significantly due to 
their distinct numerical characteristics. Although the accuracy of modal analysis is often challenging to compare in practical engineering applications, the accuracy of system 
identification can be objectively assessed based on the prediction accuracy of time-domain responses. Since modal parameter identification is derived from the identified system 
parameters, the accuracy of system identification indirectly reflects the accuracy of modal analysis.  

In another study \cite{chen5097818adaptive}, we utilized measured acceleration data from the Hangzhou Bay Bridge to compare the system identification performance of NExT+ERA, 
SSI-Data, and SSI-Cov based on the prediction accuracy of time-domain responses. The results were consistent with those presented in this study.  
As a balanced approach between SSI-Cov and SSI-Data, PCSSI effectively mitigates numerical instability while maintaining high computational flexibility and reliable modal 
identification. Consequently, PCSSI demonstrates superior performance in most practical applications.

\section{Raw Experimental Modal Data from PCSSI and SSI Methods}\label{raw_data}

This section presents the raw experimental modal analysis data obtained directly from the PCSSI, SSI-Cov, and SSI-Data methods, summarized in 
Tables \ref{tab:PCSSI_data}, \ref{tab:SSI_COV_data}, and \ref{tab:SSI_Data_data}. These tables provide the modal frequencies identified for each mode 
across multiple experimental trials. The frequency values presented in the tables capture the inherent variability among different experiments, illustrating 
the stability and accuracy of the PCSSI, SSI-Cov, and SSI-Data methods in practical modal analyses.
Analyzing these datasets allows us to observe trends, identify potential anomalies or inconsistencies, and evaluate the reliability of the respective methods 
under repeated experimental conditions. Such analysis highlights the comparative performance and robustness of each identification method.
\begin{table}[ht]
\centering
\caption{PCSSI Modal Frequencies}
\label{tab:PCSSI_data}
\begin{tabular}{|c|c|c|c|c|c|c|c|c|c|}
\hline
\# & Mode 1 & Mode 2 & Mode 3 & Mode 4 & Mode 5 & Mode 6 & Mode 7 & Mode 8 & Mode 9 \\
\hline
1  & 98.09  & 153.37  & 227.41  & 320.19  & 322.33  & 416.84  & 512.25  & 602.57  & 728.63  \\
2  & 98.33  & 153.46  & 227.87  & 319.93  & 512.40  & 601.63  & 729.61  & 818.19  & 921.58  \\
3  & 98.20  & 153.19  & 229.49  & 319.65  & 415.96  & 513.48  & 601.07  & 729.42  & 813.67  \\
4  & 98.98  & 154.13  & 227.62  & 318.22  & 405.95  & 511.32  & 608.63  & 732.82  & 821.43  \\
5  & 99.26  & 154.17  & 228.15  & 318.55  & 510.82  & 608.41  & 730.63  & 822.90  & 921.15  \\
6  & 98.45  & 153.57  & 227.54  & 319.41  & 408.52  & 511.01  & 609.74  & 731.02  & 820.57  \\
7  & 98.51  & 153.56  & 227.77  & 319.34  & 408.63  & 510.85  & 608.44  & 730.07  & 822.82  \\
8  & 103.46 & 157.50  & 228.52  & 318.15  & 511.25  & 511.29  & 608.16  & 728.31  & 817.55  \\
9  & 99.32  & 155.42  & 227.97  & 318.14  & 318.20  & 407.39  & 511.58  & 607.65  & 730.63  \\
10 & 98.64  & 153.87  & 227.64  & 319.34  & 320.63  & 409.47  & 511.69  & 607.78  & 728.94  \\
11 & 100.19 & 156.18  & 228.48  & 318.18  & 407.22  & 511.78  & 607.78  & 730.10  & 818.44  \\
12 & 99.53  & 156.57  & 227.50  & 318.21  & 318.23  & 511.16  & 731.06  & 823.03  & 918.21  \\
13 & 99.78  & 154.80  & 228.29  & 318.71  & 412.79  & 511.62  & 609.13  & 729.58  & 822.49  \\
14 & 99.45  & 154.03  & 227.78  & 317.73  & 510.54  & 510.81  & 609.18  & 731.09  & 825.82  \\
15 & 98.26  & 153.37  & 227.44  & 319.17  & 512.50  & 606.55  & 728.81  & 820.11  & 918.52  \\
16 & 98.78  & 153.79  & 228.00  & 319.49  & 511.22  & 729.58  & 822.61  & 919.43  & 1037.53 \\
17 & 98.46  & 153.76  & 227.47  & 319.54  & 408.20  & 511.03  & 607.99  & 729.39  & 821.86  \\
18 & 101.74 & 155.77  & 227.09  & 317.52  & 318.93  & 505.98  & 603.42  & 720.48  & 811.85  \\
19 & 100.96 & 157.37  & 227.41  & 317.38  & 318.16  & 410.54  & 507.01  & 600.55  & 723.19  \\
20 & 101.51 & 227.34  & 317.33  & 318.12  & 410.18  & 507.14  & 846.23  & 846.37  & 1137.69 \\
\hline
\end{tabular}
\end{table}

\begin{table}[ht]
\centering
\caption{SSI-Cov Modal Frequencies}
\label{tab:SSI_COV_data}
\begin{tabular}{|c|c|c|c|c|c|c|c|c|c|}
\hline
\# & Mode 1 & Mode 2 & Mode 3 & Mode 4 & Mode 5 & Mode 6 & Mode 7 & Mode 8 & Mode 9 \\
\hline
1  & 96.85  & 152.68  & 222.95  & 317.07  & 412.14  & 513.26  & 513.33  & 611.57  & 725.83  \\
2  & 97.19  & 152.91  & 223.75  & 317.15  & 410.76  & 493.90  & 511.76  & 513.54  & 513.54  \\
3  & 96.99  & 152.80  & 223.18  & 317.11  & 411.24  & 512.47  & 513.30  & 612.26  & 724.88  \\
4  & 223.43 & 317.09  & 411.34  & 411.89  & 513.13  & 612.18  & 724.78  & 809.60  & 815.10  \\
5  & 153.00 & 223.51  & 317.10  & 411.84  & 513.05  & 612.20  & 724.60  & 809.52  & 919.19  \\
6  & 97.06  & 152.77  & 223.30  & 317.02  & 412.10  & 496.52  & 511.26  & 513.10  & 513.17  \\
7  & 97.15  & 152.80  & 223.31  & 317.04  & 411.88  & 511.25  & 513.03  & 611.59  & 724.78  \\
8  & 223.58 & 317.02  & 410.91  & 512.97  & 612.36  & 723.95  & 809.45  & 814.83  & 919.08  \\
9  & 316.99 & 512.96  & 611.89  & 724.30  & 809.50  & 814.77  & 919.03  & 1148.75 & 1225.05 \\
10 & 97.26  & 152.82  & 223.14  & 317.01  & 412.21  & 512.95  & 611.36  & 724.46  & 809.34  \\
11 & 223.36 & 316.99  & 411.80  & 512.84  & 611.83  & 724.23  & 809.63  & 918.92  & 1148.44 \\
12 & 222.99 & 316.91  & 412.45  & 512.82  & 611.05  & 724.62  & 810.69  & 814.39  & 918.49  \\
13 & 223.07 & 316.92  & 413.11  & 512.69  & 610.58  & 724.20  & 809.49  & 918.73  & 1025.41 \\
14 & 222.99 & 316.87  & 412.66  & 512.53  & 611.25  & 724.47  & 809.65  & 918.73  & 1024.94 \\
15 & 96.89  & 152.60  & 222.99  & 316.93  & 412.11  & 497.58  & 511.03  & 512.77  & 611.42  \\
16 & 223.23 & 316.96  & 411.91  & 512.73  & 611.72  & 724.34  & 809.51  & 814.70  & 919.03  \\
17 & 152.68 & 223.02  & 316.94  & 412.12  & 512.77  & 512.79  & 611.28  & 724.27  & 809.29  \\
18 & 97.65  & 153.49  & 154.34  & 224.33  & 317.26  & 317.33  & 513.36  & 513.46  & 613.54  \\
19 & 97.55  & 153.37  & 153.82  & 224.63  & 224.77  & 317.25  & 409.46  & 494.02  & 512.06  \\
20 & 153.42 & 153.68  & 224.93  & 317.28  & 408.94  & 409.82  & 492.65  & 512.17  & 513.30  \\
\hline
\end{tabular}
\end{table}

\begin{table}[ht]
\centering
\caption{SSI-Data Modal Frequencies}
\label{tab:SSI_Data_data}
\begin{tabular}{|c|c|c|c|c|c|c|c|c|c|}
\hline
\# & Mode 1 & Mode 2 & Mode 3 & Mode 4 & Mode 5 & Mode 6 & Mode 7 & Mode 8 & Mode 9 \\
\hline
1  & 97.88  & 151.27  & 223.74  & 317.36  & 410.97  & 511.95  & 609.23  & 918.46  & 1148.38  \\
2  & 98.29  & 152.00  & 224.74  & 317.20  & 317.85  & 410.84  & 511.04  & 610.37  & 919.46  \\
3  & 98.19  & 151.90  & 224.34  & 317.31  & 410.70  & 411.04  & 511.72  & 610.20  & 918.88  \\
4  & 224.23 & 317.16  & 317.71  & 411.17  & 610.50  & 725.05  & 817.73  & 1147.95 & 1148.12  \\
5  & 97.86  & 317.17  & 317.63  & 411.56  & 511.13  & 610.65  & 724.79  & 816.39  & 816.73  \\
6  & 97.71  & 224.30  & 317.11  & 317.74  & 411.56  & 510.94  & 609.35  & 725.31  & 919.30  \\
7  & 98.06  & 317.10  & 317.70  & 411.31  & 510.59  & 609.52  & 725.16  & 817.73  & 919.31  \\
8  & 317.07 & 409.98  & 511.02  & 609.22  & 724.02  & 724.47  & 816.63  & 918.88  & 1147.99  \\
9  & 97.75  & 151.83  & 317.05  & 317.66  & 411.06  & 510.63  & 609.95  & 724.70  & 817.07  \\
10 & 97.64  & 151.48  & 317.11  & 317.65  & 411.22  & 511.19  & 609.76  & 817.53  & 919.05  \\
11 & 97.69  & 152.06  & 317.07  & 317.66  & 411.02  & 511.08  & 609.86  & 918.76  & 1024.39  \\
12 & 97.74  & 317.06  & 317.71  & 411.24  & 511.14  & 608.96  & 918.65  & 1022.27 & 1065.06  \\
13 & 98.08  & 317.17  & 317.62  & 412.07  & 510.99  & 609.28  & 817.55  & 918.23  & 1065.46  \\
14 & 97.60  & 151.41  & 316.97  & 317.47  & 411.77  & 510.75  & 609.83  & 918.29  & 1147.81  \\
15 & 151.53 & 317.01  & 317.66  & 411.22  & 510.86  & 609.36  & 725.13  & 818.93  & 919.03  \\
16 & 97.71  & 151.42  & 317.22  & 411.63  & 510.33  & 609.81  & 724.57  & 816.75  & 1147.92  \\
17 & 97.84  & 151.52  & 224.02  & 317.04  & 317.62  & 411.25  & 511.02  & 609.32  & 816.78  \\
18 & 154.65 & 225.42  & 317.32  & 317.66  & -       & -       & -       & -       & -  \\
19 & 154.53 & 154.63  & 225.35  & 225.99  & 317.35  & 509.69  & 612.54  & 725.30  & 918.98  \\
20 & 154.74 & 225.97  & 317.36  & 408.26  & 484.98  & 510.27  & 513.18  & 611.35  & 724.39  \\
\hline
\end{tabular}
\end{table}

\bibliographystyle{elsarticle-num} 
\bibliography{refers}

\end{document}